\documentclass[a4paper,11pt]{article}
\pdfoutput=1 

\usepackage{jheppub} 
\usepackage{amsmath}
\usepackage{MnSymbol}
\allowdisplaybreaks[4]        
\usepackage{amssymb}
\usepackage{euscript}     
\usepackage{color}         
\usepackage{tensor}        
\usepackage{amsthm}
\usepackage{graphicx,subcaption}
 \usepackage{mathtools}	
\usepackage{braket,stmaryrd}
\usepackage{dsfont}
\usepackage{mathrsfs}

\usepackage{hyperref}
\hypersetup{colorlinks=true,allcolors=blue}

\usepackage[most]{tcolorbox}
\usepackage[normalem]{ulem}

\usepackage[header,title,page,titletoc]{appendix}  
\usepackage[T1]{fontenc} 
\usepackage[numbers]{natbib}

\makeatletter
\newsavebox{\@brx}
\newcommand{\lla}[1][]{\savebox{\@brx}{$\m@th{#1\langle}$}%
  \mathopen{\copy\@brx\mkern2mu\kern-0.9\wd\@brx\usebox{\@brx}}}
\newcommand{\rra}[1][]{\savebox{\@brx}{$\m@th{#1\rangle}$}%
  \mathclose{\copy\@brx\mkern2mu\kern-0.9\wd\@brx\usebox{\@brx}}}
\makeatother

\definecolor{browna}{rgb}{0.76,0.72,0.65}
\definecolor{brownb}{rgb}{0.71,0.69,0.65}
\definecolor{SpringGreen}{rgb}{0.95,0.97,0.95}
\definecolor{OliverGreen}{rgb}{0.09,0.34,0.09}
\definecolor{LeftGreen}{rgb}{0.13,0.54,0.13}
\definecolor{lightblue}{rgb}{0.1,0.7,1.0}

\newtcbtheorem[no counter]{theorem}{Theorem}{
	enhanced,
	sharp corners,
	attach boxed title to top left={
		yshifttext=-1mm
	},
	colback=white,
	colframe=browna,
	fonttitle=\bfseries,
	boxed title style={
		sharp corners,
		size=small,
		colback=browna,
		colframe=browna,
	} 
}{thm}

\newtcbtheorem[no counter]{Example}{Example}{
	enhanced,
	sharp corners,
	attach boxed title to top left={
		yshifttext=-1mm
	},
	colback=white,
	colframe=brownb,
	fonttitle=\bfseries,
	coltitle=white,
	boxed title style={
		sharp corners,
		size=small,
		colback=brownb,
		colframe=brownb,
	} 
}{exm}

\newtcbtheorem{myclaim}{Claim -- }
{    coltitle=OliverGreen,    
	colback=SpringGreen,    colframe=LeftGreen,    detach title,    boxrule=0pt,    leftrule=2pt,    attach title to upper,    sharp corners,    left=1mm,}
{claim}

\definecolor{browna}{rgb}{0.76,0.72,0.65}

\subheader{\today}

\title{Inner Product of Closed Universe from Gravitational Path-integral} 

\author{Masamichi Miyaji$^1$, Sunil Sake$^2$ and Kotaro Shinmyo$^3$}
\affiliation{$^1$RIKEN Center for Interdisciplinary Theoretical and Mathematical Sciences (iTHEMS),
RIKEN, 2-1 Hirosawa, Wako, Saitama 351-0198, Japan}
\affiliation{$^2$University of California Santa Cruz, Physics Department,
1156 High Street, Santa Cruz, CA 95064, USA}
\affiliation{$^3$Yukawa Institute for Theoretical Physics, Kyoto University,
Kyoto 606-8267, Japan}
\emailAdd{masamichi.miyaji@gmail.com, sunilsake1@gmail.com, kotaro.shinmyo@yukawa.kyoto-u.ac.jp}
\abstract{\ndsabstract}

\abstract{We construct the Hilbert space of a closed universe of fixed topology in anti-de Sitter (AdS) and de Sitter (dS) Jackiw–Teitelboim (JT) gravity from the gravitational path integral. Quantizing gravity involves several subtleties, including the choice of lapse contour, path-integral measure, operator ordering, global structure of the gauge group, the implementation of gravitational constraints and the gauge symmetries at initial and final timeslices. We develop a formalism that addresses these issues and determines the inner product of physical states, providing a framework to assign probabilities to cosmological wavefunctions. For a fixed topology and a lapse contour, we compute the inner product exactly and obtain the corresponding resolution of the identity in JT gravity. We demonstrate that the co-invariant and invariant Hilbert spaces in refined algebraic quantization arise from particular gauge fixing at the initial and final timeslices. In particular, the formalism provides precise meaning for a gravitational wavefunction with fixed geometric data as a co-invariant state, including its integral measure. These constructions can be formulated within the Hamiltonian BFV–BRST formalism. We also discuss the global structure of the Hilbert space, which differs from that of BF theory and depends on the prescription adopted near the big bang/crunch. Our results imply that the cylinder geometry captures the gravitational inner product for all perturbative states, including the Hartle–Hawking no-boundary state.}

\numberwithin{equation}{section}

\begin{document}
    \begin{flushright}
    RIKEN-iTHEMS-Report-26\\
    YITP-26-134
    \end{flushright}
	\maketitle
	\flushbottom


\section{Introduction}
\label{sec:introduction}

A quantum theory of a spatially closed universe has many properties that are different compared to universe with asymptotic boundary.  Due to the absence of boundary, the time translation is a gauge symmetry, and thus the Hamiltonian annihilates the physical state, namely the Wheeler--DeWitt equation \cite{DeWitt:1967yk} . On the other hand, the Wheeler--DeWitt equation alone does not by itself determine which solutions are physical states or what positive inner
product they carry \cite{DeWitt:1967yk,Hartle:1983ai}.  These questions are
especially sharp when the initial and final spatial slices are themselves
closed universes: time translation of such a slice is a gauge transformation rather than
an evolution generated by an asymptotic charge, and the gravitational path
integral between two slices should consequently be interpreted as a physical
pairing rather than as an ordinary transition amplitude in an external time.

A useful canonical language in constructing Hilbert space is refined algebraic
quantization \cite{Giulini:1998rk,Marolf:2000iq}. One first constructs an auxiliary Hilbert space
$\mathcal D_{\rm aux}$ before imposing the gravitational constraints.  For a
single constraint $\hat{\mathcal H}$, the physical information may then be
described either by invariant, generally distributional, states satisfying
$\hat{\mathcal H}\Psi=0$, or by co-invariant states
$[f]\in\mathcal D_{\rm aux}/\hat{\mathcal H}\mathcal D_{\rm aux}$.
The inner product in the co-invariant state is induced by \cite{Marolf:1995cn, Hartle:1997dc, Giulini:1998rk, Marolf:2000iq, Held:2024rmg, Held:2025mai, Alonso-Monsalve:2025lvt}
\begin{equation}
    \langle\langle f_2|f_1\rangle\rangle_{\rm co-inv}=\langle f_2|\eta|f_1\rangle_{\rm aux},~~~\eta\equiv \int_{\mathcal C}dT\,
    e^{-iT\hat{\mathcal H}},
    \label{eq:intro-rigging-map}
\end{equation}
after quotienting null states and taking the completion, where $\mathcal C$ is the appropriate group-averaging contour. The rigging map $\eta$ sends co-invariant classes to
generally distributional invariant states, although it does not preserve the
auxiliary inner product. In generic interacting theory, the auxiliary Hilbert space is described by the Hamiltonian BRST/BFV formulation \cite{Fradkin:1975cq, Batalin:1977pb, Henneaux:1992ig}. Physical states are described by BRST cohomology classes \cite{Kugo:1977zq,Kugo:1979gm}. 
The invariant and co-invariant descriptions correspond to \emph{distinct} cohomology classes \cite{Witten:2022xxp,Held:2025mai}. The formulation is also used in the reference frame context \cite{Hoehn:2019fsy}.

The primary objective of this paper is to derive the algebraic quantization from the manifestly generally covariant gravitational path-integral, which suffers a variety of mysteries and ambiguities. In gravitational path-integral, it is important to specify and handle the path-integral measure, which is crucial in maintaining the covariance, and is not apparent in the canonical/Dirac quantization which has the operator ordering ambiguity \cite{Halliwell:1988wc} and does not possess manifest covariance. It is also important to specify the contour of integration for the field, especially the lapse contour. Finally, we have to clarify what sort of boundary condition we should impose at the future and past boundaries, which are often imposed without careful treatments. 

In this paper, we will address these issues using the
Jackiw--Teitelboim (JT) gravity \cite{Teitelboim:1983ux, Jackiw:1984je}.  The theory has no local propagating
gravitational degrees of freedom, yet admits nontrivial global phase-space
as well as topology-changing process.  It can be formulated both as a first-order gauge-theory
description or in familiar second-order dilaton-gravity framework
\cite{Teitelboim:1983ux,Jackiw:1984je,Witten:1991we}. Setup with open but no boundary term at asymptotic infinity was investigated in \cite{Henneaux:1985nw, Louis-Martinez:1993bge, Grumiller:2002nm}. With asymptotically
AdS boundary conditions, JT gravity captures the universal nearly-$AdS_2$
dynamics and has a genus expansion generated by a double-scaled matrix
integral \cite{Almheiri:2014cka, Maldacena:2016upp, Engelsoy:2016xyb, Kitaev:2017awl, Saad:2019lba, Moitra:2019bub, Moitra:2021uiv}.  Its phase space, canonical Hilbert spaces
and observable algebras have simple structure and were studied in 
\cite{Henneaux:1985nw, Harlow:2018tqv, Penington:2023dql}.  For positive cosmological constant, its perturbative and matrix model description were investigated in \cite{Maldacena:2019cbz, Cotler:2019nbi, Cotler:2024xzz}, and the structure of classical phase space on spatial
circle topology has been classified by \cite{Alonso-Monsalve:2024oii, Held:2024rmg, Maxfield:2026ynw}, and its canonical quantization except for $n-$dS BH sector has been established in \cite{Held:2024rmg}. The solutions for the Wheeler-deWitt equation for the big-bang and big-crunch sectors have been worked out in \cite{Moitra:2022glw, Nanda:2023wne, Dey:2025osp}. 
Lorentzian quantum cosmology has been studied using minisuperspace path integrals, with particular attention to lapse contours and boundary conditions \cite{Feldbrugge:2017kzv,Feldbrugge:2017fcc,Feldbrugge:2017mbc,DiazDorronsoro:2017hti,DiazDorronsoro:2018wro,Janssen:2019sex,Vilenkin:2018dch,Vilenkin:2018oja,Halliwell:2018ejl,DiTucci:2019dji,DiTucci:2019bui,Matsui:2023tkw,Ailiga:2023wzl,Honda:2024Resurgence,ChouNishimura:2024MonteCarlo,Honda:2026Dimensional,HondaShinmyo:2026Stokes,Matsui:2026pax}.
For JT gravity and related cosmological models, these methods have been applied in \cite{FanarasVilenkin:2021JTKS,FanarasVilenkin:2022TunnelingKS,Honda:2024hdr,Matsui:2025Determinants}.

\subsection*{Summary of results}

Our objective is to derive the physical Hilbert space directly from the
Lorentzian JT path integral on a cylinder whose boundaries are closed,
unmarked spacelike circles defining the initial and final conditions.  We discuss the treatment for both negative and positive
cosmological constants. The important feature of gravity in closed universe is that the theory possesses the gauge symmetry which is non-vanishing and not constrained at the initial and final
boundaries; this is the crucial difference from the standard asymptotic JT path
integral, in which boundary reparametrizations that act on a marked boundary
can carry physical information. This boundary gauge symmetry is crucial in the understanding of the Hilbert space of closed universe, as will be described in the following. 
\begin{figure}[t]
	\begin{center}
		\includegraphics[width=3cm,clip]{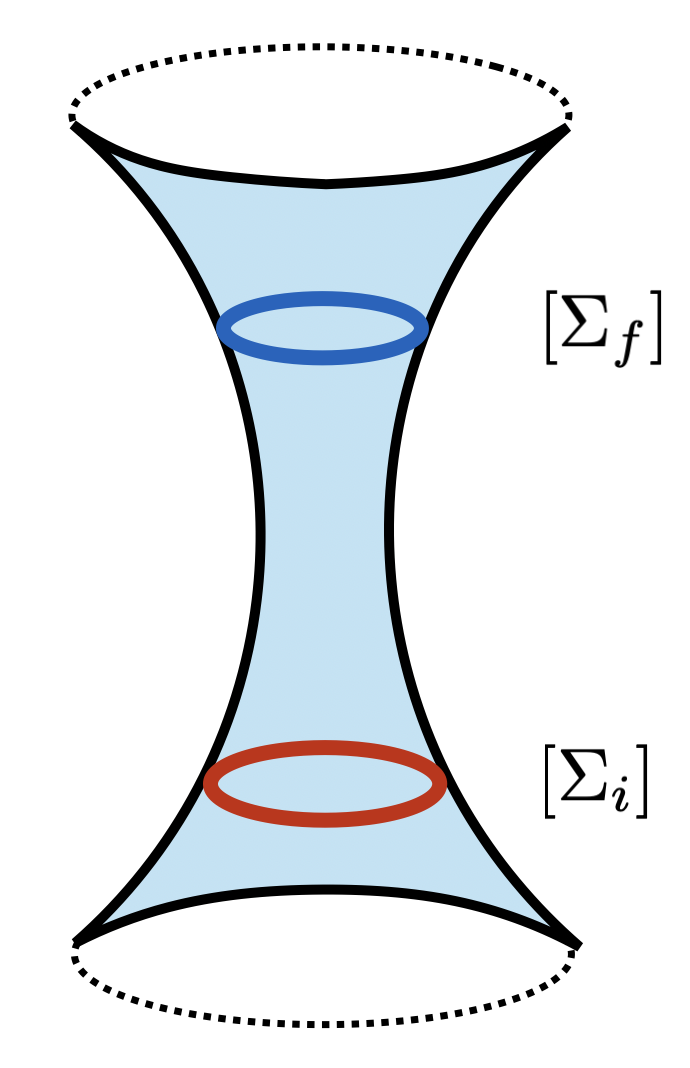}
        \caption{Spacetime geometry on which gravitational path-integral is performed for the inner product between initial state $[\Sigma_i]$ and the final state $[\Sigma_f]$. }
	    \label{fig:lapsecontour}
    \end{center}
\end{figure}

\paragraph{Off-shell boundary gauge symmetry.}
We begin with the first-order formalism with the functional measure from
ultralocal metrics on the gravitational fields. In this theory, the density of the field-space and of the gauge-orbit density cancel with each other, so that the measure simplifies. This allows us to show that the canonically quantized theory with standard phase space Liouville measure is indeed \emph{covariant}. On top of this, we first show that after gauge-fixing the local Lorentz boost, the first-order theory is equivalent to the second-order phase space path-integral including the path-integral measure. The first-order theory can be written in terms of BF theory variables. Although the BF theory has globally distinct gauge degree of freedom compared to JT gravity, it is useful to analyze the infinitesimal gauge symmetry, in particular the boundary off-shell gauge symmetry is manifest in BF language.

We impose a
gauge in which the spatial metric and extrinsic curvature are
constant on each slice. With this choice of gauge we explicitly confirm that the spatial nonzero-mode Faddeev--Popov determinant cancels
the Jacobian generated by the spatial nonzero-mode constraint delta functions.  On a
regular gauge patch, away from the exceptional $n$-dS loci where the
Faddeev--Popov operator develops zero modes, all nonzero Fourier modes can
therefore be integrated out exactly. Thus the full path-integral can be performed exactly, without relying on assumptions such as mini-super space nor semi-classical approximation. 

We clarify that the apparent two formulations of states, namely co-invariant and invariant Hilbert spaces, comes from distinct gauge fixing of the boundary gauge symmetry.
\paragraph{Co-invariant Hilbert space from boundary gauge fixing.}

 The co-invariant gauge-fixing first set the diffeomorphism vanishing at the boundary. In this case the fields at the boundary are completely fixed, and the resulting inner product is
\begin{equation}
    K_{\rm co}(a_f,k_f;a_i,k_i)
    =
    \left\langle a_f,k_f\right|
    \int_{\mathbb{R}-i\epsilon}dN_0\,
    e^{-iN_0\hat{\mathcal H}_0}
    \left|a_i,k_i\right\rangle_{\rm aux},
    \label{eq:intro-group-averaged-kernel}
\end{equation}
with $\hat{\mathcal H}_0$ the homogeneous Hamiltonian constraint, and $N_0$ the homogeneous lapse. $a$ is length of the spatial slice and $k$ is related to the extrinsic curvature of the slice by a factor of $a$. This is the group-averaging inner product of \cite{Giulini:1998rk, Marolf:2000iq}, which has been derived in dS JT gravity from canonical quantization \cite{Held:2024rmg, Held:2025mai}. Here we directly derive this result from the gravitational path-integral, clarifying the importance of boundary gauge-fixing. 

We will argue the lapse contour $\mathcal{C}$ which is slightly deformed from $\mathbb{R}$, such as $\mathbb{R}-i\epsilon$ is plausible from the absence of degenerate point, implementation of the Hamitonian constraint \cite{Halliwell:1988wc}, interchange of order of path-integral \cite{Banihashemi:2024aal} as well as the Kontsevich-Segal-Witten condition \cite{Kontsevich:2021dmb, Witten:2021nzp}. See also \cite{Casali:2021ewu}.

We note the important difference of the co-invariant gauge fixing of boundary fields from direct imposition of boundary condition \cite{Halliwell:1988wc}. The difference comes from the Hamiltonian constraint and the path-integral measure. The Hamiltonian constraint dictates timeslices are gauge-equivalent, so naive specification of boundary conditions for all fields is over-complete. Furthermore, the correct measure consistent with the gauge-invariant path-integral measure needs to be specified for boundary condition, which is necessary to identify the resolution of identity and is often overlooked. By taking into account of the boundary gauge symmetry, we can capture the correct measure for the boundary condition, therefore the resolution of the identity of the Hilbert space.

\paragraph{Invariant Hilbert space and the resolution of identity.}

We can instead take the boundary gauge-fixing where the remaining gauge symmetry has vanishing time derivative. By construction, it is clear that the resulting inner product and therefore the Hilbert space are equivalent to those of the co-invariant one. Introducing the gauge-invariant coordinate
$G_1=(a^2\pm k^2)/2$ with $+$ for AdS and $-$ for dS, and the discrete label $\alpha$ for connected orbit, we find the physical inner product
\begin{equation}
    \langle\langle G_1',\alpha'|G_1,\alpha\rangle\rangle_{\rm inv}
    =\delta(G_1'-G_1)\delta_{\alpha,\alpha'},
    \label{eq:intro-reduced-kernel}
\end{equation}
where $|G_1,\alpha\rangle\rangle_{\rm inv}$ is the invariant physical state.
Equivalently, in a length coordinate this becomes
$\delta(b-b')/|b|$ on each connected component. This expression for the inner product defines the physical Hilbert space. Although the rigging map does not preserve the
auxiliary inner product, we show that $\eta$ maps the invariant physical inner product to the co-invariant group-averaging inner product, providing an isometry.

The global reduced phase spaces depend on the sign of the cosmological constant
and on whether the spatial vielbein is regarded as signed, whether spatial
orientation reversal is gauged, or whether one restricts to positive scale
factor from the outset.  We spell out these alternatives for closed AdS, the
global-dS sector, and the dS big-bang/big-crunch sectors.  The last
prescription requires additional boundary data whenever a Hamiltonian orbit
reaches the singular locus $a=0$.  

\paragraph{Lessons for gravitational inner product.}
While our gravitational path-integral is over smooth spacetimes, the Hilbert space contains the Hartle--Hawking state \cite{Hartle:1983ai} naturally without including disconnected disk geometry. When the Lorentzian geometry can be continued to Euclidean half sphere, regularity selects a delta function supported at the unit-holonomy
configuration, which is indeed contained as a
special state in the Hilbert space we constructed.

\paragraph{Organization.}
The paper is organized as follows.  In
Section~\ref{sec:first-second-order} we develop the first- and second-order
Lorentzian formulations, including boundary conditions, gauge transformations,
and functional measures.  Section~\ref{sec:gauge-fixed-path-integral}
performs the gauge-fixed path integral. Section ~\ref{sec:zero-mode-brst} explains the Hamiltonian BRST description of the system. In
Section~\ref{sec:perturbative-hilbert} we construct the fixed-topology
physical Hilbert spaces in the AdS and dS sectors and discuss the
Hartle--Hawking state. Several technical derivations, including the worldline analogue and the
complete cancellation of nonzero-mode determinants, are collected in the
appendices.


\section{First and second order formalisms of Lorentzian JT gravity}\label{sec:first-second-order}

We begin with the first order formalism of JT gravity in closed universe with cylinder topology. We explicitly keep track of path-integral measures and show the equivalence; ultra-local measure of the first / second order formalism and the configuration space, and the flat canonical measure of the phase space.  

\subsection{First order formalism}

We consider Lorentz signature spacetime with cylinder topology, with spatial section given by a circle. The cylinder has two spacelike boundaries, the initial and the final boundary. The zweibeins and the spin connection are
\begin{equation}
    \begin{split}
        e^a&=e^{a}_\mu dx^\mu,~\omega^{ab}=\epsilon^{ab}\omega,~\epsilon_{01}=1,~\eta_{00}=-\eta_{11}=-1.
    \end{split}
\end{equation}
Then the JT gravity action with fixed boundary length and extrinsic curvature in the first order formalism is
\begin{equation}\label{eq:actionfirstorder}
    I_\pm=\frac{1}{4\pi}\int \Big(2\Phi(-d\omega\pm e^0\wedge e^1)+2X^0(de^0-\omega\wedge e^1)+2X^1(de^1-\omega\wedge e^0)\Big),
\end{equation}
where $+$ is for AdS and $-$ for dS. In the rest of the paper, the top sign corresponds to AdS and the bottom one to dS unless otherwise specified. Integrating out $X^0,~X^1$ yields the torsion free condition
\begin{equation}
    \begin{split}
        de^a+\omega^a{}_b\wedge e^b=0.
    \end{split}
\end{equation}
Since $d\omega=-\frac{R}{2}~\text{Vol}$, we can see the action is JT gravity action without boundary term. 
Further integrating out dilaton $\Phi$ fixes the spacetime geometry to be
\begin{equation}
    R\pm 2=0.
\end{equation}

The action can conveniently cast into the form of a $PSL(2,\mathbb{R})$ BF theory, in terms of gauge field $A$, field strength $F=dA+A\wedge A$ and multiplier $B$. For AdS, we take
\begin{equation}
    A=
    \frac{1}{2}\begin{pmatrix}
    e^1 & e^0+\omega\\
    -e^0+\omega & -e^1
    \end{pmatrix},
    ~B=
    \begin{pmatrix}
    X^1 & -X^0-\Phi\\
    X^0-\Phi & -X^1
    \end{pmatrix},
\end{equation}
while for dS, we take
\begin{equation}
    A=
    \frac{1}{2}\begin{pmatrix}
    e^0 & e^1+\omega\\
    -e^1+\omega & -e^0
    \end{pmatrix},
    ~B=
    \begin{pmatrix}
    X^0 & -X^1-\Phi\\
    X^1-\Phi & -X^0
    \end{pmatrix}.
\end{equation}
In either case, we have \footnote{For AdS, we have
\begin{equation}
    F=
    \frac{1}{2}
    \begin{pmatrix}
    (de^1-\omega\wedge e^0) & (de^0-\omega\wedge e^1)+(d\omega-e^0\wedge e^1)\\
    -(de^0-\omega\wedge e^1)+(d\omega-e^0\wedge e^1) & -(de^1-\omega\wedge e^0)
    \end{pmatrix}.
\end{equation}
For dS, we have
\begin{equation}
    F=
    \frac{1}{2}
    \begin{pmatrix}
    (de^0-\omega\wedge e^1) & (de^1-\omega\wedge e^0)+(d\omega+e^0\wedge e^1)\\
    -(de^1-\omega\wedge e^0)+(d\omega+e^0\wedge e^1) & -(de^0-\omega\wedge e^1)
    \end{pmatrix}.
\end{equation}
}
\begin{equation}\label{eq:bfaction}
    \frac{1}{2\pi}\int \text{Tr}[BF]=I_\pm,
\end{equation}
where $+$is for AdS and $-$ for dS. Integrating out $B$ yields the flatness condition $F=0$, which is equivalent to the torsion free condition and $R\pm2=0$. This form of theory was proposed and its canonical quantization was studied in \cite{Isler:1989hq, Chamseddine:1989yz}.

The BF theory gauge field has the field-space metric
\begin{equation}
    G(\sigma,\eta)=2\int\text{Tr}[\sigma\wedge*\eta],
\end{equation}
where $*$ is the Hodge-star \footnote{Here we avoid the use of $T$ which flips the sign of the negative eigenvalue of the metric. This is because such transformation will break the local Lorentz boost symmetry. $T$ operation is allowed only after we fix the local Lorentz.} \cite{Witten:1991we, Saad:2019lba}. We emphasize, however, the gauge symmetry of the gravitational theory matches with that of $PSL(2,\mathbb{R})$ BF theory only in small neighborhood of identity. The global structure of the two gauge theories do not coincide, as we will explain later in this section.

\subsubsection*{ADM variables}

Let us introduce ADM variables \cite{Arnowitt:1962hi}. We use $2\pi$-periodic spatial coordinate $\theta$
\begin{equation}
    e^0=Ndt+Md\theta,~e^1=a(d\theta+N_\perp dt),~\omega=vdt+kd\theta,~\theta\sim \theta+2\pi,
\end{equation}
here $N,~N_\perp$ are lapse and shift, respectively. We also denote
\begin{equation}
    p:=X^1.
\end{equation}
The JT action in terms of ADM variables is
\begin{equation}\label{eq:firstorderaction}
\begin{split}
    I_\pm&
    =\frac{1}{2\pi}\int dtd\theta 
    \Big(\Phi(-\dot{k}+v'\pm (Na-aMN_\perp))
    +X^0(-av+N_\perp ak-N'+\dot{M})
    +p(\dot{a}-(N_\perp a)'+Nk-Mv)\Big)\\
    &=\frac{1}{2\pi}\int dtd\theta \Big(X^0\dot{M}+p\dot{a}-\Phi\dot{k}-v\mathcal{H}_s-N\mathcal{H}-N_\perp\mathcal{H}_\perp\Big),
    \end{split}
\end{equation}
with the gravitational constraints
\begin{equation}
    \begin{cases}
        \mathcal{H}&=-pk\mp a\Phi-X'^0,\\
        \mathcal{H}_\perp&=-X^0ak-p'a\pm a\Phi M,\\
        \mathcal{H}_s&=\Phi'+aX^0+pM.
    \end{cases}
\end{equation}
Using the Hodge dual $*e^0=e^1,~*e^1=e^{0}$, the metric 
on the space of gauge field fluctuations is \footnote{If we write in terms of components, we have \begin{equation}
\begin{split}
    G^A_\mathcal{F}(\delta_1A,\delta_2A)
    &=\int dtdx\Big(\mp\frac{(a^2-M^2)\delta_1N\delta_2N+(MN-a^2N_\perp)(\delta_1M\delta_2N+\delta_2M\delta_1N)+(a^2N_\perp^2-N^2)\delta_1M\delta_2M}{a(N-MN_\perp)}\\
    &\pm
    \frac{a^2(a^2-M^2)\delta_1N_\perp\delta_2N_\perp+Ma(N-MN_\perp)(\delta_1a\delta_2N_\perp+\delta_1N_\perp\delta_2a)-(N-MN_\perp)^2\delta_1a\delta_2a}{a(N-MN_\perp)}\\
    &
    +
    \frac{(a^2-M^2)\delta_1v\delta_2v+(MN-a^2N_\perp)(\delta_1k\delta_2v+\delta_2k\delta_1v)+(a^2N_\perp^2-N^2)\delta_1k\delta_2k}{Na-MaN_\perp}   \Big).
\end{split}
\end{equation}}
\begin{equation}\label{eq:metricA}
\begin{split}
    G^A_\mathcal{F}(\delta_1A,\delta_2A)
    &= \int (\mp\delta_1 e^0\wedge*\delta_2e^0\pm\delta_1 e^1\wedge*\delta_2e^1+\delta_1 \omega\wedge*\delta_2\omega)
\end{split}
\end{equation}
The inner product for field $B$ is similarly defined as
\begin{equation}\label{eq:metricX}
\begin{split}
    G^X_\mathcal{F}(\delta_1X,\delta_2X)
    &=\frac{1}{2}\int dtd\theta (Na-MaN_\perp)\Big(\mp(\delta_1 X^0)(\delta_2 X^0)\pm(\delta_1 p)(\delta_2 p)+(\delta_1 \Phi)(\delta_2 \Phi) \Big).
\end{split}
\end{equation}
We note that these metric are distinct from the metric induced from the symplectic form associated with the action \eqref{eq:firstorderaction}, as the former induces non-trivial volume density while the latter gives a flat line-element. We note only physical phase space measure should be flat, but not necessarily the extended field space measure before the reduction.

\subsubsection*{Diffeomorphism}

The infinitesimal diffeomorphism is given by the infinitesimal gauge transformation  $A\rightarrow g^{-1}(A+d)g,~B\rightarrow g^{-1}Bg$ with
\begin{equation}
    g_{\text{AdS}}=1+
    \frac{1}{2}\begin{pmatrix}
    a\xi_\perp & \xi-s\\
    -\xi-s & -a\xi_\perp
    \end{pmatrix},
    ~g_{\text{dS}}=1+
    \frac{1}{2}\begin{pmatrix}
    \xi & a\xi_\perp-s\\
    -a\xi_\perp-s & -\xi
    \end{pmatrix}.
\end{equation}
The corresponding transformation of various fields is given by 
\begin{equation}\label{eq:diffgeneral}
\begin{cases}
    &\delta a=-k\xi+\Big(a\xi_\perp\Big)'-Ms
    ,~~~\delta p=X^0s\pm\Phi\xi,
    \\    
    &\delta \Phi=-p\xi-aX^0\xi_\perp,~~~~~~~~~~
    \delta k=-s'\pm a\xi\mp aM\xi_\perp,
    \\
    &\delta X^0=sp\mp a\Phi\xi_\perp,~~~~~~~~~~~~
    \delta M=\xi'-as-ak\xi_\perp,\\
    \end{cases}
\end{equation}
and for the Lagrange-multiplier fields
\begin{equation}\label{eq:diffgeneralmultipliers}
\begin{cases}
    \delta v&=-\dot{s}\mp a(N\xi_\perp-N_\perp\xi),
    \\ 
    \delta N&=\dot{\xi}-va\xi_\perp-aN_\perp s,
    \\
    \delta N_\perp&=\Big(\dot{\xi}_\perp-N_\perp\xi_\perp'+\xi_\perp\frac{\dot{a}-N_\perp a'}{a}\Big)+\frac{-v+N_\perp k}{a}\xi-\frac{N-N_\perp M}{a}s.
    \end{cases}
\end{equation}
We note that the BF action \eqref{eq:bfaction} is manifestly gauge-invariant, \emph{for arbitrary gauge transformation parameters, which can be non-trivial even at boundary}. Under this gauge transformation, the gravitational constraints transform as
\begin{equation}
      \begin{cases}
         \delta \mathcal{H}_s&=\Big(\frac{\mathcal{H}_\perp}{a}\Big)\xi+\mathcal{H}(a\xi_\perp),\\
         \delta \mathcal{H}&=\Big(\frac{\mathcal{H}_\perp}{a}\Big)s\pm \mathcal{H}_s(a\xi_\perp),\\
         \delta \Big(\frac{\mathcal{H}_\perp}{a}\Big)&=\mathcal{H}s\mp \mathcal{H}_s\xi.
      \end{cases}
\end{equation}
Furthermore, the path-integral measure is gauge-invariant. Since the Hodge star leaves the inner products \eqref{eq:diffgeneral} and \eqref{eq:diffgeneralmultipliers} invariant. We can see that $\xi$ is a local time-translation, $\xi_\perp$ a local spatial-translation, and $s$ is a local Lorentz boost. Since spatial direction is $2\pi$ periodic, only $\xi_{0\perp}:=\int\frac{d\theta}{2\pi}\xi_\perp$ is constrained to be $2\pi$-periodic.

\paragraph{Global structure of gauge group: BF theory vs gravity.}

Let us emphasize that the global structures of gauge group of $PSL(2,\mathbb{R})$ BF theory and gravitational theory are different, although infinitesimal structure is common. In $PSL(2,\mathbb{R})$ group, the only compact direction is $\exp(\frac{ix}{2}\sigma_2)$ and is $2\pi$-periodic. In both AdS and dS gravity, such periodicity is absent. Furthermore, in gravity there is a periodicity $\xi_{0\perp}\sim\xi_{0\perp}+2\pi$ in the gauge transformation, whose counterpart in BF theory is a non-trivial gauge transformation.

\subsection{Boundary condition, gauge-fixing and path-integral measure}


\subsubsection*{Boundary condition and unmarked boundary}

At each boundary, we fix the values of the first-order fields, denoted collectively $\varphi$, only up to $PSL(2,\mathbb{R})$ gauge transformations.
Equivalently, we specify gauge-invariant data $O(\varphi)$ that distinguish the boundary gauge orbits by inserting
\begin{equation}
\begin{split}
    &\delta\!\left(
    O(\varphi|_{t=0,1})-O(\varphi_{i,f})
    \right)
    \\
    &\qquad\equiv
    \delta\!\left(
    O(\varphi|_{t=0})-O(\varphi_i)
    \right)
    \delta\!\left(
    O(\varphi|_{t=1})-O(\varphi_f)
    \right).
\end{split}
\end{equation}
A boundary gauge choice can then be used to select convenient representatives of these orbits. To specify the orbit, we use the  gauge-invariant data $(G_1,\alpha)$. Here $\alpha$ labels connected components of the gauge-orbit with the same $G_1$. To define $G_1$, we first consider holonomy
\begin{equation}
\begin{split}
    U&\equiv \mathcal{P}e^{-\int_\mathcal{C} A},
\end{split}
\end{equation}
which transforms as $gUg^{-1}$. Thus 
\begin{equation}
    G_1\equiv \pm\frac{\text{Tr}[(\log(U))^2]}{4\pi^2},
    \label{G1def}
\end{equation}
is a gauge invariant quantity.

Let us consider sewing two path-integrals to construct one path-integral. First, the two boundary conditions need to match with each other, thus in particular they are in the same gauge-orbit. On each gauge-orbit, we gauge fix and select one representative. Then we integrate over gauge-orbits.

Let \(M_{21}=M_2\cup_\Sigma M_1\), where the two copies of the
gluing surface \(\Sigma\) carry opposite orientations.  The
restrictions of the two field configurations onto \(\Sigma\) must
be identical.  If \(O\) is a
complete set of gauge-invariant labels for the boundary gauge
orbits and the corresponding physical states are
delta-normalized with respect to \(d\mu_\Sigma(O)\), sewing is
defined by
\begin{equation}
    K_{M_{21}}(O_2,O_0)
    =
    \int d\mu_\Sigma(O)\,
    K_{M_2}(O_2,O)\,
    K_{M_1}(O,O_0),
    \label{eq:sewing-over-boundary-orbits}
\end{equation}
for amplitude $K_M(O_1,O_2)$. Equivalently, on a Gribov-free gauge slice, the quotient measure
may be represented by the Faddeev--Popov measure
\begin{equation}
    d\mu_\Sigma(O)
    =
    D\phi_\Sigma\,
    \delta[\chi(\phi_\Sigma)]\,
    \left|
        \det\frac{\delta\chi}{\delta\epsilon}
    \right|,
\end{equation}
where the diagonal gauge symmetry acting on the gluing interface
is divided out only once. For BV-BFV treatment, see
\cite{Cattaneo:2015vsa}.

Because the boundary circle has a $U(1)$ isometry, it is important to distinguish marked and unmarked boundaries.
The moduli space of an unmarked hyperbolic cylinder is $\mathbb{R}_{>0}$, whereas a marked cylinder has the additional angular modulus and hence the moduli space $\mathbb{R}_{>0}\times S^1$.
In our gravitational path integral, the initial and final boundaries are unmarked, as required by diffeomorphism invariance.
The boundary $U(1)$ isometry is therefore gauged, and its volume is divided out.
Consequently, a boundary state is specified only by its length, $a$, and the values of $k$ and $M$.
For a marked boundary, the position of the mark is additional physical data, and the corresponding $U(1)$ volume is not divided out.


\subsubsection*{Gauge-fixed path-integral measure in gravity}

Let us denote collectively all first-order fields by $\varphi^I=(a,k,v,M,N,N_\perp,X^0,p,\Phi).$
The formal gravitational path integral is
\begin{equation}\label{eq:formalpathintegral}
K(O(\varphi|_{t=1}),O(\varphi|_{t=0}))=\int \frac{[D\varphi]\sqrt{|\det G_{\cal F}[\varphi]|}}
{\operatorname{Vol}({\cal G}[\varphi])}
\,e^{iI[\varphi]}\delta(O(\varphi|_{t=0,1})-O(\varphi_{i,f})),
\end{equation}
where $[D\varphi]\sqrt{|\det G_{\cal F}[\varphi]|}$
is the field-space measure induced from \eqref{eq:metricA} and
\eqref{eq:metricX}. Here we denote the flat measure as $[D\varphi]$, more explicitly
\begin{equation}
    [D\varphi]\equiv [DN][DN_\perp][Dv][Da][Dp][D\Phi][Dk][DM][DX^0].
\end{equation}
The gauge group is locally parametrized by
$\epsilon^\alpha=(\xi,\xi_\perp,s)$. Note $\epsilon^\alpha$ is not subject to any boundary condition. The invariant metric of the gauge group ${\cal G}[\varphi]$ at the identity element is
\begin{equation}
\gamma(\epsilon_1,\epsilon_2)
=
\int dtd\theta \,(Na-MaN_\perp)
\left[\mp\xi_1\xi_2\pm(a\xi_{1\perp})(a\xi_{2\perp})+s_1s_2\right].
\end{equation}
We denote this metric by $\gamma_{\alpha\beta}[\varphi]$. Note that this metric is that of $\mathfrak{sl}(2,\mathbb{R})$, and is different from substitution of \eqref{eq:diffgeneral} into \eqref{eq:metricA} which contains derivatives. In the
coordinate system $(\xi,\xi_\perp,s)$, the corresponding local density in the bulk is
formally
\begin{equation}
\sqrt{|\det\gamma|}_{(t,\theta )}
=
|a|\,|Na-MaN_\perp|^{3/2}.
\end{equation}
Only the rigid spatial translation zero mode is compact, with period $2\pi$;
the local gauge parameters are otherwise noncompact infinitesimal coordinates
on the gauge orbit.  Note this structure is different from the $PSL(2,\mathbb{R})$ BF theory.

We impose the gauge-fixing conditions $(M,G)=0$.
With the invariant measure on ${\cal G}$, the gauge-fixing identity
is
\begin{equation}
1
=\frac{\Delta_{\rm FP}^{(M,G)}[\varphi]}{\sqrt{|\det{}\gamma[\varphi]|}}
\int_{{\cal G}}
[D\epsilon]\sqrt{|\det{}\gamma_{\rm }[\varphi^\epsilon]|}\,
\delta[M(\varphi^\epsilon)]\delta[G(\varphi^\epsilon)]\,
,
\end{equation}
where we define the Faddeev-Popov determinant
$\Delta_{\rm FP}^{(M,G)}[\varphi]:=
|\det{}{\left.
\frac{\delta(M,~G)^A[\varphi^\epsilon]}
{\delta\epsilon^\alpha}
\right|_{\epsilon=0}}|.$

Inserting this identity to \eqref{eq:formalpathintegral} gives
\begin{equation}\label{eq:fullpathintegral}
\begin{split}
K
&=
\int
[D\varphi] \frac{\sqrt{|\det G_{\cal F}[\varphi]|}}{\sqrt{|\det{}\gamma[\varphi]|}}\,
\delta[M]\delta[G]\,
\Delta_{\rm FP}^{(M,G)}[\varphi]
e^{iI[\varphi]}\delta(O(\varphi|_{t=0,1})-O(\varphi|_{i,f})) \\
&=\int
[D\varphi]
\delta[M]\delta[G]\,
\Delta_{\rm FP}^{(M,G)}[\varphi]
e^{iI[\varphi]}\delta(O(\varphi|_{t=0,1})-O(\varphi|_{i,f})) ,
\end{split}
\end{equation}
where we used that the density
$\sqrt{|\det G_{\cal F}|}$ agrees with the gauge-orbit density
$\sqrt{|\det{}\gamma|}$
\begin{equation}\label{eq:cancellation}
\sqrt{|\det G_\mathcal{F}|}_{(t,\theta )}
=|a|\,|Na-MaN_\perp|^{3/2}=\sqrt{|\det\gamma|}_{(t,\theta )},
\end{equation}
as well as there is no boundary condition imposed, so that the cancellation of boundary measure has no ambiguity. Thus we see that the path-integral measure is \emph{flat} except for FP determinant, observed in a different form in \cite{Witten:1991we}. We can utilize this fact to cast the system into the canonical formalism. The uplift of this theory to Hamiltonian BFV/BRST formalism is done in appendix \ref{appendix:BRST}.

\subsection{Reduction to second-order formalism from the first-order formalism}

We will fix the local Lorentz symmetry by the gauge-fixing $M=0$. By integrating out auxiliary variables $X^0,~p$ and Lagrange multiplier $v$, we obtain the second order formalism for both configuration and phase space without losing manifest diffeomorphism invariance. 

\subsubsection*{Phase space path-integral for the second-order formalism}

In order to obtain the canonical formalism for the second order formalism, we consider the gauge fixing for $s$
\begin{equation}
    M=0,
\end{equation}
and integrate out $X^0$ and $v$. Integrating out $X^0$ introduces the factor
\begin{equation}
    \prod_{t,\theta }\Big(\delta(-av+N_\perp ak-N'))\Big).
\end{equation}
Integrating out $v$ yields the factor
\begin{equation}
    \prod_{t,\theta }\frac{1}{a},
\end{equation}
and the exact equation of motion
\begin{equation}\label{eq:exacteomforV}
    -av+N_\perp ak-N'=0.
\end{equation}
Note that if we integrate out $v$ first, we arrive at the constraint
\begin{equation}\label{eq:exacteomforX0}
    \mathcal{H}_s=\Phi'+X^0a=0.
\end{equation}
The invariant metric at the identity element is now
\begin{equation}
\gamma^P(\epsilon_1,\epsilon_2)
=
\int dtd\theta \,(Na-MaN_\perp)
\left[\mp\xi_1\xi_2\pm(a\xi_{1\perp})(a\xi_{2\perp})\right].
\end{equation}
the superscript $P$ indicating that it is being computed in phase space.
Combining with the  FP determinant and the above factor, we have
\begin{equation}
\begin{split}
    \Delta_{FP}^{(M,G)}\prod_{t,\theta }\Big{|}\frac{1}{a}\Big{|}    
    &=
    \prod_{t,\theta }
    \Big{|}\text{det}
    \begin{pmatrix}
    \frac{\delta M}{\delta s} & \frac{\delta G}{\delta s}\\
    \frac{\delta M}{\delta (\xi,\xi_\perp)}  & \frac{\delta G}{\delta (\xi,\xi_\perp)}
    \end{pmatrix}
    \frac{1}{a}\Big{|} 
    =
    \prod_{t,\theta }
    \Big{|}
    \text{det}
    \begin{pmatrix}
    -a & \frac{\delta G}{\delta s}\\
    \frac{\delta M}{\delta (\xi,\xi_\perp)}  & \frac{\delta G}{\delta (\xi,\xi_\perp)}
    \end{pmatrix}
    \frac{1}{a}\Big{|} 
    \\
    &=
    \prod_{t,\theta }
    \Big{|}\text{det}\Big(\frac{\delta G}{\delta (\xi,\xi_\perp)}-\frac{\delta G}{\delta s}\frac{1}{a}\frac{\partial (\xi'-ak\xi_\perp)}{\partial (\xi,\xi_\perp)}\Big)\Big{|}\equiv\Delta_{\text{FP}}^{(G)}.
\end{split}
\end{equation}
Since we are fixing the $s$ modes completely by the gauge-fixing $M=0$, and $s$ modes have nothing to do with the boundary condition, we have
\begin{equation}
    \sqrt{|\det{}\gamma
    [\varphi]|}_{(t,\theta )}=\sqrt{|\det{}\gamma^P
    [\varphi]|}_{(t,\theta )}|Na|^{1/2},
\end{equation}
everywhere including at the boundary. We also arrive at the phase space action
\begin{equation}
\begin{split}
    I_{\pm,\text{phase}}=\int dt\frac{d\theta}{2\pi}\Big[-\Phi\dot{k}+p\dot{a}-N\mathcal{H}-N_\perp\mathcal{H}_\perp \Big].
    \end{split}
\end{equation}
Here the phase-space constraints are
\begin{equation}
    \mathcal{H}=-kp\mp a\Phi+\left(\frac{\Phi'}{a}\right)',~\mathcal{H}_\perp=k\Phi'-ap',
\end{equation}
which can also be obtained by substitution of the constraint $\mathcal{H}_s$ in  $\mathcal{H}$ and $\mathcal{H}_\perp$. We thus arrive at the phase space path-integral in the second order formalism
\begin{equation}\label{eq:secondorderZ}
\begin{split}
     K&=\int[DaDkDpD\Phi DNDN_\perp]\delta(G)
     \Delta_{\text{FP}}^{(G)}~e^{iI_{\pm,\text{phase}}}\delta(O(\varphi|_{t=0,1})-O(\varphi|_{i,f})).
\end{split}
\end{equation}

The diffeomorphism in the phase space can be obtained from that of the first order formulation \eqref{eq:diffgeneral}\eqref{eq:diffgeneralmultipliers}, by substituting the gauge-fixing condition $M=0$, the exact equation of motions \eqref{eq:exacteomforV} and \eqref{eq:exacteomforX0}, and condition on gauge parameters $s=\frac{\xi'}{a}-k\xi_\perp$. The transformation is given by
\begin{equation}\label{eq:diffphasespace}
\begin{cases}
    \delta a&=-k\xi+\Big(a\xi_\perp\Big)'
    ,~~\delta p=\pm\Phi\xi-\frac{\Phi'}{a^2}\xi'+(p'+\frac{\mathcal{H}_\perp}{a})\xi_\perp,\\
    \delta \Phi&=-p\xi+\Phi'\xi_\perp,
    ~~~~\delta k=\Big(\pm a\xi-(\frac{\xi'}{a})'\Big)+(k\xi_\perp)',\\
    \delta N&=\dot{\xi}-N_\perp\xi'+N'\xi_\perp,
    \\
    \delta N_\perp&=\Big(\frac{-N\xi'+N'\xi}{a^2}\Big)+\Big(\dot{\xi}_\perp+N_\perp'\xi_\perp-N_\perp\xi_\perp'\Big)+\xi_\perp\frac{\dot{a}-(N_\perp a)'+Nk}{a}.    
    \end{cases}
\end{equation}
Note that the substitution is allowed because $v$ and $X^0$ appear at most linearly in the action, making the equation of motions \eqref{eq:exacteomforV} \eqref{eq:exacteomforX0} valid whenever evaluated in the path-integral with no insertions depending on $v$ nor $X^0$ \footnote{In the main text, we begin with the Hamiltonian BRST first order action with canonical variables containing $\tilde{N}_\perp=aN_\perp$, instead of $N_\perp$ itself, as explained in the appendix \ref{appendix:BRST}. The current gauge transformation \eqref{eq:diffphasespace} originates from reduction from such canonical variables. Instead, we can proceed with alternative canonical variables containing $N_\perp$ with redefined momentum $p_{\rm new}$. The same diffeomorphism transformation takes a different form as 
\begin{equation}
\begin{cases}
    \delta a&=-k\xi+\Big(a\xi_\perp\Big)'
    ,~~\delta p_{\rm new}=\pm\Phi\xi-\frac{\Phi'}{a^2}\xi'+p_{\rm new}'\xi_\perp,\\
    \delta \Phi&=-p_{\rm new}\xi+\Phi'\xi_\perp,
    ~~~~\delta k=\Big(\pm a\xi-(\frac{\xi'}{a})'\Big)+(k\xi_\perp)',\\
    \delta N&=\dot{\xi}-N_\perp\xi'+N'\xi_\perp,
    \\
    \delta N_\perp&=\Big(\frac{-N\xi'+N'\xi}{a^2}\Big)+\Big(\dot{\xi}_\perp+N_\perp'\xi_\perp-N_\perp\xi_\perp'\Big).    
    \end{cases}
\end{equation}
Since it is the same transformation in a different canonical coordinate, the Fadeev-Popov determinant is equivalent, provided that gauge-fixing functions are expressed in the new coordinate. 
}
\footnote{The exact equation of motion is not invariant under diffeomorphism,
\begin{equation}
    \delta(\Phi'+aX^0)=\frac{\mathcal{H}_\perp}{a}\xi+a\mathcal{H}\xi_\perp,
\end{equation}
whose right hand side is vanishing whenever evaluated in the path-integral. Thus, the exactness of these equation of motion are true only when evaluated in the path-integral, not by themselves. }. 

Let us examine the boundary condition for the gauge transformation. We can show that the action and the path-integral measure are invariant for any value of $\xi,\xi_\perp$. First we note that constraints transform as
\begin{equation}
\begin{split}
    \delta\mathcal{H}&=\frac{2\mathcal{H}_\perp\xi'}{a^2}+\Big(\frac{\mathcal{H}_\perp}{a^2}\Big)'\xi-\frac{k\mathcal{H}_\perp}{a}\xi_\perp+(\mathcal{H}\xi_\perp)',\\
    \delta\mathcal{H}_\perp&=\mathcal{H}\xi'+\frac{a'}{a}\mathcal{H}_\perp\xi_\perp+\mathcal{H}_\perp\xi'_\perp.
\end{split}
\end{equation}
Using these expressions, we can explicitly show that regardless of the boundary condition of $\xi,~\xi_\perp$, we have
\begin{equation}\label{eq:boundaryinvariance}
    \delta I_{\text{phase space}}=0.
\end{equation}


\subsubsection*{Conserved charge in phase space}

Let us consider conserved charges in the phase space. We define
\begin{equation}
    G_2\equiv\int\frac{d\theta }{2\pi}\frac{1}{2}\left(p^2-\frac{\Phi'^2}{a^2}\pm\Phi^2\right),
\end{equation}
following \cite{Louis-Martinez:1993bge}. This quantity transforms as
\begin{equation}
    \delta G_2=\int\frac{d\theta }{2\pi}\left(-\frac{\Phi'}{a^3}\mathcal{H}_\perp\xi-\frac{\Phi'}{a}\mathcal{H}\xi_\perp\right),
\end{equation}
which vanishes when the constraints are satisfied. Therefore, $G_2$ is gauge invariant when evaluated in the path-integral. 

\subsubsection*{Holonomy}

In BF theory, since $F=0$, the trace of holonomy is conserved for cylinder amplitude. Let us take a gauge-fixing $M=a'=k'=0$, and consider the holonomy for constant $t$ timeslice. The holonomy is given by
\begin{equation}
\begin{split}
    U&=
    \begin{pmatrix}
    e^{\pi\sqrt{k_0^2\pm a_0^2}} & 0 \\
    0 & e^{-\pi\sqrt{k_0^2\pm a_0^2}}
    \end{pmatrix}.
\end{split}
\end{equation}
Thus the conserved quantity $G_1$ defined in \eqref{G1def} becomes
\begin{equation}
    G_1= \pm\frac{k_0^2\pm a_0^2}{2}.
\end{equation}

\subsubsection*{Reduction to configuration space path-integral from the phase space}

Let us further integrate out momenta $p,~k$ to obtain the expression for configuration space gauge-fixed path-integral. We assume that $\delta(G) \Delta_{\text{FP}}^{(G)}$ does not depend on the momenta. Then one can integrate out $p,~k$ to obtain the exact equations of motion that are valid when evaluated in the path-integral
\begin{equation}\label{eq:exacteom}
    Nk+(\dot{a}-(N_\perp a)')=Np+\dot{\Phi}-N_\perp\Phi'=0,
\end{equation}
and the factor from the delta function
\begin{equation}
    \frac{1}{\prod_{t,\theta }|N|}.
\end{equation}
Thus, the full path-integral is given by
\begin{equation}\label{eq:pathintegralconfig}
     \int
     [DaD\Phi DNDN_\perp]\delta(G) \frac{1}{\prod_{t,\theta }|N| }~e^{iI_{\pm,\text{conf}}}\Delta_{\text{FP}}^{(G)} \delta(O(\varphi|_{t=0,1})-O(\varphi_{i,f})),
 \end{equation}
with configuration space action
\begin{equation}
\begin{split}
    I_{\pm\text{conf}} 
    &
    = \frac{1}{2}\int dt\frac{d\theta}{2\pi}\sqrt{-g}(R\pm 2)\Phi\\
    &=- \int \mathrm{d}t \frac{\mathrm{d}\theta}{2\pi} 
    \left[ N \left( \Big(\frac{\Phi'}{a}\Big)'\mp a \Phi \right) 
    + N^{-1} (\dot{\Phi} - N_\perp \Phi') (\dot{a} - (N_\perp a)') \right]
    +\Big[\int\frac{d\theta}{2\pi}aK
    \Phi\Big]^{t=1}_{t=0},\label{eq:JTactionND}
\end{split}
\end{equation}
which is the JT gravity action with fixed boundary metric and extrinsic curvature $K=\frac{\dot{a}-(N_\perp a)'}{Na}$ \cite{Goel:2020yxl}. 

Let us examine the diffeomorphism in the configuration space. Using the exact equation of motion \eqref{eq:exacteom} which is valid when evaluated in the path integral, we have  
\begin{equation}\label{eq:diffconfig}
\begin{cases}
    \delta a&=\frac{\dot{a}-(N_\perp a)'}{N}\xi+\Big(a\xi_\perp\Big)'
    ,\\
    \delta \Phi&=\frac{\dot{\Phi}-N_\perp\Phi'}{N}\xi+\Phi'\xi_\perp,\\
    \delta N&=\dot{\xi}-N_\perp\xi'+N'\xi_\perp,
    \\
    \delta N_\perp&=\Big(\frac{-N\xi'+N'\xi}{a^2}\Big)+\Big(\dot{\xi}_\perp+N_\perp'\xi_\perp-N_\perp\xi_\perp'\Big).    
    \end{cases}
\end{equation}
This is precisely the geometric-diffeomorphism induced from the coordinate transformation
\begin{equation}\label{eq:diffgeometric}
    (\delta t,~\delta \theta)=\xi(t,\theta) \left(\frac{1}{N},-\frac{N_\perp}{N}\right)+a\xi_\perp(t,\theta) \left(0,~\frac{1}{a}\right).
\end{equation}
One important difference of the diffeomorphism \eqref{eq:diffconfig} in configuration space from the phase space \eqref{eq:diffphasespace} is that, since the geometric-diffeomorphism \eqref{eq:diffgeometric} moves the boundary timeslices beyond the original spacetime geometry, it transforms the action nontrivially. Indeed, there is a boundary contribution 
\begin{equation}\label{eq:actionvar}
    \delta I_{\text{conf}}=\frac{1}{4\pi}\Big[\int d\theta \, \sqrt{-g}\,\Phi\,(R\pm2)\frac{\xi}{N}\Big]_{t=0}^{t=1},
\end{equation}
which is proportional to the exact equation of motion $R\pm2=0$. This not-trivial transformation property seems to be in conflict with the gauge invariance of the path-integral, which can be arbitrary at the boundary. 

To understand this point, let us examine the relation of the above path-integral measure with the standard measure from the ultra-local inner product in configuration space. We note that all the fields do not have boundary conditions. The ultra-local inner product is given by
\begin{equation}
    G^{g,\text{conf}}_\mathcal{F}(\delta_1 g,~\delta_2 g)=\frac{1}{2}\int dtd\theta ~\sqrt{-g}\Big(g^{\mu\alpha}g^{\nu\beta}+cg^{\mu\nu}g^{\alpha\beta}\Big)\delta_1g_{\mu\nu}\delta_2 g_{\alpha\beta}.
\end{equation}
Here, the parameter $c$ tunes relative measure between the Weyl transformation and the symmetric traceless deformation of the metric, which does not affect the gravitational path-integral of JT theory, so we can set $c=0$. The metric is then
\begin{equation}\label{eq:innerproduct}
    G^{g,\text{conf}}_\mathcal{F}(\delta_1 g,~\delta_2 g)=\frac{1}{2}\int dtd\theta~(Na)\Big(\frac{2(\delta_1 a)(\delta_2 a)}{a^2}+\frac{2(\delta_1 N)(\delta_2 N)}{N^2}-\frac{a^2(\delta_1 N_\perp)(\delta_2 N_\perp)}{N^2}\Big).
\end{equation}
For the dilaton, we have the scalar inner product
\begin{equation}
    G^{\Phi,\text{conf}}_\mathcal{F}(\delta_1 \Phi,~\delta_2 \Phi)=\int dtd\theta~(Na)(\delta_1 \Phi)(\delta_2 \Phi).
\end{equation}
The total metric is $G^{\text{conf}}_\mathcal{F}=G^{g,\text{conf}}_\mathcal{F}+G^{\Phi,\text{conf}}_\mathcal{F}$. The invariant metric at the identity element is now
\begin{equation}
\gamma^C(\epsilon_1,\epsilon_2)
=
\int dtd\theta \,(Na)
\left[-\xi_1\xi_2+(a\xi_{1\perp})(a\xi_{2\perp})\right].
\end{equation}
where the superscript $C$ indicates that the measure is being computed in the configuration space. 

The invariant measure for gauge transformation is same as that in phase space. Thus the gauge-fixed configuration space path-integral is
\begin{equation}
\begin{split}
K&=
\int \frac{[DaD\Phi DNDN_\perp]}
{\operatorname{Vol}({\cal G}^{\text{conf}})}
\sqrt{|\det G_\mathcal{F}^{\text{conf}}[\varphi]|}
\,e^{iI_\text{conf}[\varphi]}\delta(O(\varphi|_{t=0,1})-O(\varphi_{i,f}))
\\
&
=\int [DaD\Phi DNDN_\perp]\,
\frac{\sqrt{|\det G_\mathcal{F}^{\text{conf}}[\varphi]|}}{\sqrt{|\det{}\gamma^C [\varphi]|}}
\delta[G]\,
\Delta_{\rm FP}^{(G)}
e^{iI_\text{conf}}\delta(O(\varphi|_{t=0,1})-O(\varphi_{i,f})) \\
&=\int [DaD\Phi DNDN_\perp]\,
\frac{1}{\prod_{(t,\theta )}|N|}
\delta[G]\,
\Delta_{\rm FP}^{(G)}
e^{iI_\text{conf}}\delta(O(\varphi|_{t=0,1})-O(\varphi_{i,f})),
\end{split}
\end{equation}
where we used
\begin{equation}
\frac{\sqrt{|\det G_\mathcal{F}^{\text{conf}}[\varphi]|_{(t,\theta )}}}{\sqrt{|\det{}\gamma^C [\varphi]|_{(t,\theta )}}}=\frac{1}{|N|}.
\end{equation}
Thus we have shown the equivalence between the configuration space and phase space path-integral. However, notice the boundary condition $\delta(O(\varphi|_{t=0,1})-O(\varphi_{i,f}))$ and the gauge symmetry which is free even at the boundary, despite \eqref{eq:actionvar}. The gauge invariance of the action is manifest only after imposing the exact equation of motion $R\pm 2=0$.

\subsubsection*{Boundary gauge transformation and comparison with JT gravity with asymptotic boundary condition}

We emphasize the difference of the path-integral from the familiar JT gravity path-integral where the asymptotic boundary has a preferred coordinate and therefore boundaries are marked \footnote{When there is a preferred coordinate, the diffeomoprohim is vanishing. When there is a marked point, the diffeomorphism is constrained to be vanishing on that point.}. In particular, the gauge symmetry is small diffeomorphism which vanishes at the boundaries, and thus the gauge group volume is that of small diffeomorphism. On the other hand, in our closed universe phase space gravitational path-integral, the gauge transformation is non-vanishing at the boundaries. In the next section, we will use this gauge redundancy to take the gauge fixing condition $a'=k'=0$ which is similar to taking the fixed constant $g_{uu}$, extrinsic curvature $K=\frac{\dot{a}-(N_\perp a)'}{Na}$ boundary condition at asymptotic boundary. However, we have an extra measure from the Fadeev-Popov determinant, which render the boundary condition gauge-invariant. Such boundary measure factor is not included in standard asymptotic boundary. In particular, the path-integral is distinct from the bra-ket wormholes due to this boundary condition \cite{Chen:2020tes, Miyaji:2021lcq, Fumagalli:2024msi}.


\subsection{Classical solutions}

We describe general solutions of equation of motion under the (rather strong) assumptions that $N=1,N_\perp=0,a'=0$ case. 

\subsubsection*{AdS big-bang-crunch solution}
For AdS, we only have AdS big-bang-crunch solution
\begin{equation}
    ds^2=-dt^2+b^2\cos^2(t) d\theta^2,
\end{equation}
which can also be continued to the Euclidean double trumpet
\begin{equation}
    ds^2=d\tau^2+b^2\cosh^2(\tau) d\theta^2,
\end{equation}
via analytical continuation $\tau=it$. ADM variables are
\begin{equation}
    \begin{split}
        a&=b\cos t,~
        \Phi=\Phi_0\sin t,~
        G_2=\Phi_0^2,
    \end{split}
\end{equation}
which experiences big-crunch and big-bang periodically in time. The holonomy is
\begin{equation}
\begin{split}
    U&=
    \begin{pmatrix}
    e^{\pi\sqrt{k^2+a^2}} & 0 \\
    0 & e^{-\pi\sqrt{k^2+a^2}}
    \end{pmatrix},~\text{Tr}[U]=2\cosh(\pi\sqrt{k^2+a^2}).
\end{split}
\end{equation}

\subsubsection*{Global dS}

In dS, we have three types of solutions. The first solution is the global dS, which has future and the past asymptotic boundaries. We have $-2G_1=k^2-a^2=-b^2$ and
\begin{equation}
    \begin{split}
        a&=b\cosh t,~
        \Phi=\Phi_0\sinh t,~
        G_2=\frac{\Phi_0^2}{2}.
    \end{split}
\end{equation}
The size of universe is minimal at $t=0$.
The holonomy is
\begin{equation}
\begin{split}
    U&=
    \begin{pmatrix}
    e^{i\pi b} & 0 \\
    0 & e^{-i\pi b}
    \end{pmatrix},~\text{Tr}[U]=2\cos(\pi b).
\end{split}
\end{equation}
When $b$ is an integer, the holonomy is trivial $\text{Tr}[U]=\pm 2$ in $PSL(2,\mathbb{R})$ sense and can originate from the no-boundary state which also has trivial holonomy by definition. We emphasize that states with trivial holonomy are special and are not generic.

\subsubsection*{dS big-bang and big-crunch solution}
This solution also has future and the past asymptotic boundaries. It has conical singularity at initial or final timeslice, and can be continued beyond. We have $k^2-a^2=\beta^2$ and
\begin{equation}
    \begin{split}
        a&=\beta\sinh t,~
        \Phi=\Phi_0\cosh t,~
        G_2=-\frac{\Phi_0^2}{2}.
    \end{split}
\end{equation}
This solution can be obtained as a portion of Milne patch with periodic identification. The holonomy is
\begin{equation}
\begin{split}
    U&=
    \begin{pmatrix}
    e^{\pi\beta} & 0 \\
    0 & e^{-\pi\beta}
    \end{pmatrix},~\text{Tr}[U]=2\cosh(\pi\beta).
\end{split}
\end{equation}
\subsubsection*{$n$-dS BH $(n\in\mathbb{N})$ solution}
We have $k^2-a^2=-n^2$ and
\begin{equation}
    \begin{split}
        a&=\pm n\cosh t,~
        \Phi=\Phi_0\cosh t\sin(n \theta),~
        G_2=-\frac{\Phi_0^2}{2}.
    \end{split}
\end{equation}
The holonomy is
\begin{equation}
\begin{split}
    U&=
    \begin{pmatrix}
    e^{i\pi n} & 0 \\
    0 & e^{-i\pi n}
    \end{pmatrix},~\text{Tr}[U]=2\cos(\pi n).
\end{split}
\end{equation}
These classification of spacetimes will continue to be useful even when we perform path-integral analysis.


\section{Gauge-fixed path-integral}\label{sec:gauge-fixed-path-integral}

We first integrate out the nonzero spatial modes  while leaving the zero modes unfixed.
We then evaluate the same gauge-redundant zero-mode system in the constant-lapse and invariant gauges.
The BFV--BRST construction identifies the corresponding boundary states {obtained in these two different ways of zero-mode gauge fixing} and relates their physical inner products.

Fields with Dirichlet boundary conditions can be expanded as 
\begin{equation}
    a(t,\theta )=\sum_{n\in\mathbb{Z}}e^{in\theta }\Big(a_n(t=0)+(a_n(t=1)-a(t=0))t+\sum_{m>0}a_n(m)\sqrt{2}\sin(m\pi t)\Big),
\end{equation}
while fields with free boundary condition as 
\begin{equation}
    N(t,\theta )=\sum_{n\in\mathbb{Z}}e^{in\theta }\Big(N_n(0)+\sum_{m>0}N_n(m)\sqrt{2}\cos(m\pi t)\Big).
\end{equation}
The convergence is in $L^2[0,1]$, and we further assume that their derivatives also converge in $L^2[0,1]$, thus they are in Sobolev space $W^{1,2}[0,1]$ so that they are weakly differentiable. We call homogeneous mode $n=0$ as zero-mode, while inhomogeneous one $n\neq0$ as non-zero mode.

\subsection{Nonzero modes and reduction to the zero-mode phase space}
\label{subsec:nonzero-mode-reduction}

We separate the spatial gauge fixing for the nonzero modes from the choice of gauge for the zero modes. By performing the former, the path-integral can be reduced to the zero-mode phase space path-integral.

Let us consider the foliation of spacetime whose extrinsic curvature of timeslice is constant on each timeslice. For each timeslice, we can associate $\theta $ coordinate up to its origin so that it satisfies $a'=0$. In this way, we can construct coordinate that satisfies the gauge condition
\begin{equation}\label{eq:gaugecondition}
    a'=k'=0.
\end{equation}
This is the gauge for non-zero modes we will use in this paper. We emphasize this gauge is not general enough as it does not apply to $n$-dS BH. The condition \eqref{eq:gaugecondition} does not fix the zero-mode gauge symmetry and we will further impose a gauge conditions for zero-modes, whose contribution to the Fadeev-Popov determinant factorizes with the one for above non-zero mode contribution $\Delta_{FP}^{n\neq 0}$, for the reason that $\frac{\delta a'(\theta)}{\delta\xi_0}=\frac{\delta a'(\theta)}{\delta\xi_{0\perp}}=\frac{\delta k'(\theta)}{\delta\xi_0}=\frac{\delta k'(\theta)}{\delta\xi_{0\perp}}=0$. The non-zero mode contribution is then, 
\begin{equation}
\begin{split}
    \Delta_{FP}^{n\neq 0}&=
    \prod_t\Big{|}
    \text{det}
    \begin{pmatrix}
    \frac{(\delta a)'_n}{\delta\xi_m} & \frac{(\delta k)'_n}{\delta\xi_{m}} \\
    \frac{(\delta a)'_n}{\delta\xi_{m\perp}} & \frac{(\delta k)'_n}{\delta\xi_{m\perp}}
    \end{pmatrix}_{n,m\neq0}\Big{|}
    = 
    \prod_t\Big{|}
    \prod_{n\neq 0}\text{det}
    \begin{pmatrix}
    -ink & \pm ina+i\frac{n^3}{a} \\
    -an^2 & -kn^2
    \end{pmatrix}\Big{|}
    \\
    &\propto\prod_{t}\Big{|}\prod_{n\neq0}(n^2+k_0^2\pm a_0^2)\Big{|}
    .\end{split}
    \label{fpnz}
\end{equation}
Note here that the FP determinant vanishes at $n$-dS BH, invalidating the application of this gauge for this spacetime. On the other hand, integrating out non-zero modes of lapse $N_n$ and shift $N_{n\perp}$ in the path-integral gives
\begin{equation}
\begin{split}
    \int &[\prod_{n\neq0}DN_nDN_{n\perp}]~e^{-i\int dt\Big[N_n\mathcal{H}_{-n}+N_{n\perp}\mathcal{H}_{-n\perp}\Big]}\\&
    \propto\prod_t\prod_{n\neq0}\delta(\mathcal{H}_n)\delta(\mathcal{H}_{n\perp})
    =\prod_t\prod_{n\neq0}\frac{\delta(p_n)\delta(\Phi_n)}{|k_0^2\pm a_0^2+n^2||n|}.
\end{split}
\label{NnNpnint}
\end{equation}
 {Combining \eqref{fpnz} and \eqref{NnNpnint}, it can be seen easily that the $k_0,a_0$ dependent factors cancel leaving just the delta functions, upto a constant factor. These delta functions can further be used to perform the integrals for non-zero modes for $p_n$ and $\Phi_n$}. Thus we arrived at the zero-mode phase space path-integral. The left-over zero mode integrals are over the fields
\begin{equation}
    (a_0(t),~p_0(t),~\Phi_0(t),~k_0(t),~N_0(t),~N_{0\perp}(t)).
\end{equation}
The remaining gauge symmetry is
\begin{equation}
    (\xi_0(t),~\xi_{0\perp}(t)),~\text{no boundary conditions}.
\end{equation}
The action is now
\begin{align}
    I_{0}
    &=\int_0^1dt\,
    \left[
        p_0\dot a_0-\Phi_0\dot k_0-N_0\mathcal H_0
    \right],
    \notag\\
    \mathcal H_0&=-k_0p_0\mp a_0\Phi_0.\label{zeromodes}
\end{align}
Due to the absence of $N_{0\perp}(t)$ in the action, we can gauge fix it to
\begin{equation}
    N_{0\perp}=0,
\end{equation}
using 
\begin{equation}
\delta N_{\perp0}=\dot{\xi}_{0\perp}+\frac{\dot{a}_0+N_0k_0}{a_0}\xi_{0\perp},
\end{equation}
 combined with the zero-mode of the exact equation of motion, \eqref{eq:exacteom}, given by  $\dot{a}_0+N_0k_0=0$. Note that the associated Fadeev-Popov determinant is constant when evaluated in the path-integral, as long as there is no $p$ insertions so that the exact equation of motion is valid. 

For the gauge symmetry corresponding to time-independent $\xi_{0\perp}$  , we explicitly divide the gauge volume in the path-integral instead of gauge-fixing. The volume density at each timeslice is proportional to the volume itself
\begin{equation}
    \text{Vol}(U(1)_{\xi_{0\perp}})|_t=\int_0^{2\pi}d\xi_{0\perp}\sqrt{N_0a_0}|a_0|=2\pi|N_0^{1/2}a_0^{3/2}|=2\pi\sqrt{\gamma_{U(1)_{\xi_{0\perp}}}}|_t.
\end{equation}
This leads to the identity $\text{Vol}(U(1)_{\xi_{0\perp}})|_t\sqrt{\gamma_{\mathcal{G}/U(1)}}|_t=2\pi\sqrt{\gamma_\mathcal{G}}|_t$, which cancels with the path-integral measure as established in \eqref{eq:cancellation}.

Thus the remaining variables are
\begin{equation}
    (a_0(t),~p_0(t),~\Phi_0(t),~k_0(t),~N_0(t)),
\end{equation}
and the remaining gauge symmetry is
\begin{equation}
    \xi_0(t),~\text{free boundary condition},
\end{equation}
which induces the following transformation
\begin{equation}\label{eq:zeromodetransformation}
    \delta a=-k\xi
    ,~~\delta p=\pm\Phi\xi,~~
    \delta \Phi=-p\xi,
    ~~\delta k=\pm a\xi,~~
    \delta N=\dot{\xi}.    
\end{equation}
The full path-integral is now
\begin{align}
    K
    &=
    \int
    \frac{[Da_0Dp_0D\Phi_0Dk_0DN_0]}{\operatorname{Vol}(\mathcal G_{BR})}
    \exp\left(
        iI_0
    \right)\Delta_{FP}^{BR}\delta(B)\Delta(R),
\end{align}
where $\delta(B)$ is the boundary gauge-fixing, $\delta(R)$ is the bulk gauge-fixing and  $\Delta_{FP}^{BR}$ is corresponding Fadeev-Popov determinant.

\subsection{Canonical zero-mode variables.}
\label{subsec:canonical-zero-mode-variables}

Let us introduce canonical variables $(q,-\mathcal{H}_0,G_1,\Pi_{G_1})$ for zero-mode phase space. The canonical one-form takes the form
\footnote{In the AdS sector, we introduce
\begin{equation}
    a_0=b\cos u,
    \qquad
    k_0=-b\sin u,
    \qquad
    b>0,
    \qquad
    G_1=\frac{b^2}{2},
    \qquad
    \Pi_{G_1}
    =
    \frac{p_0\cos u+\Phi_0\sin u}{b}.
    \label{AdSu}
\end{equation}
In the global dS sector, we introduce
\begin{equation}
    a_0=b\cosh u,
    \qquad
    k_0=b\sinh u,
    \qquad
    b\neq0,
    \qquad
    G_1=\frac{b^2}{2},
    \qquad
    \Pi_{G_1}
    =
    \frac{p_0\cosh u-\Phi_0\sinh u}{b}.\label{gbdsu}
\end{equation}
In the dS big-bang/crunch sector, we use the signed parameter $\beta$ and introduce
\begin{equation}
    a_0=\beta\sinh v,
    \qquad
    k_0=\beta\cosh v,
    \qquad
    \beta\neq0,
    \qquad
    G_1=-\frac{\beta^2}{2},
    \qquad
    \Pi_{G_1}
    =
    \frac{\Phi_0\cosh v-p_0\sinh v}{\beta}.\label{dsbgbcv}
\end{equation}}
\begin{equation}\label{eq:complete-gauge-canonical-form}
    p_0\,da_0-\Phi_0\,dk_0
    =
    \Pi_{G_1}\,dG_1-\mathcal{H}_0\,dq,
\end{equation}
where $q=u$ in the AdS and global dS sectors and $q=v$ in the dS big-bang/crunch sector see \eqref{AdSu}-\eqref{dsbgbcv}.
The phase-space measure is preserved
\begin{equation}
    da_0\,dk_0\,dp_0\,d\Phi_0
    =
    dG_1\,dq\,d\Pi_{G_1}\,d\mathcal{H}_0.
\end{equation}
Using the canonical pairs $(G_1,\Pi_{G_1})$ and $(q,-\mathcal H_0)$,
the zero-mode action becomes
\begin{align}
    I_0
    &={}
    \int_0^1 dt\,
    \left[
        \Pi_{G_1}\dot G_1
        -\mathcal H_0(\dot q+N_0)
    \right].
    \label{eq:zero-mode-fp-common-action}
\end{align}
The zero-mode gauge transformations are
\begin{align}
    \delta q
    &=-\xi,
    &
    \delta N_0
    &=\dot\xi,
    &
    \delta\mathcal H_0&=
    \delta G_1
    =\delta\Pi_{G_1}=0.
    \label{eq:zero-mode-fp-common-transformations}
\end{align}
In particular, $\dot q+N_0$ is gauge invariant. 

\paragraph{Periodicity in time and gauge group.}

We emphasize that the above coordinate transformation is local, and there can be difference between two coordinates globally. Indeed, the coordinate $(q,G_1)$ provides an universal cover of $(a_0,k_0)$ coordinate with the range of $q$ being $\mathbb{R}$. This property will play an important role in Lorentzian AdS, since the Hamiltonian evolution is periodic in $(a_0,k_0)$ but not in $(q,G_1)$. In such case, we can introduce $n$-th period in $(a_0,k_0)$ coordinate as $(a_0,k_0,n)$. Then, there will be no periodicity any more, as the originally periodic gauge transformation acts as $n\rightarrow n+1$. We can partially gauge fix the boundary value of $n$ to a fixed value $n_0$.

Alternatively, without introducing $n$, we can perform the path-integral with $(a_0,k_0)$ variables, starting from the $(q,G_1)$ variables as follows. Let us denote the gauge group for $(q,G_1)$ as $\mathcal{G}$, then the gauge group acting on $(a_0,k_0)$ faithfully via 
\begin{equation}
    \mathcal{G}'=\mathcal{G}/2\pi\mathbb{Z},
\end{equation}
whose total volume is given by
\begin{equation}
    \text{Vol}(\mathcal{G})=|\mathbb{Z}|\text{Vol}(\mathcal{G}/2\pi\mathbb{Z}).
\end{equation}
Thus when we perform path-integral in $(a_0,k_0)$ coordinate, the gauge group volume of $\mathcal{G}'$ is canceled by the Fadeev-Popov determinant, while leaving extra $|\mathbb{Z}|$ in the denominator.

\subsection{Boundary zero-mode gauge-fixing}
\label{subsec:zero-mode-fp-gauges}

In this section, we will gauge-fix the boundary gauge degree of freedom of $\xi_0$. Depending on how we fix this zero-mode gauge, we will encounter seemingly different Hilbert spaces, namely co-invariant and invariant Hilbert spaces. Understanding of these different Hilbert spaces in the BFV--BRST formalism will be described in Section~\ref{sec:zero-mode-brst}.

{Before we turn to the boundary gauge fixing, let us also mention that }the bulk gauge symmetry can be fixed in various ways, and in this paper we will work in the gauge
\begin{equation}
\dot{N}_0=0.
\end{equation}
Associated Fadeev-Popov determinant will be factorized from that of the boundary gauge-fixing in the following, and gives only constant contribution
\begin{equation}
\begin{split}
    \Delta_{FP}^{\dot{N}_0=0}&=
    \prod_t\Big{|}
    \frac{\delta \dot{N}_0}{\delta\xi_0}\Big{|}=
    \prod_t\Big{|}
    \frac{\delta \ddot{\xi}_0}{\delta\xi_0}\Big{|}=\text{const}.
    \end{split}
\end{equation}
We can also choose $q(t)=q_{\rm gf}(t)$ gauge as in Appendix~\ref{subsec:complete-zero-mode} for example. Note that this gauge condition prohibits global time-folds where $N_0$ changes its sign, while locally the lapse can change its sign. Such local time-fold in the path-integral introduces multi-valued fields in general, which we discuss in more detail in the next section. 

\subsubsection{Gauge-fixing for co-invariant state: group-averaging}
\label{subsubsec:zero-mode-fp-group-averaging}

In this section, we use the boundary zero-mode gauge symmetry to fix boundary values of fields to specific values. Namely, combined with the physical boundary consition $G_1=\frac{a_0^2\pm k_0^2}{2}$, we can fix 
\begin{equation}
    a_0|_{\Sigma_{i/f}}=a_{0i/f} \quad\text{or}
    \quad
    k_0|_{\Sigma_{i/f}}=k_{0i/f}.
    \label{bdyak}
\end{equation}
The remaining gauge transformations obey
\begin{equation}
    \xi|_{\partial M}=0.
\end{equation}
Thus the above gauge fixing procedure is equivalent to imposing boundary conditions on the gauge transformation. In the BRST/BFV description, this corresponds to Dirichlet boundary
conditions for the corresponding ghost fields.  The resulting boundary
states correspond to the co-invariant states, and the path-integral corresponds to the inner product of co-invariant states. We will find that the Hilbert space inner product is given by the group-averaging inner product. 

\paragraph{Gauge-fixing.}
Let us gauge-fix the boundary values of $k_0$, in addition to the physical, gauge-invariant boundary condition $G_1=\frac{a_0^2\pm k_0^2}{2}$. The Fadeev-Popov determinant for this gauge-fixing, combined with the gauge-fixing delta function and boundary condition fixing delta function is, at $t=0$
\begin{equation}\label{eq:boundaryconditionmeasure}
       \Big{|}\frac{\delta_\xi k_0}{\delta\xi_{0i}}\Big{|}\delta(k_0-k_{0i})\delta(G_1(a_0,k_0)-G_1(i))=\delta(a_0-a_{0i})\delta(k_0-k_{0i}),
\end{equation}
similarly for $t=1$. Here we implicitly assumed $a_0>0$. Thus we see that this gauge-fixing is equivalent to directly inserting delta functions  for $a_0$, $k_0$ at the boundary. In terms of alternative coordinate $(q,\mathcal{H},G_1,\Pi_{G_1})$, this gauge fixing corresponds to fixing boundary values of
\begin{equation}
       q,
\end{equation}
with delta function insertion and trivial Fadeev-Popov determinant
\begin{equation}
      \delta(q-q_i)\delta(G_1-G_{1i}).
\end{equation}
This corresponds to preparing $|G_{1},q\rangle$ state in co-invariant Hilbert space, which we will see later. 

We thus impose the boundary zero-mode gauge-fixing conditions $a_0(t=0)=a_i$, $a_0(t=1)=a_f$ and \eqref{eq:gaugecondition}. Then we obtain, in terms of two coordinates,
\begin{equation}\label{eq:physicalpathintegral}
\begin{split}
    K^{\rm FP}_{\rm co}(G_{1f},G_{1i})&=
    \int_{\Gamma'} dN_0\prod_{t}\Big(Da_0Dk_0D\Phi_0 Dp_0\Big)
    ~\exp\left\{i\int dt\Big[p_0\dot{a}_0-\dot{k}_0\Phi_0-N_0\mathcal{H}_0\Big]\right\}\\
    &
    \times \delta(a_0(t=0)-a_{0i})\delta(k_0(t=0)-k_{0i})\delta(a_0(t=1)-a_{0f})\delta(k_0(t=1)-k_{0f})\\
    &=
    \int_\Gamma dN_0\prod_{t}\Big(DG_1D\Pi_{G_1}Dq D\mathcal{H}_0\Big)
    ~\exp\left\{
        i\int_0^1 dt\,
        \left[
            \Pi_{G_1}\dot G_1
            -\mathcal H_0(\dot q+N_0)
        \right]
    \right\}
    \\
    &
    \times \delta(q(t=0)-q_i)\delta(G_1(t=0)-G_{1i})\delta(q(t=1)-q_f)\delta(G_1(t=1)-G_{1f}).
        \end{split}
\end{equation}
Here the lapse integration contour $\Gamma'$ in $(a_0,k_0)$ coordinate depends on the global structure of the gauge group and periodicity, see subsection \ref{subsec:canonical-zero-mode-variables}. The lapse contour for $(q,G_1)$ coordinate is $\Gamma=\mathbb{R}$, while $\Gamma'=\mathbb{R}$ or $\mathbb{R}/2\pi\mathbb{Z}$ depending on whether the gauge group is $\mathcal{G}$ or $\mathcal{G}/2\pi\mathbb{Z}$. 
In the above formula, we first integrate out $\Pi_{G_1}$ and $\mathcal{H}_0$ to get constant $G_1$ and $\dot{q}+N_0=0$. Then the constant $q$ mode and $N_0$ are fixed by the boundary delta functions of $q$. Finally, integrating constant $G_1$ mode, we can conclude
\begin{equation}\label{eq:co-invariant-result}
    K^{\rm FP}_{\rm co}(G_{1f},G_{1i})=\delta(G_{1f}-G_{1i}),
\end{equation}
assuming initial and final boundary conditions are in the same connected component.

From the expression \eqref{eq:physicalpathintegral}, we clearly have a canonical operator realization of our path-integral. The operators $p_0,~\Phi_0$ on the auxiliary Hilbert space $H_{\text{aux}}$ are conjugate momenta of $a_0,~-k_0$
\begin{equation}
    \begin{split}
    [\hat{a}_0,\hat{p}_0]=i,~[-\hat{k}_0,\hat{\Phi}_0]=[\hat{\Phi}_0,\hat{k}_0]=i,
    \end{split}
\end{equation}
whose eigenstates give an orthonormal basis
\begin{equation}
    \int da_0dk_0~|a_0,k_0\rangle\langle a_0,k_0|= \int dqdG_1~|q,G_1\rangle\langle q,G_1|= \mathbb{I}_{H_{\text{aux}}}.
\end{equation}
We arrive at operator expression of the amplitude
\begin{equation}
     K^{\rm FP}_{\rm co}((a_f,k_f),(a_i,k_i))=\langle a_f,k_f|\int_{\mathbb{R}} dN_0~e^{-iN_0\hat{\mathcal{H}}_0}|a_i,k_i\rangle.
\end{equation}
We can now consider the projection map $\pi$ from auxiliary Hilbert space $H_{\text{aux}}$ to the co-invariant Hilbert space via
\begin{equation}\label{eq:coinvariant}
    \pi:H_{\text{aux}}\rightarrow H_{\text{co}}\equiv H_{\text{aux}}/\text{Im}[\mathcal{H}_0].
\end{equation}
The inner product of this quotient space is given by the group averaging \cite{Held:2024rmg}
\begin{equation}
    \langle\langle a_f,k_f|a_i,k_i\rangle\rangle= \langle a_f,k_f|\int_{\mathbb{R}} dN_0~e^{-iN_0\hat{\mathcal{H}}_0}|a_i,k_i\rangle,
\end{equation}
where $|a,~k\rangle$ is a representative of $|a,~k\rangle\rangle$ in $H_{\text{aux}}$. This Hilbert space $H_{\text{co}}$ is called co-invariant Hilbert space. We note that the group averaging map 
\begin{equation}\label{eq:groupaveraging}
    \eta\equiv\int_{\mathbb{R}} dN_0~e^{-iN_0\hat{\mathcal{H}}_0}:H_{\text{co}}\rightarrow H_{\text{inv}}\equiv \Big\{|\Psi\rangle\in H_{\text{aux}}\Big{|}\mathcal{H}_0|\Psi\rangle=0 \Big\}.
\end{equation}
is called in \emph{rigging map} \cite{Giulini:1998rk, Marolf:2000iq}.
However, the naive inner product induced from auxiliary Hilbert space is divergent due to massive over-counting and the rigging map above is not an isomorphism if such induced inner product were used. This issue will be fixed in the next subsection by introducing correct inner product on the invariant subspace above, called invariant Hilbert space.

Thus we have shown that our gravitational path-integral which connects the initial and final boundary conditions, gives the inner product in the co-invariant Hilbert space, and the co-invariant Hilbert space can be used as the physical Hilbert space.


\subsubsection{Gauge-fixing for invariant state: lapse and diagonal gauge fixing}
\label{subsubsec:zero-mode-fp-invariant}

{We now outline a different gauge fixing procedure for the boundary gauge symmetry that leads to invariant Hilbert space}. We will fix the lapse at the boundaries along with fixing a linear combination of $q_i,q_f$. 
The
Wheeler--DeWitt equation will follow from the integrations over the
unfixed components of $q$.

As before, the physical boundary condition we impose is
\begin{align}
    G_1(0)&=G_{1i},
    &
    G_1(1)&=G_{1f}.
    \label{eq:zero-mode-fp-invariant-boundary-data}
\end{align}
This corresponds to preparing $|G_{1},\mathcal{H}=0\rangle$ state in the invariant Hilbert space. We then gauge-fix the boundary values 
\begin{align}
    N_*&=N_{0i}=N_{0f}=1,
    &
    N_{0\perp i}&=N_{0\perp f}=0.
    \label{eq:zero-mode-fp-invariant-boundary-data}
\end{align}

Then, after the gauge-fixing $a'=k'=0$, the boundary time derivatives are constrained \eqref{eq:zeromodetransformation}
\begin{align}
    \dot\xi(0)&=0,
    &
    \dot\xi(1)&=0.
\label{eq:zero-mode-fp-invariant-gauge-domain}
\end{align}
As before, we impose the bulk constant-lapse gauge
\begin{align}
    F_N[N_0](t)
    &:={}
    \dot N_0(t)
    =0.
\label{eq:zero-mode-fp-invariant-lapse-gauge}
\end{align}
Together with the endpoint values of $N_0$, this gives
\begin{equation}
    N_0(t)=N_\star.
\end{equation}
The condition $F_N=0$ does not fix the constant transformation
$\xi(t)=\xi_0$.  
We remove this residual symmetry by the symmetric diagonal
gauge condition
\begin{align}
    \chi_{\rm diag}[q]
    &:={}
    q_\star-q_+
    =0,
    &
    q_+
    &:={}
    \frac{q_i+q_f}{2}.
\label{eq:zero-mode-fp-invariant-diagonal-gauge}
\end{align}
This is a gauge-fixing insertion necessary for the
invariant--invariant path-integral. Note that it is not an independent boundary condition on either endpoint state separately. The other orthogonal combination of $q(0)-q(1)$  will be integrated over.

To compute the Faddeev--Popov determinant, decompose the gauge parameter as
\begin{align}
    \xi(t)=\xi_++\xi_b(t),
    ~~~~~
    \xi_+:=\frac{\xi(0)+\xi(1)}{2},~~~~~\frac{\xi_b(0)+\xi_b(1)}{2}=0.
\end{align}
Then
\begin{align}
    \delta_\xi F_N(t)
    &=\ddot\xi_b(t),
    &
    \delta_\xi\chi_{\rm diag}
    &=\xi_+.
\label{eq:zero-mode-fp-invariant-variations}
\end{align}
The operator $\partial_t^2$ is invertible on the domain
\begin{equation}
\label{xibconds}
    \dot\xi_b(0)=\dot\xi_b(1)=0,
    \qquad
    \frac{\xi_b(0)+\xi_b(1)}{2}=0.
\end{equation}
Thus the determinant factorizes as
\begin{align}
    \Delta_{\rm inv}
    &:={}
    \left|\det{}'_{NN}\partial_t^2\right|
    \left|
      \frac{\partial\chi_{\rm diag}}{\partial\xi_+}
    \right|
    =
    \left|\det{}'_{NN}\partial_t^2\right|.
\label{eq:zero-mode-fp-invariant-determinant}
\end{align}
Here $NN$ denotes Neumann-Neumann boundary condition on $\xi_b$. The prime removes the constant mode, which is fixed separately
by $\chi_{\rm diag}=0$.  In a time-sliced regularization,
$\Delta_{\rm inv}$ is a nonzero field-independent constant.

The gauge-fixed inner product is therefore
\begin{align}
K_{\rm inv}^{\rm FP}(G_{1f},G_{1i})
&={}
\mathcal N_{\rm inv}
\int
\mathcal DG_1\,
\mathcal D\Pi_{G_1}\,
\mathcal Dq\,
\mathcal D\mathcal H_0\,
\mathcal DN_0\,
\Delta_{\rm inv}\,
\delta_{nz}[\dot N_0]\,
\delta(\chi_{\rm diag})
\notag\\
&\qquad\times
\exp\left\{
 i\int_0^1dt\,
 \left[
   \Pi_{G_1}\dot G_1
   -\mathcal H_0(\dot q+N_0)
 \right]
\right\}\nonumber\\
&\qquad\times\delta(N_0(0)-N_{0i})\delta(G_1(0)-G_{1i})\delta(N_0(1)-N_{0f})\delta(G_1(1)-G_{1f}).
\label{eq:zero-mode-fp-invariant-kernel}
\end{align}
Here $\delta_{nz}$ denotes the delta functional on the non-constant modes satisfying $\int_0^1dt f=0$. In particular, $q_i$, $q_f$, and the endpoint
values of $\mathcal H_0$ remain integrated. 

For later comparison with the BFV first-order form, the two gauge-fixing
delta functions may be exponentiated separately:
\begin{align}
\delta[\dot N_0] \, \delta(\chi_{\rm diag})
&={}
\mathcal C_\Pi
\int\mathcal D\Pi_0^{nz}
\int_{\mathbb{R}}\frac{d\lambda_0}{2\pi}
\exp\left\{i
 \int_0^1dt\,(\Pi_0^{nz}\dot N_0
 +\lambda_0\chi_{\rm diag})
\right\},
\label{eq:zero-mode-fp-invariant-delta-representation}
\end{align}
where
\begin{equation}
    \int_0^1dt\,\Pi_0^{nz}(t)=0,
\end{equation}
is imposed. 

The corresponding ghost-free first-order action is
\begin{align}
I_{\rm inv}^{\rm FP}
&={}
\int_0^1dt\,
\left[
  \Pi_{G_1}\dot G_1
  -\mathcal H_0(\dot q+N_0)
  +\Pi_0^{nz}\dot N_0
\right]
+\lambda_0\left(q_\star-\frac{q_i+q_f}{2}\right).
\label{eq:zero-mode-fp-invariant-ghost-free-action}
\end{align}

Accordingly,
\begin{align}
K_{\rm inv}^{\rm FP}(G_{1f},G_{1i})
&={}
\mathcal N_{\rm inv}
\int
\mathcal DG_1\,
\mathcal D\Pi_{G_1}\,
\mathcal Dq\,
\mathcal D\mathcal H_0\,
\mathcal DN_0\,
\mathcal D\Pi_0^{nz}\,
\frac{d\lambda_0}{2\pi}\,
 e^{iI_{\rm inv}^{\rm FP}}\nonumber\\
 &\qquad\times\delta(N_0(0)-N_{0i})\delta(G_1(0)-G_{1i})\delta(N_0(1)-N_{0f})\delta(G_1(1)-G_{1f})
\label{eq:zero-mode-fp-invariant-ghost-free-kernel}
\end{align}

We now perform the $q$-sector integrations.  Decompose
\begin{align}
    q(t)
    &=q_+
      +\left(t-\frac12\right)q_-
      +\tilde q(t),
    &
    q_-&:=q_f-q_i,
    &
    \tilde q(0)&=\tilde q(1)=0.
\label{eq:zero-mode-fp-invariant-q-decomposition}
\end{align}

The functional measure becomes, up to a field-independent Jacobian,
\begin{equation}
    \mathcal Dq
    =\mathcal J_q\,dq_+\,dq_-\,\mathcal D\tilde q.
\end{equation}
The part of the action involving $q$, $\mathcal H_0$, and $N_0$ is
\begin{align}
-\int_0^1dt\,
\mathcal H_0(\dot q+N_0)
&={}
\int_0^1dt\,\dot{\mathcal H}_0\,\tilde q
-q_-\int_0^1dt\,\mathcal H_0
-\int_0^1dt\,N_0\mathcal H_0.
\label{eq:zero-mode-fp-invariant-q-reduction}
\end{align}
The variable $q_+$ does not enter the action and parametrizes the
residual diagonal gauge orbit.  The insertion
$\delta(q_\star-q_+)$ removes precisely this gauge volume.  The integration
over $\tilde q$ gives
\begin{equation}
    \delta[\dot{\mathcal H}_0],
\end{equation}
and hence
\begin{equation}
    \mathcal H_0(t)=h.
\end{equation}
Furthermore, $\delta_0[\dot N_0]$ together with the endpoint values of
$N_0$ gives $N_0(t)=N_\star$.  The remaining finite-dimensional factor is
therefore
\begin{align}
    \int_{\mathbb R}dh
    \int_{\mathbb R}dq_-\,
    \exp\left[-ih(q_-+T_\star)\right],
    \qquad
    T_\star:=\int_0^1dt\,N_0(t)=N_\star.
\label{eq:zero-mode-fp-invariant-fourier-pair}
\end{align}
For a compact gauge orbit, the $q_-$ integral over $\mathbb R$ is understood
as the integral on a lifted fundamental domain together with the sum over
winding sectors.
Integrating first over $q_-$ gives
\begin{equation}
    \int_{\mathbb R}dq_-\,
    e^{-ih(q_-+T_\star)}
    =2\pi\delta(h),
\end{equation}
so that
\begin{equation}
    \mathcal H_0(t)=0
\end{equation}
throughout the interval.  Equivalently, integrating first over $h$ gives
\begin{equation}
    2\pi\delta(q_-+T_\star),
\end{equation}
which is the integrated equation of motion
$q_f-q_i=-\int_0^1dt\,N_0$.  These are two orders of evaluation of the same
Fourier integral and must not be imposed as two independent delta
functions.

The remaining $(G_1,\Pi_{G_1})$ integral imposes $\dot G_1=0$.  After a
field-independent normalization is chosen, one obtains
\begin{align}
    K^{\rm FP}_{\rm inv}(G_{1f},G_{1i})
    &=\delta(G_{1f}-G_{1i}).
\label{eq:zero-mode-fp-invariant-reduced-kernel}
\end{align}
Thus the boundary wavefunctions satisfy a version of the Wheeler-deWitt equation
\begin{equation}
    \hat{\mathcal H}_0\Psi_{\rm inv}=0,
\end{equation}
and the path integral computes the inner product in the invariant Hilbert space.

In BRST/BFV language, the nonconstant normal ghost modes obey Neumann
boundary conditions \eqref{xibconds},
\begin{equation}
    \dot c_0(0)=\dot c_0(1)=0.
\end{equation}

The constant ghost zero mode left unfixed by the lapse gauge is removed, or
equivalently saturated, by the Faddeev--Popov factor associated with
$\chi_{\rm diag}$.  The lapse and diagonal gauge determinants are both
field independent.  Since the transformation
$(a_0,p_0;k_0,-\Phi_0)\mapsto(G_1,\Pi_{G_1};q,-\mathcal H_0)$ is canonical,
this gauge choice introduces no additional field-dependent zero-mode
measure and preserves the measure cancellation established above.
\footnote{Appendix~\ref{app:alternative-invariant-fp} evaluates the same invariant inner product using an alternative gauge condition.}

Since the inner product are equal
\begin{equation}
    K_{\rm inv}^{\rm FP}(G_{1f},G_{1i})
    =K_{\rm co}^{\rm FP}(G_{1f},G_{1i}),
\end{equation}
the rigging map $\eta$, \eqref{eq:groupaveraging}, provides an \emph{isomorphism}, not only an one-to one mapping.



\subsection{Comparison with JT with asymptotic boundary}

As we have explained earlier in \ref{sec:introduction}, in the standard JT gravity gravitational path-integral, the boundary has preferred coordinate and the boundary reparameterization is not gauged. Thus the cylinder partition function with fixed length and constant extrinsic curvature boundary condition, differs from what we have computed by the Fadeev-Popov determinant. Apart from the extra FP determinant for the boundary gauge symmetry, we also have to take into account the measure for the boundary fields. In the standard path-integral, the gauge transformation parameters $\xi,~\xi_\perp$ vanish at the boundary, whereas in our case we gauge-fix the boundary diffeomorphism for which we get an additional FP determinant. Thus we need to divide by the FP determinant for the gauge fixing condition $a'=k'=0$. We assume the regularization by mode expansion in time. Then it is enough to multiply the factor
\begin{equation}
    \frac{1}{\Big[\prod_{n\neq0}(n^2+k_0^2\pm a_0^2)\Big]}\propto\frac{k_0^2\pm a_0^2}{\sinh^2(\pi\sqrt{k_0^2\pm a_0^2})},
\end{equation}
\emph{only once}. This is because the number of modes for Dirichlet boundary condition is smaller than that of free boundary condition by one, see Appendix. \ref{sec:worldline}. We also impose the boundary condition, fixing $a_0$ and $k_0$, by inserting delta function $\delta(a_0-a_{0i})\delta(k_0-k_{0i})$ in the usual JT calculation, which coincides with our boundary gauge fixing determinant as in \eqref{eq:boundaryconditionmeasure}.

Furthermore, for the boundary being a marked one, there is additional physical degree of freedom, the twist mode measuring the relative twist between the two boundaries. For this reason we cannot fix to $N_{0\perp}=0$ gauge, and instead we have to take $\dot{N}_{0\perp}=0$ gauge. However, this difference only contributes as a constant multiplicative factor. With our convention
$e^1=a(d\theta+N_\perp dt)$, the homogeneous normal flow gives
\begin{equation}
    X(1)=\theta+\Theta,
    \qquad
    \Theta:=-\int_0^1dt\,N_{0\perp}(t).
\end{equation}
 Quotienting by boundary-fixing Dehn twists identifies
$\Theta\sim\Theta+2\pi$. This periodicity does not, by itself,
make the twist contribution constant: for a geodesic of length
$\lambda$, the length-valued twist satisfies
\begin{equation}
    \tau=\frac{\lambda}{2\pi}\Theta,
    \qquad
    d\tau=\frac{\lambda}{2\pi}\,d\Theta
    \quad\text{at fixed }\lambda.
\end{equation}
The twist measure must therefore be included in the comparison.

In the AdS sector with  $k_0^2+a_0^2=2G_1>0$, setting
$\lambda=2\pi\sqrt{\mathcal 2G_1}$ gives
\begin{equation}
    \delta(G_{1f}-G_{1i})
    =
    \frac{4\pi^2}{\lambda_i}
    \delta(\lambda_f-\lambda_i).
\end{equation}
If the combined endpoint and twist factors introduce no further
$G_1$-dependent normalization, the proposed correction
yields, up to an overall constant,
\begin{equation}
    Z_{\rm ND}^{\rm cyl}
    \propto
    \frac{G_{1i}}{2\sinh^2(\pi\sqrt{2G_{1i}})}
    \delta(G_{1f}-G_{1i})
    =
    \frac{\lambda_i\,\delta(\lambda_f-\lambda_i)}
    {4\sinh^2(\lambda_i/2)}.
\end{equation}
After Wick rotation, through $k=-aK_L=iaK_E$, we have $\lambda=L\sqrt{1-K_{\rm E}^2}$ in the
hyperbolic sector, and the last expression agrees with the
Euclidean ND cylinder amplitude in (3.49) of
Ref.~\cite{Goel:2020yxl}. This agreement is a consistency check,
rather than an independent derivation of the boundary measure.


\subsection{Lapse contour}

When the real part of $N$ is positive, then $t$ is future-directed. When it is negative, $t$ is past-directed and therefore the "final" timeslice may not be the future of the "initial" timeslice. Thus if the lapse changes sign, our coordinate chart covers the same point multiple times. The same issue exists for spatial direction, if we allow $a$ to take \emph{negative} values.
\begin{figure}[t]
	\begin{center}
		\includegraphics[width=5cm,clip]{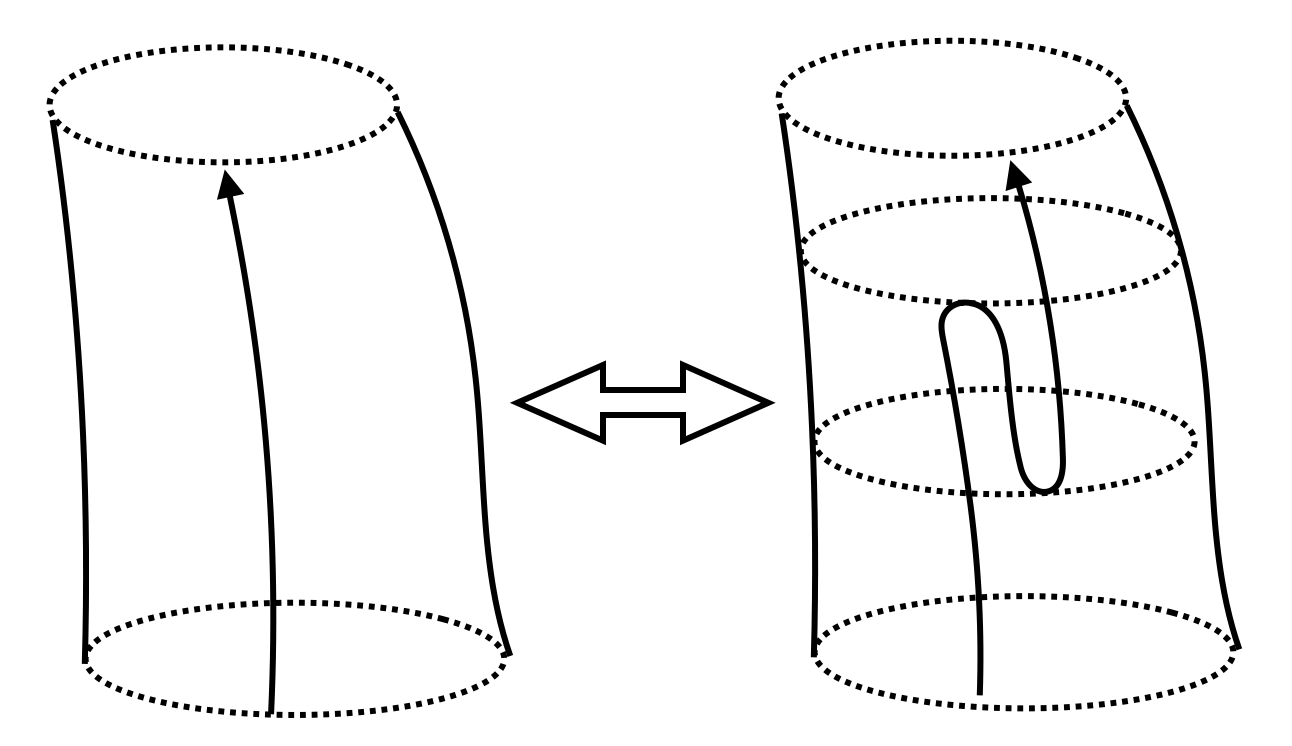}
        \caption{When the lapse can alternate its sign, the value of fields can be multi-valued on the spacetime manifold. Such multi-valuedness is merely apparent, and is removed via the Hamiltonian gauge constraint.}
	    \label{fig:lapsecontour}
    \end{center}
\end{figure}
In general, we should be worried about such multi-valued fields may lead to inconsistent with causality. To address this problem, restriction to positive $N$ was considered in  \cite{Teitelboim:1983fh}. However, the Hamiltonian constraint ensures that such multi-valuedness is just a gauge redundancy. In the co-invariant inner product, as long as $\mathcal{H}_0$ is Hermitian and time-independent, the time-folds in the operator $\mathcal{T}e^{-i\int dt N_0\hat{\mathcal{H}_0}}$ cancel with each other.

We emphasize that having negative lapse is natural from the viewpoint of inner product, in order for the complex conjugate being consistent. For example, if $(a_f,k_f)$ is in the future of $(a_i,k_i)$ and the range of $N_0$ is restricted to be positive, the following equality would not hold
\begin{equation}
    \langle\langle a_f,k_f|a_i,k_i\rangle\rangle^*=\langle\langle a_i,k_i|a_f,k_f\rangle\rangle=\langle a_i,k_i|\int_\Gamma dN_0~e^{-iN_0\hat{\mathcal{H}}_0}|a_f,k_f\rangle.
\end{equation}
Here $\Gamma$ is the integration contour of $N_0$, which is now going to be generalized from $\mathbb{R}$ to $\mathbb{R}-i\epsilon$. The Hermicity of $\int_\Gamma dN_0e^{-iN_0\hat{\mathcal{H}_0}}$ is crucial in order to construct co-invariant Hilbert space from extended Hilbert space which requires $\epsilon$ to be infinitesimal. 

From the Kontsevich-Segal-Witten condition \cite{Kontsevich:2021dmb, Witten:2021nzp}, the lapse $N$ should not be purely real, but needs to have at least small imaginary part. Furthermore, the sign of the imaginary part cannot change, and thus it never crosses the real line. In our path-integral, we considered the contour $\Gamma$ to be $-\infty$ to $+\infty$ with infinitesimal constant imaginary term similar to $i\epsilon$ prescription. When we consider a Hamiltonian that is bounded from below, then we should take negative imaginary part $-i\epsilon$, so that the operator $e^{-i\int_{\mathbb{R}-i\epsilon} N_0\mathcal{H}_0}$ is well-defined \cite{Banihashemi:2024aal}. On the other hand, in our current model, the Hamitonian is neither bounded from above nor below. Thus this argument does not determine the sign of the imaginary part.

We note that the different integration contour $[0,\infty]$ ensures the final timeslice is the future of the initial timeslice, while the Hamiltonian constraints are not enforced in such contour \cite{Teitelboim:1983fh}. We refer the reader to \cite{Casali:2021ewu} for detailed explanation on the connection between the lapse contour and group-averaging.

\paragraph{Positive lapse contour: propagator.}
For comparison, restricting to positive lapse selects future-directed propagation \cite{Teitelboim:1983fh}.
Appendix~\ref{app:positive-lapse-reduction} evaluates the phase-space path integral \eqref{eq:secondorderZ} with the pointwise restriction $N(t,\theta )>0$, using the same phase-space measure and fixed boundary representatives, and obtains the kernel \eqref{eq:app-positive-lapse-kernel}.
We emphasize that now the \emph{global structure} of the gauge group is modified, although they are locally the same away from $N=0$.
On a regular connected component of the gauge orbit with $q\in\mathbb R$, this gives the same zero-mode integral \eqref{eq:physicalpathintegral} with $N_0>0$.
Using $\dot q=-N_0$ and $\hat{\mathcal H}_0=i\partial_q$, we obtain
\begin{align}
    &K_+
    :=
    \delta(G_{1f}-G_{1i})
    \int_0^\infty dN_0\,\delta(q_f-q_i+N_0)
    =
    \delta(G_{1f}-G_{1i})\Theta(q_i-q_f),
    \\
    &\hat{\mathcal H}_{0,f}K_+
    =
    -i\,\delta(G_{1f}-G_{1i})\delta(q_f-q_i).
\end{align}
Thus $K_+$ is a Wheeler--DeWitt Green function with a contact source from $N_0=0$, rather than the physical inner-product kernel \cite{Casali:2021ewu}.
It satisfies the homogeneous constraint away from this source.
The step function records only the ordering of the boundary representatives; $q$ remains a gauge coordinate.
Defining $K_-$ by the same integral over $-\infty<N_0<0$, their sum gives
\begin{align}
    K_++K_-
    &=
    \delta(G_{1f}-G_{1i})
    \bigl[\Theta(q_i-q_f)+\Theta(q_f-q_i)\bigr]
    \nonumber\\
    &=
    \delta(G_{1f}-G_{1i}),
\end{align}
reproducing the physical inner-product kernel \eqref{eq:co-invariant-result}.
These kernel identities are understood distributionally.

\subsection{Relation to mini-superspace approximation}

Starting from the partially gauge-fixed path-integral given by \eqref{eq:physicalpathintegral}, integrating out momenta yields
\begin{equation}
\begin{split}
    Z&=\int \frac{dN_0\prod_{t}\Big(da_0d\Phi_0\Big)}{\prod_t|N_0|}
    ~e^{i\int dt\Big(-\frac{\dot{a}_0\dot{\Phi}_0}{N_0}\pm N_0a_0\Phi_0\Big)+i[\frac{\dot{a}_0}{N_0}\Phi_0]^1_0}\\
    &
    \times \delta(a_0(t=0)-a_{0i})\delta(k_0(t=0)-k_{0i})\delta(a_0(t=1)-a_{0f})\delta(k_0(t=1)-k_{0f})
    .
    \end{split}
\end{equation}
Using the zeta function regularization for mode expansion in time, using the fact that $N_0$ has free boundary condition before the gauge fixing $\dot{N}_0=0$, $\frac{1}{\prod_t|N_0|}$ can be replaced by $\frac{1}{N_0^{1/2}}$. Thus the path-integral measure diverges at the origin unless the lapse contour is in complex plane, though the path-integral itself does not diverge. We emphasize that our approach is exact without truncation like mini-superspace. The solvability of this model comes from the fact that most of degrees of freedom are dependent due to constraints.


\section{Zero-mode BFV--BRST description}
\label{sec:zero-mode-brst}

We give a BFV--BRST description of the reduced path integrals of
Section~\ref{sec:gauge-fixed-path-integral}, obtained after the nonzero spatial modes
have been integrated out.  We work on a regular connected
component on which the remaining gauge orbit is non-compact. The additional global identifications are discussed in Section~\ref{sec:perturbative-hilbert}. Important references include \cite{Batalin:1994rd, Held:2025mai}.

Let us use the canonical coordinates
\begin{equation}\label{eq:zero-mode-canonical-variables}
    (G_1,\Pi_{G_1};q,-\mathcal H_0),
    \qquad
    p_0da_0-\Phi_0dk_0=\Pi_{G_1}dG_1-\mathcal H_0dq.
\end{equation}
The reduced action is
\begin{equation}\label{eq:zero-mode-reduced-action}
    I_0=\int_0^1dt\,
    \left[\Pi_{G_1}\dot G_1-\mathcal H_0(\dot q+N_0)\right],
\end{equation}
and the residual gauge transformation gives
\begin{equation}\label{eq:zero-mode-geometric-gauge-transformations}
    \delta_\xi q=-\xi,
    \qquad
    \delta_\xi N_0=\dot\xi.
\end{equation}
We first set up the langauge of BRST formalism, defining various quantities. Following this, we will discuss the two different boundary conditions mentioned in the previous section and show how they can naturally be understood in this BRST language thereby giving the BRST way of understanding the co-invariant and invariant states. For full BRST/BFV setups, see appendix \ref{appendix:BRST}.


\subsection{The zero-mode BRST complex}
\label{subsubsec:zero-mode-full-complex}
In the BFV formalism, a canonical conjugate is introduced for every field
\begin{equation}
    (N_0,\Pi_0),\qquad(c_0,\rho_0),
    \qquad(\bar c_0,\bar\rho_{\,0}),
\end{equation}
with unit graded Poisson brackets.  Their operator realizations satisfy
\begin{align}
    [\hat q,\hat{\mathcal H}_0]&=-i,
    & [\hat N_0,\hat\Pi_0]&=i,
    \notag\\
    [\hat c_0,\hat\rho_0]_+&=i,
    & [\hat{\bar c}_0,\hat{\bar\rho}_{\,0}]_+&=i,
    \label{eq:zero-mode-canonical-brackets}
\end{align}
where the subscript $_+$ implies it is an anti-commutator. We could also define $b$-fields $b_0\equiv-i\rho_0,~\bar{b}_0\equiv-i\bar{\rho}_0$. The BRST charge and the
left-acting BRST differential are
\begin{equation}\label{eq:zero-mode-brst-charge}
    Q_{B,0}=-c_0\mathcal H_0-\Pi_0\bar\rho_{\,0},
    \qquad
    sF:=\{Q_{B,0},F\}_{\rm P},
\end{equation}
with 
\begin{equation}\label{eq:zero-mode-brst-Leibniz}
    s(FG)=(sF)G+(-1)^{|F|}F(sG).
\end{equation}
In particular,
\begin{align}
    &sq=-c_0,~~~~~s\rho_0=-\mathcal H_0,
    ~~~~~sN_0=\bar\rho_{\,0},~~~~~s\bar c_0=-\Pi_0,
    \notag\\
    &sc_0=s\bar\rho_{\,0}=0, ~~~~~~~s\mathcal H_0=s\Pi_0=sG_1=s\Pi_{G_1}=0.
    \label{eq:zero-mode-brst-transformations}
\end{align}
These transformations are nilpotent as it should be, see more explanation in Appendix \ref{appendix:BRST}. After the ghost
momenta are integrated out, $\bar\rho_{\,0}=\dot c_0$, so that
$(sq,sN_0)=(-c_0,\dot c_0)$ reproduces
Eq.~\eqref{eq:zero-mode-geometric-gauge-transformations}.

\paragraph{Ghost doublets.}
Choose the state $|\downarrow\downarrow\rangle$ of ghost number $\text{gh}=-1$, annihilated
by $\rho_0$ and $\bar c_0$ as $\rho_0|\downarrow\downarrow\rangle=\bar c_0|\downarrow\downarrow\rangle=0$, and define 
\begin{equation}\label{eq:zero-mode-ghost-basis}
\begin{aligned}
    |\uparrow\downarrow\rangle
    :=-c_0|\downarrow\downarrow\rangle,&~~
    |\downarrow\uparrow\rangle
    :=-\bar\rho_{\,0}|\downarrow\downarrow\rangle,~~&(\text{gh}=0),
    \\
    |\uparrow\uparrow\rangle
    &:=c_0\bar\rho_{\,0}|\downarrow\downarrow\rangle~~&(\text{gh}=1).
\end{aligned}
\end{equation}
The relative signs in this basis are part of the convention.  They
allow the BRST complex and the mixed boundary pairing,  with the
following simple forms :
\begin{equation}\label{eq:zero-mode-full-brst-complex}
    |\downarrow\downarrow\rangle
    \overset{\hat Q_{B,0}}{\longrightarrow}
    |\downarrow\uparrow\rangle\oplus|\uparrow\downarrow\rangle
    \overset{\hat Q_{B,0}}{\longrightarrow}
    |\uparrow\uparrow\rangle,
\end{equation}
\begin{equation}
\begin{split}
    \hat{Q}_{B,0}\Big(|f\rangle_{\rm aux}\otimes|r\rangle_N
    \otimes|\downarrow\downarrow\rangle_{\rm gh}\Big)&=
    \hat{\mathcal{H}}_{0}|f\rangle_{\rm aux}\otimes|r\rangle_N
    \otimes|\uparrow\downarrow\rangle_{\rm gh}+
    |f\rangle_{\rm aux}\otimes\hat{\Pi}_{0}|r\rangle_N
    \otimes|\downarrow\uparrow\rangle_{\rm gh}
    ,\\
    \hat{Q}_{B,0}\Big(|f\rangle_{\rm aux}\otimes|r\rangle_N
    \otimes|\uparrow\downarrow\rangle_{\rm gh}\Big)&=-|f\rangle_{\rm aux}\otimes\hat{\Pi}_{0}|r\rangle_N\otimes|\uparrow\uparrow\rangle_{\rm gh},
    \\
    \hat{Q}_{B,0}\Big(|f\rangle_{\rm aux}\otimes|r\rangle_N
    \otimes|\downarrow\uparrow\rangle_{\rm gh}\Big)&=
    \hat{\mathcal{H}}_{0}|f\rangle_{\rm aux}\otimes|r\rangle_N
    \otimes|\uparrow\uparrow\rangle_{\rm gh}.
\end{split}
\label{qbconds}
\end{equation}
Here $|r\rangle_N$ denote an arbitrary state in $(N_0,\Pi_0)$-sector. Using this nilpotent BRST charge, we can construct cohomology class which are closed under the BRST operator and equivalent under BRST-exact state shift, which was used in the operator construction of physical states \cite{Kugo:1977zq, Kugo:1979gm, Henneaux:1992ig}.
\paragraph{Co-invariant and invariant BRST cohomology.}
Let us construct the representatives of the two types of BRST cohomology, the co-invariant and invariant BRST cohomology, using representation in terms of subspace in the extended Hilbert space. The co-invariant representatives are
\begin{align}
    |{\rm co};f\rangle_{\rm BFV}
       &=|f\rangle_{\rm aux}\otimes|\Pi_0=0\rangle_N
          \otimes|\uparrow\downarrow\rangle_{\rm gh}.
          \label{eq:zero-mode-co-state}
\end{align}
These co-invariant representatives are BRST closed, thus the subspace of co-invariant representative gives a set of BRST cohomology class. The representation above is not yet unique due to the equivalence relation among co-invariant representatives. The equivalence relation, which can be understood easily from \eqref{qbconds}, is 
\begin{equation}\label{eq:equivalence}
    |f\rangle_{\rm aux}\sim |f\rangle_{\rm aux}+\hat{\mathcal H}_0|g\rangle_{\rm aux}.
\end{equation}
The invariant representatives are
\begin{align}
    |{\rm inv};f;r\rangle_{\rm BFV}
       &=-|f\rangle_{\rm inv}\otimes|r\rangle_N
          \otimes|\downarrow\uparrow\rangle_{\rm gh},~~(\hat{\mathcal{H}}_0|f\rangle_{\rm inv}=0).
          \label{eq:zero-mode-inv-state}
\end{align}
The invariant representative is BRST closed because
$\hat{\mathcal H}_0|f\rangle_{\rm inv}=0$ and
$\bar\rho_{\,0}|\downarrow\uparrow\rangle=0$, thus the subspace of invariant representative gives a set of BRST cohomology class. The equivalence relation among invariant representatives is
\begin{equation}
    |r\rangle_N\sim |r\rangle_N+\hat{\Pi}_0|s\rangle_N.
\end{equation}
Thus we can further take representatives within $(N_0,\Pi_0)$-sector as
\begin{equation}
    |N_0=\nu\rangle_N,
\end{equation}
the $\nu$-eigenstate of $\hat{N}_0$.

The rigging map $\eta$ defines one-to-one mapping between the co-invariant and invariant cohomology. Indeed, the extended map from co-invariant cohomology class to invariant one
\begin{equation}
    \eta: [|{\rm co};f\rangle_{\rm BFV}]_{\rm co}\rightarrow[|{\rm inv}; -\hat{\eta}f;\nu\rangle_{\rm BFV}]_{\rm inv},
\end{equation}
where
\begin{equation}
    \eta\equiv\int_{\Gamma} dTe^{-iT\hat{\mathcal{H}_0}}
\end{equation}
respects the equivalence relation \eqref{eq:equivalence}, with $ [|{\rm co};f\rangle_{\rm BFV}]_{\rm co}$ etc representing cohomology class. Note that
\begin{equation}
    \eta\int d\mathcal{H}_0f(\mathcal{H}_0)|G_1,\mathcal{H}_0\rangle_{\rm aux}=2\pi f(0)|G_1,\mathcal{H}_0=0\rangle_{\rm aux},
\end{equation}
show that the rigging map is indeed one-to-one mapping.


\paragraph{Indefinite inner product in the extended Hilbert space.}

We take $\hat c_0,\hat{\bar c}_0$ to be Hermitian and
$\hat N_0,\hat{\Pi}_{0}, \hat\rho_0,\hat{\bar\rho}_{\,0}$ to be anti-Hermitian with respect to the following indefinite inner product. As a consequence, the BRST operator $\hat{Q}_{B,0}$ is Hermitian. The negativity in the indefinite inner product will disappear in the physical inner product
\footnote{We can also construct indefinite inner product where $N_0$ and $\Pi_0$ are anti-Hermitian, but BRST charge is Hermitian. This is merely an issue of convention.}. 

Following \cite{Held:2025mai}, we can set
\begin{equation}\label{eq:zero-mode-lapse-ghost-form}
\begin{aligned}
    \langle g_2|g_1\rangle_N
       &=\int_{\mathbb R}dN_0\,
          \overline{g_2(-N_0)}g_1(N_0),\\
    \langle\downarrow\uparrow|\uparrow\downarrow\rangle_{\rm gh}
       &=\langle\uparrow\downarrow|\downarrow\uparrow\rangle_{\rm gh}
        =-1,\\
    \langle\downarrow\downarrow|\uparrow\uparrow\rangle_{\rm gh}
       &=\langle\uparrow\uparrow|\downarrow\downarrow\rangle_{\rm gh}
        =-1,
\end{aligned}
\end{equation}
with all other ghost basis pairings zero and the bar means usual complex conjugate.  With this form,
$\hat N_0$ and $\hat\Pi_0=-i\partial_{N_0}$ are anti-Hermitian but have real eigenvalues.
We normalize {their eigenstates} as
\begin{equation}
\begin{split}
    \langle N_0=\nu|\Pi_0=0\rangle_N=1,
    \qquad
    \mathbb{I}_N=\int d\nu|\nu\rangle\langle-\nu|_N=
    \int \frac{d\Pi_0}{2\pi}|\Pi_0\rangle\langle -\Pi_0|_N,
    \end{split}
\end{equation}
Consequently, a coordinate path integral with
$N_0(0)=N_0(1)=N_\star$ corresponds to a ket lapse label $+N_\star$ and a bra lapse
label $-N_\star$.  This is a consequence of the lapse form, not a change in the coordinate endpoint values. In particular, the value of the lapse remain constant and real although $\hat{N}_0$ is anti-Hermitian.

With the phase in \eqref{eq:zero-mode-inv-state}, the BFV
inner product between invariant and co-invariant states is
\begin{equation}\label{eq:zero-mode-canonical-pairing}
\begin{aligned}
    \mathcal B(\psi,[f])
       &:={}_{\rm BFV}\langle{\rm inv};\psi;\nu
                           |{\rm co};f\rangle_{\rm BFV}\\
       &=\int_{\mathcal I}dG_1\int_{\mathbb R}dq\,
                    \overline{\psi(G_1)}f(G_1,q).
\end{aligned}
\end{equation}
It is independent of $\nu$ and of the co-invariant representative,
since the shift $f\rightarrow f+i\partial_q\lambda(G_1,q)$ does not affect the inner product.


\subsection{Co-invariant realization: group averaging}
\label{subsubsec:zero-mode-co-realization}

For co-invariant state, we impose the boundary condition for $G_1,q,\Pi_0,c_0,\bar c_0$,
with
\begin{equation}\label{eq:co-invariant-bc}
    \Pi_0=c_0=\bar c_0=0\qquad (t=0,1),
\end{equation}
in the path-integral in BRST/BFV formalism. The lapse and the conjugate ghost momenta are free at the endpoints.

Let us first show the equivalence between the original gauge-fixed gravitational path-integral in co-invariant gauge with that of BRST/BFV formalism. The BRST/BFV action for the gauge-fixing in subsection \ref{subsubsec:zero-mode-fp-group-averaging} is given by 
\begin{align}
 I_{\rm co}={}&\int_0^1dt\,
 \Big[
 \Pi_{G_1}\dot G_1-\mathcal H_0(\dot q+N_0)
 +\Pi_0 \dot N_0
 +\rho_0\dot c_0+\bar\rho_{\,0}\dot{\bar c}_0
 -\rho_0\bar\rho_{\,0}
 \Big].
 \label{eq:zero-mode-co-action}
\end{align}
The BRST variation of this action is $sI_{\rm BRST/BFV}=[\Pi_0\bar \rho_0]^1_0$,  which is vanishing due to the boundary condition. Integrating the ghost momenta $\rho_0$ gives $\bar\rho_{\,0}=\dot c_0$. Since the remaining ghost action is non-interacting with imposed Dirichlet boundary condition \eqref{eq:co-invariant-bc}, integrating out ghosts gives the Dirichlet determinant. Integrating out $\Pi_0$, combined with \eqref{eq:co-invariant-bc}, gives delta function $\delta(\dot{N}_0)$ everywhere in the bulk.  Thus we have
\begin{equation}\label{eq:proposal-co-fp-factor}
 \int \mathcal D\Pi_0\,\mathcal D({\rm ghosts})\,
 e^{i\int_0^1dt\,
 \Big[
 \Pi_0 \dot N_0
 +\rho_0\dot c_0+\bar\rho_{\,0}\dot{\bar c}_0
 -\rho_0\bar\rho_{\,0}
 \Big]}
 \ \propto\ 
 \det\nolimits_D(-\partial_t^2)\,\delta[\dot N_0].
\end{equation}
This exactly reproduces the FP prescription of
Subsection~\ref{subsubsec:zero-mode-fp-group-averaging}.

Let us now write the BRST/BFV path-integral in operator representation. For the gauge-fixing fermion density $\Psi_N$, we have
\begin{equation}
    \hat\Psi_N=\hat\rho_0\hat N_0,
    \qquad
    [\hat Q_{B,0},\hat\Psi_N]_+
    =-i\hat N_0\hat{\mathcal H}_0
      -i\hat\rho_0\hat{\bar\rho}_{\,0}.
\end{equation}
Thus the extended phase space path-integral including ghosts can be rewritten as 
\begin{equation}
 K^{\rm BFV}_{\rm co}(f_2;f_1)={}_{\rm BFV}\langle{\rm co};f_2|
 e^{[\hat Q_{B,0},\hat\Psi_N]_+}
 |{\rm co};f_1\rangle_{\rm BFV}.
\end{equation}
Note that we have
\begin{equation}
 {}_{\rm gh}\langle\uparrow\downarrow|
 (1-i\hat\rho_0\hat{\bar\rho}_{\,0})
 |\uparrow\downarrow\rangle_{\rm gh}=1.
\end{equation}
Consequently the BFV matrix element gives the group average
\begin{equation}\label{eq:proposal-co-operator-pairing}
  K^{\rm BFV}_{\rm co}(f_2;f_1)={}_{\rm BFV}\langle{\rm co};f_2|
 e^{[\hat Q_{B,0},\hat\Psi_N]_+}
 |{\rm co};f_1\rangle_{\rm BFV}
 =\int_{\mathbb R}dN_0\,
 \langle f_2|e^{-iN_0\hat{\mathcal H}_0}|f_1\rangle_{\rm aux}.
\end{equation}
This is the zero-mode version of the BFV group-averaging construction
in \cite{Held:2025mai}. We thus have
\begin{align}
 K^{\rm BFV}_{\rm co}(G_{1f},q_f;G_{1i},q_i)=
 K^{\rm FP}_{\rm co}(G_{1f},q_f;G_{1i},q_i)=\delta(G_{1f}-G_{1i}),
 \label{eq:zero-mode-co-kernel}
\end{align}
using \eqref{eq:co-invariant-result}.


\subsection{Invariant realization: lapse and diagonal gauge fixing}
\label{subsubsec:zero-mode-inv-realization}

We now represent the second gauge fixing condition mentioned in 
Subsection~\ref{subsubsec:zero-mode-fp-invariant}. We impose the boundary condition for $G_1,N_0,\rho_0,\bar \rho_0$ in addition to diagonal gauge-fixing $(q_i+q_f)/2$ \footnote{Appendix~\ref{app:alternative-invariant-fp} evaluates the same invariant inner product using an alternative gauge condition.},
\begin{equation}\label{eq:zero-mode-invariant-boundary-data}
 N_0=N_\star\ne0,
 \qquad
 \rho_0=\bar\rho_{\,0}=0,~~(t=0,1).
\end{equation}

The BRST/BFV action for the invariant gauge-fixing in subsection \ref{subsubsec:zero-mode-fp-invariant} is given by
\begin{align}
 I_{\rm inv}={}&I_{\rm co}+s\Psi_{\rm diag}.
 \label{eq:zero-mode-co-action}
\end{align}
To understand this expression, we first use the general decomposition
\begin{equation}
\begin{split}
 &f(t)=f_{\rm av}+f_0^{\rm nz}(t),
 \qquad \int_0^1dt\,f_0^{\rm nz}(t)=0,
 \end{split}
\end{equation}
in addition to the boundary values
\begin{equation}
 q_+=\frac{q(0)+q(1)}2,
 \qquad c_+=\frac{c_0(0)+c_0(1)}2,
 \qquad \chi_{\rm diag}=q_\star-q_+.
\end{equation}
The BRST transformation gives
\begin{equation}
 sq_+=-c_+,\qquad
 s\bar c_{\rm av}=-\Pi_{\rm av},\qquad
 s\Pi_{\rm av}=sc_+=0.
\end{equation}
Using them, the additional gauge-fixing fermion $\Psi_{\rm diag}$ can be understood as
\begin{equation}\label{eq:zero-mode-inv-diagonal-fermion}
 \Psi_{\rm diag}=\bar c_{\rm av}\chi_{\rm diag},
 \qquad
 s\Psi_{\rm diag}
 =-\Pi_{\rm av}\chi_{\rm diag}-\bar c_{\rm av}c_+.
\end{equation}
The BRST variation of this action is $sI_{\rm inv}=[\Pi_0\bar \rho_0]^1_0$, which is vanishing due to the boundary condition. 

We will now show the equivalence between the original gauge-fixed gravitational path-integral in invariant gauge with that of BRST/BFV formalism. Since the lapse endpoint values agree,
$\int\Pi_0\dot N_0dt=\int\Pi_0^{\rm nz}\dot N_0dt$.
Thus integrating $\Pi_0^{\rm nz}$ gives $\delta_{\rm nz}[\dot N_0]$. Next, integrating the ghost momenta gives
$\bar\rho_{\,0}=\dot c_0$ and $\rho_{\,0}=-\dot{\bar{c}}_0$, imposing the Neumann boundary condition $\dot{c}_0=\dot{\bar{c}}_0=0$ at the endpoints.
The remaining ghost terms in the action, including the diagonal insertion, are therefore
\begin{equation}
 \int_0^1dt\, \dot c_0\dot{\bar{c}}_0
 -\bar c_{\rm av}c_+.
\end{equation}
With $\int d\bar c_{\rm av}\,dc_+\,c_+\bar c_{\rm av}=1$, the integral gives
\begin{equation}
 -i\int\,d\bar c_{\rm av}\,dc_+\,
 e^{-i\bar c_{\rm av}c_+}
 =1.
\end{equation}
Consequently, integrating $c_0,\bar c_0,\rho_0,\bar \rho_0, $ gives
\begin{equation}\label{eq:proposal-invariant-fp-factor}
 \Delta_{\rm inv}=\det{}'_{NN}(-\partial_t^2),
\end{equation}
The resulting action is therefore exactly
\begin{equation}
 I_0+\int_0^1dt\,\Pi_0^{\rm nz}\dot N_0
       -\Pi_{\rm av}\left(q_\star-\frac{q_i+q_f}{2}\right),
\end{equation}
which is the invariant-gauge FP action \eqref{eq:zero-mode-fp-invariant-ghost-free-action}, with identification $\Pi_{\rm av}=-\lambda_0$. The integrations over the
unfixed components of $q$, already evaluated in
Subsection~\ref{subsubsec:zero-mode-fp-invariant}, impose
$\dot{\mathcal H}_0=0$. 

Thus the path-integral can be rewritten as 
\begin{equation}
 K^{\rm BFV}_{\rm inv}(f_2;f_1)={}_{\rm BFV}\langle{\rm inv};f_2;\nu|
 e^{[\hat Q_{B,0},\hat\Psi_N+\hat\Psi_{\rm diag}]_+}
 |{\rm inv};f_1;\nu\rangle_{\rm BFV}.
\end{equation}
Thus we have
\begin{equation}\label{eq:zero-mode-invariant-kernel}
 K_{\rm inv}^{\rm BFV}(G_{1f},G_{1i})
 =K_{\rm inv}^{\rm FP}(G_{1f},G_{1i})
 =\delta(G_{1f}-G_{1i}),
\end{equation}
using \eqref{eq:zero-mode-fp-invariant-reduced-kernel}.
Thus we see that the rigging map $\eta$ provides an isomorphism between the invariant and co-invariant Hilbert spaces.


\section{Global structure of the Hilbert space}\label{sec:perturbative-hilbert}


In this section, we investigate the Hilbert space derived from the Lorentzian gravitational path-integral in the previous sections. The local structure has been extensively discussed in \cite{Held:2024rmg}. New materials we add are its global structure due to the possible singularity at $a=0$, the gauging reflection and the overlap with the Hartle-Hawking state. 

We will omit subscript $0$ in $a_0,k_0,\Phi_0,p_0$ in this section. Let us consider the extended Hilbert space $H_{\text{aux}}$, and its states $|\psi\rangle=\int dadk\psi(a,k)|a,k\rangle\in H_{\text{aux}}$. The action of the Hamiltonian constraint $\mathcal{H}_0$ on the wavefunction $\psi(t,a,k)$ of  $e^{-i\mathcal{H}_0t}|\psi\rangle=\int dadk\psi(t,a,k)|a,k\rangle$ is, as described in \cite{Held:2024rmg},
\begin{equation}
    \hat{\mathcal{H}}_0=i\left(k\frac{\partial}{\partial a}\mp a\frac{\partial}{\partial k}\right),~
    i\frac{\partial}{\partial t}\psi(t,a,k)= \hat{\mathcal{H}}_0\psi(t,a,k).
\end{equation}
Using the method of characteristics, the general solution with the initial condition $\psi(0,a,k)=\psi_0(a,k)$ is found to be, for AdS and dS
\begin{align}
    \psi_t(a,k)_{AdS}= \psi_0\Bigl(a\cos t + k\sin t,\;  k\cos t-a\sin t \Bigr),\nonumber\\
    \psi_t(a,k)_{dS}= \psi_0\Bigl(a\cosh t + k\sinh t,\; a\sinh t + k\cosh t\Bigr).
\end{align}
Here \(a\) is regarded as a signed vielbein variable rather
than the positive metric scale factor \(\sqrt{g_{\theta \theta }}\), therefore $(a,k)\in\mathbb R^2$. 

The parameter $u$ and $v$ appearing in the following also provide the explicit realizations of $q$. 


\subsection{Lorentzian closed AdS}

Let us introduce polar coordinates
\begin{align}
    a=b\cos(-u),~
    k=b\sin(-u),
    ~
    b>0,~
    u\sim u+2\pi .
\end{align}
The Hamiltonian constraint and the time-evolved
wavefunction are
\begin{align}
    \hat{\mathcal H}_0
    =
    i\frac{\partial}{\partial u},
    ~
    \psi_t(b,u)=\psi_0(b,u+t).
\end{align}
Correspondingly, the generalized position eigenstates
transform with the inverse shift,
\begin{equation}
\begin{split}
    e^{-it\hat{\mathcal H}_0}
    |b,u\rangle
    =
    |b,u-t\rangle .
\end{split}
\end{equation}
Thus the Hamiltonian gauge orbit at fixed \(b\) is a
circle of period \(2\pi\) in $(a,k)$-extended Hilbert space. This period is independent from $b$. As explained in \ref{subsec:canonical-zero-mode-variables}, we must divide the path-integral by an extra $|\mathbb{Z}|$, which is equivalent to using the group averaging map
\begin{equation}
    \eta'=\int_0^{2\pi}dN_0\,
    e^{-iN_0\hat{\mathcal H}_0}.
\end{equation}
Alternatively, we can consider $(q,G_1)$-extended Hilbert space with the integration contour $\Gamma=\mathbb{R}$.

The orbit contains the representative 
\begin{equation}
    a=b,~k=0,
\end{equation}
and hence the corresponding co-invariant state may be
denoted by
\begin{equation}
    |G_1=b^2/2\rangle\!\rangle.
\end{equation}
Then the inner product is
\begin{equation}
\begin{split}
    \left\langle\!\left\langle
    G_1
    \middle|
    G_1'
    \right\rangle\!\right\rangle
    &=
    \langle a=b,k=0|
    \int_0^{2\pi}dN_0\,
    e^{-iN_0\hat{\mathcal H}_0}
    |a=b',k=0\rangle
    \\
    &=
    \frac{\delta(b-b')}{b}
    =
    \delta(G_1-G_1').
\end{split}
\end{equation}
The resolution of the identity is
\begin{equation}
    \mathbb I_{\rm AdS}
    =
    \int_0^\infty b\,db\,
    |b\rangle\!\rangle
    \langle\!\langle b|
    =
    \int_0^\infty dG_1\,
    |G_1\rangle\!\rangle
    \langle\!\langle G_1|,
\end{equation}
for the physical Hilbert space
\begin{equation}
    \mathcal H_{\rm phys}
    \simeq
    L^2(\mathbb R_+,b\,db).
\end{equation}


\subsection{Three global prescriptions for the sign of \(a\)}\label{subseq: three sign of a}

The sign of \(a\) requires a global choice.  If \(a\) is
defined as the positive metric scale factor,
\(a=\sqrt{g_{\theta \theta }}\), then \(a<0\) has no geometric
meaning.  Negative values can be introduced only after
regarding \(a\) as a signed vielbein or connection
component.  In that formulation there are three
logically distinct prescriptions.\\
\begin{itemize}
\item
\textbf{Option I. $a\in\mathbb{R}$ and unoriented theory:}
One extends the configuration space to \(a\in\mathbb R\)
and includes the orientation-reversing diffeomorphism $r:\theta \longmapsto -\theta $ in the gauge group. Its action is
$R:(a,k)\longmapsto(-a,-k)$. States related by \(R\) are identified.

\item\textbf{Option II. $a\in\mathbb{R}$ and oriented theory:}
One again allows \(a\in\mathbb R\), but retains the
orientation of the spatial circle as physical data and
does not quotient by \(R\). 

\item\textbf{Option III. $a>0$ theory:}
One restricts the configuration space from the outset to $a>0$. The Hamiltonian flow may then reach the boundary \(a=0\) in finite parameter time, where we have a singularity. The quantum theory is not defined until a boundary prescription at \(a=0\) is specified. The prescription is not determined from the JT theory, but rather by unknown UV theory describing the singularity $a=0$.

\end{itemize}

The gauging of an anti-unitary time-reflection or CRT
symmetry discussed in
\cite{Harlow:2023hjb,Harlow:2025pvj}
is conceptually related, but it should not be identified
with the unitary spatial-orientation reflection \(R\)
considered here.


\subsubsection*{Consequences for closed AdS}

For closed AdS, the Hamiltonian evolution on the signed
plane satisfies
\begin{equation}
    e^{-i\pi\hat{\mathcal H}_0}
    |a,k\rangle
    =
    |-a,-k\rangle .
\end{equation}
Therefore, the reflection is already contained in the Hamiltonian
gauge orbit
\begin{equation}
    R=e^{-i\pi\hat{\mathcal H}_0}.
\end{equation}
Consequently, Options I and II give the
same reduced Hilbert space
\begin{align}
    \mathcal H_{\rm phys}^{\rm(I)}\simeq
    \mathcal H_{\rm phys}^{\rm(II)}
    \simeq
    L^2(\mathbb R_+,b\,db).
\end{align}
Option III gives the same Hilbert space as Option I only for an
appropriate boundary condition at \(a=0\).


\subsection{Lorentzian closed dS}


\subsubsection*{Global dS: \(a^2-k^2>0\)}

The global-dS has two connected components.
We use
\begin{align}
    a=b\cosh u,
    ~
    k=b\sinh u,
    ~
    b\neq0,
    ~
    u\in\mathbb R.
\end{align}
Then we have
\begin{equation}
\begin{split}
    \hat{\mathcal H}_0
    =
    i\frac{\partial}{\partial u},&~G_1=\frac{b^2}{2},\\
    e^{-it\hat{\mathcal H}_0}
    |b\cosh u,b\sinh u\rangle
    &=
    |b\cosh(u-t),b\sinh(u-t)\rangle .
    \end{split}
\end{equation}
A convenient
representative is $a=b,~k=0$. Thus, before imposing the reflection quotient, the
physical states are labeled by $b$, thus the states can be written as $|b\rangle\!\rangle$. Their inner product is
\begin{equation}
    \left\langle\!\left\langle
    b
    \middle|
    b'
    \right\rangle\!\right\rangle
    =
    \frac{\delta(b-b')}{|b|}.
\end{equation}
The three prescriptions now have the following effects:

\begin{itemize}

\item \textbf{Option I:} The reflection $R$ maps
\[
b\longmapsto -b
\]
and the two components are identified.  The physical
Hilbert space is one copy of
\[
L^2(\mathbb R_+,dG_1).
\]

\item \textbf{Option II:} The two components are retained:
\begin{equation}
    \mathcal H_{\rm global}^{\rm(II)}
    \simeq
    L^2(\mathbb R_+,dG_1)
    \oplus
    L^2(\mathbb R_+,dG_1).
\end{equation}
In this theory \(G_1=\frac{b^2}{2}\) alone is not a complete label;
the additional sign of \(b\) is required.

\item \textbf{Option III:} \(a>0\) selects \(b>0\), and hence
only one copy remains.  No boundary condition at
\(a=0\) is needed in this sector because
\[
a=b\cosh u\geq b>0
\]
along the entire orbit.

\end{itemize}
For Options I and III, the resolution of the identity is
\begin{equation}
    \mathbb I_{\rm dS:global}
    =
    \int_0^\infty bdb\,
    |b\rangle\!\rangle
    \langle\!\langle b|.
\end{equation}
For Option II we have
\begin{equation}
    \mathbb I_{\rm dS:global}
    =
    \int_{-\infty}^\infty |b|db\,
    |b\rangle\!\rangle
    \langle\!\langle b|.
\end{equation}


\subsubsection*{Big-bang and big-crunch:
\(k^2-a^2>0\)}

In this region we use
\begin{align}
    a=\beta\sinh v,
    ~
    k=\beta\cosh v,
    ~
    \beta\neq0,
    ~
    v\in\mathbb R.
\end{align}
The sign of \(\beta\) labels the two disconnected
Hamiltonian orbits.  We have
\begin{equation}
\begin{split}
    \hat{\mathcal H}_0
    =
    i\frac{\partial}{\partial v},
    &~
    G_1=-\frac{\beta^2}{2},\\
    e^{-it\hat{\mathcal H}_0}
    |\beta\sinh v,\beta\cosh v\rangle
    &=
    |\beta\sinh(v-t),\beta\cosh(v-t)\rangle .
\end{split}
\end{equation}
A convenient representative of each full signed orbit is $a=0,~k=\beta$, and we denote the corresponding co-invariant state by $|\beta\rangle\!\rangle$. The inner product is
\begin{equation}
    \langle\!\langle\beta|\beta'\rangle\!\rangle
    =
    \frac{\delta(\beta-\beta')}{|\beta|}.
\end{equation}
The three global prescriptions are now genuinely
different.

\begin{itemize}

\item \textbf{Option I:} The reflection acts as
\begin{equation}
    R:\beta\longmapsto-\beta.
\end{equation}
The two branches are identified. The Hilbert space and the resolution of identity are 
\begin{equation}
    \mathcal H_{\rm bang/crunch}^{\rm(I)}
    \simeq
    L^2(\mathbb R_+,\beta\,d\beta),~\mathbb I_{\rm bang/crunch}^{\rm(I)}
    =
    \int_0^\infty \beta\,d\beta\,
    |\beta\rangle\!\rangle
    \langle\!\langle\beta|.
\end{equation}

\item \textbf{Option II:} \(\beta>0\) and \(\beta<0\) remain
independent:
\begin{equation}
    \mathcal H_{\rm bang/crunch}^{\rm(II)}
    \simeq
    L^2(\mathbb R\setminus\{0\},|\beta|\,d\beta).
\end{equation}
Equivalently, this is a direct sum of two copies of
\(L^2(\mathbb R_+,\beta\,d\beta)\).  Its resolution of the
identity is
\begin{equation}
    \mathbb I_{\rm bang/crunch}^{\rm(II)}
    =
    \int_{-\infty}^{\infty}
    d\beta\,|\beta|\,
    |\beta\rangle\!\rangle
    \langle\!\langle\beta|.
\end{equation}

\item \textbf{Option III:} The condition \(a>0\) restricts each
full orbit to a half-orbit:
\begin{equation}
    \beta>0:\quad v>0,
    \qquad
    \beta<0:\quad v<0.
\end{equation}
For \(\beta>0\), evolution toward increasing \(t\)
reaches \(a=0\), and this branch describes a big-crunch.
For \(\beta<0\), the \(a=0\) surface lies in the past and
the branch describes a big-bang. Both half-orbits end at $a=0,~k=\beta$.
Consequently, Option III is not fully defined until a boundary condition at \(a=0\) is specified.

\end{itemize}

\subsection{Comparison with the Hilbert space of BF theory}
\label{subsec:bf-hilbert-comparison}

Let us compare the Hilbert space of JT gravity and  $G=PSL(2,\mathbb R)$ BF theory, whose inifinitesimal gauge theory structure is identical, but globally distinct. Consider $G=PSL(2,\mathbb R)$ BF theory on an oriented
spatial circle, quotienting by all single-valued $G$-valued gauge
transformations, including their winding components, without imposing
nondegeneracy of the vielbein. The states in BF theory can be completely specified by conjugacy class of holonomy. The global structure of the of BF theory gauge group introduces extra discrete gauge equivalence between geometrically distinct states in JT gravity.

In global dS, the extra identification in $G=PSL(2,\mathbb R)$ BF theory is
\begin{equation}
    b\sim b+1,~~\text{induced by}~~g(\theta )\equiv\exp(\frac{i\theta }{2}\sigma_2).
\end{equation}
identifying spacetimes with different sizes. This global gauge identification disappears when we take $\tilde{G}$ BF theory, where $\tilde{G}$ is the universal cover \cite{Schaller:1993np}.

In big-bang-crunch dS, the extra identification in $G=PSL(2,\mathbb R)$ or $\tilde{G}$ BF theory gives
\begin{equation}
    \beta\sim -\beta,~~\text{induced by}~~g(\theta )\equiv\exp(\frac{i\pi}{2}\sigma_2),
\end{equation}
identifying big-bang and big crunch universe \cite{Alonso-Monsalve:2024oii}. Note that even if we take the universal cover, the difference with gravity remains.

\section{Discussion}

\subsection*{Hartle-Hawking state}

By analytically continuing Lorentzian time to Euclidean time, and considering smooth geometry which caps off, we obtain the path-integral computing the wavefunction of the Hartle-Hawking state. Such state has trivial holonomy, thus only candidates are $b\in\mathbb{N}$ in global dS and $n$-dS BH. For dS JT gravity, the Hartle--Hawking no-boundary
wavefunction is prepared by a Euclidean hemisphere which
caps off smoothly.  Consider a geodesic boundary state
with $k=0,~ a=b$. Its Euclidean continuation has metric
\begin{equation}
    ds_E^2=d\tau^2+b^2\cos^2\tau\,d\theta^2.
\end{equation}
The cap is smooth only if $b=1$. The Hartle--Hawking wavefunction is therefore the distributional state
\begin{equation}
    \psi_{\rm HH}(b)\propto\delta(b-1)
\end{equation}
Similarly, in $n$-dS BH, the smooth cap is only realized when $n=1$. We note that these cases have not been properly treated in the current paper, as they have vanishing Fadeev-Popov determinant, therefore the gauge-fixing was insufficient. These issues will be treated in \cite{WIP}.

\begin{figure}[t]
	\begin{center}
		\includegraphics[width=3cm,clip]{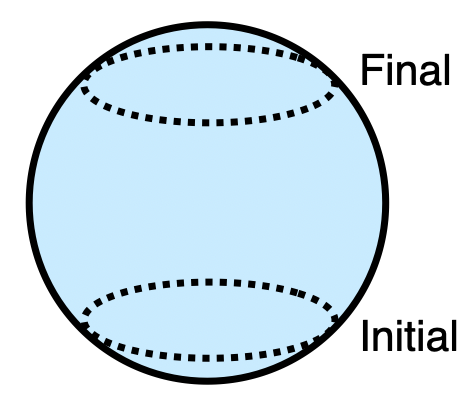}
        \caption{The leading Euclidean path-integral that computes the norm of Hartle--Hawking state.}
	    \label{fig:sphere}
    \end{center}
\end{figure}

Thus the Hartle-Hawking state is rather special state in the full physical Hilbert space. One may wonder why we have a Hartle-Hawking state without including two disconnected Euclidean disk geometry, which appears to be directly related to the no-boundary state. This is because we are interested in the equivalence classes by diffeomorphism. Timeslices that live in mutually disconected spaces can never be equivalent with each other via diffeomorphism. Thus, when we are interested in such equivalence classes, such disconnected geometries should not contribute to the overlap. The inner product of Hartle--Hawking state can be computed by considering sphere with two timeslices as initial and final timeslices as described in Fig. \ref{fig:sphere}. 

In particular, as long as we sum over smooth geometries with trivial holonomy and do not include conical defects, the Euclidean path integral does not compute the trace of the full Hilbert space. Indeed, smooth Euclidean hemisphere prepares \(|{\rm HH}\rangle\),
whereas the full sphere obtained by gluing two
hemispheres computes
\begin{equation}
    Z(S^2)=\langle{\rm HH}|{\rm HH}\rangle,
\end{equation}
which is not the trace of the identity on the full physical Hilbert
space. Note that the situation is different from the identification with the static patch partition function \cite{Banihashemi:2024weu}. A trace in the full closed universe Hilbert space requires a different sewing operation,
namely gluing the two ends of a cylinder and summing over
a complete set of intermediate states, requiring the \emph{torus} topology. This situation also clarifies the contrast with AdS/CFT.  In an eternal
AdS black hole, the Euclidean half-cylinder prepares an unnormalized
thermofield-double state, and its norm is the thermal partition function \cite{Maldacena:2001kr}
\begin{equation}
    Z(\beta)=\langle {\rm TFD}_{\beta}|{\rm TFD}_{\beta}\rangle.
\end{equation}
That equality follows from the asymptotic boundary Hamiltonian and the
standard thermal periodic identification.  In a closed gravitational universe, there is
no analogous asymptotic Hamiltonian. And so the Euclidean sphere, the norm
of a no-boundary state, and a trace over a physical Hilbert space need
not coincide. 

On top of the above clarification, we do have a subtlety in the identification of the norm of Hartle-Hawking state with the sphere partition function is the Polchinski phase, coming from the Wick rotation of conformal mode contour \cite{Polchinski:1988ua}, which can be cured by the inclusion of observer into the path-integral \cite{Ivo:2025yek, Chen:2025jqm}. 

Closely related, but independent problem is the indefiniteness of group-averaging inner product in non-compact group \cite{Louko:2003cn, Marolf:2026rzv}. It is interesting how these problem arize in gravity.

\subsection*{Relation to dS/CFT?}

It would be interesting to explore the implications of our construction for
the dS/CFT correspondence \cite{Strominger:2001pn, Maldacena:2002vr},
where the late-time gravitational wavefunction is related to the partition
function of a Euclidean CFT. See \cite{Hikida:2021ese, Hikida:2022ltr} for concrete proposals based on Chern--Simons theory with complex central charges, and the construction of bulk operator in CFT for \cite{Doi:2024nty}.
The relation between wavefunctions and CFT partition functions, however,
should be distinguished from the construction of our physical inner product, as the treatment of the boundary gauge symmetry is distinct.

Similarly, it would also be interesting to examine how the boundary gauge symmetry and
the physical inner-product measure enter path-integral optimization
\cite{Caputa:2017urj, Caputa:2017yrh}, whose holographic formulation relates
optimized geometries to the semiclassical extremization of
Hartle--Hawking wavefunctions
\cite{Boruch:2020wax, Boruch:2021hqs}.

\subsection*{Application to other gravitational theories}

A relation between the norm of HH state and sphere contribution to the partition function in the sine-dilaton gravity and a proposed matrix model dual, was found in \cite{Blommaert:2025rgw}. It was found that the Klein--Gordon norm of the
no-boundary mini-superspace wavefunction, with the leading genus-zero $O(N^2)$ term equals the logarithm of the proposed finite-cut
matrix integral partition function, which can be interpreted as a sphere contribution. Thus at leading order in large $N$, the equality is
\begin{equation}
    (\log Z_{\rm MM})|_{g=0}=
    \langle\psi_{\rm NB}|\psi_{\rm NB}\rangle_{\rm KG}.\label{logzkgnorm}
\end{equation}
Since the Klein-Gordon inner product in a "good gauge" matches with the group-averaging inner product \cite{Held:2024rmg, Held:2025mai}, it would be interesting to see how the above result,\eqref{logzkgnorm}, can be reproduced from such gauge. Furthermore, it would be interesting to apply the current formalism for inner product to variety of gravitational theories such as sine-dilaton gravity and Liouville theory, as well as theories in higher dimensions or with matter. Inclusion of matter into this theory is also very interesting. Inclusion of matter is particularly important, which requires the dual matrix model to have matter matrix, and the matrix model is given by the ETH matrix model \cite{Jafferis:2022uhu, Jafferis:2022wez}.

\subsection*{Other gauge-fixings and boundary conditions}

We have focused on special boundary conditions which fixes the extrinsic curvature of the timeslice, instead of fixing the value of dilaton. The most important failure of this gauge is that it cannot capture $n$-dS BH spacetime \cite{WIP}. In general, other gauge-fixing such as synchronous gauge, conformal gauge, de Donder gauge, Eddington-Finkelstein should also work and are interesting to work out explicitly. While the fixed extrinsic curvature boundary in hyperbolic space is rigid even at finite cut-off, fixed dilaton leads to fractal configuration to be measure one at finite cut-off \cite{Ferrari:2024kpz, Chaudhuri:2024yau, Ferrari:2024ndr}. Thus it is interesting to understand whether such fractal timeslice can be gauge equivalent to fixed extrinsic curvature slices, possibly via "fractal" diffeomorphism.  


\acknowledgments

We thank Timothy Budd, Frank Ferrari, Jesse Held, Don Marolf, Henry Maxfield, Sandip Trivedi and Gabriele di Ubaldo for helpful comments and discussions. MM is supported by JSPS KAKENHI Grant Numbers 24K17044 and 25K00997.
KS is supported by Grant-in-Aid for JSPS Fellows No.~25KJ1498. SKS is supported by the DOE grant DE-SC001010.

\newpage
\appendix


\section{Mode expansion in time}\label{sec:worldline}

\subsection{Zeta function regularization for mode expansion}

For a field $f(t)$ on the interval $t\in[0,1]$, we use the mode expansions
\begin{equation}
    \begin{split}
        f^{\rm D}(t)&=f^{\rm D}(0)+(f^{\rm D}(1)-f^{\rm D}(0))t+\sqrt{2}\sum_{n>0}f^{\rm D}_n\sin(n\pi t),~~({\rm Dirichlet~b.c.}),\\
        f^{\rm F}(t)&=f^{\rm F}_0+\sqrt{2}\sum_{n>0}f^{\rm F}_n\cos(n\pi t),~~({\rm Free~b.c.}).
    \end{split}
\end{equation}
In the free case, the endpoint values are unrestricted. The zeta function regularization of infinite product, is organized by the eigenvalues of the Laplacian $\hat{\Delta}\equiv-\partial_t^2$. The zeta-regularized infinite product is
\begin{equation}
    \begin{split}
        \sum_{n>0}e^{-\epsilon n^2}
        &=\frac{\sqrt{\pi}}{2\sqrt{\epsilon}}-\frac12
        +O\!\left(\epsilon^{-1/2}e^{-\pi^2/\epsilon}\right),\\
        \prod_{n>0}^{\rm reg}c&=c^{-1/2},
        \qquad
        \prod_{n\geq0}^{\rm reg}c=c^{1/2}.
    \end{split}
\end{equation}
It is important to note that the Dirichlet and the free boundary conditions differ only by the single constant mode $f_0$. This shows the naive expectation that there are two modes difference since there are two boundaries is wrong.

\subsection{Worldline example}

We check the consistency between measures in configuration space and phase space in the example of  worldline point particle. 

\paragraph{Configuration space.}
Let us first consider one-particle worldline following \cite{Cohen:1985sm, Held:2025mai}. The spacetime action of massive particle in Lorentzian signature with the worldline metric $ds^2=-N^2dt^2$ by
\begin{equation}
    I=\frac{1}{2}\int_0^1 dt(N^{-1}\left(\dot{X}\right)^{2}-Nm^2)+[-N^{-1}X\dot{X}]^1_0.
    \label{neumptpact}
\end{equation}
The boundary term above imposes the on-shell solution to have boundary condition $\delta (N^{-1}\dot{X})=0$, namely Neumann boundary condition.
We can parametrize the variation of $N$ in terms of difffeomorphism of the interval $f:[0,1]\rightarrow [0,1]$ and the overall length, $c=\int_0^1dt N(t)$. Consider a reparametrization of $t$ to a different coordinate $\tau$ given by $t=f(\tau)$ such that $\tilde{N}(\tau)=N(f(\tau))f'(\tau)=c$. We then have 
\begin{equation}
    \delta c=f'(\tau)\delta N+\frac{d}{d\tau}(N(f(\tau))\delta f).
\end{equation}
The metric on the space of fluctuations $\delta N$, obtained from the ultra-local inner product, can then be rewritten as
\begin{equation}
    ||\delta N||^2=\frac{1}{4}\int_0^1dt\sqrt{-g}(\delta g_{tt})^2g^{tt}g^{tt} =\int_0^1dtN^{-1}(\delta N)^2=\frac{(\delta c)^2}{c}+\frac{1}{c}\int_0^1d\tau(\frac{d}{d\tau}(N\delta f))^2.
\end{equation}
For the reparameterization $\delta\tau$, $\delta f$ is given by $\delta f=f'(\tau)\delta\tau=c\delta\tau/N$. Thus
\begin{equation}
    ||\delta N||^2=\frac{(\delta c)^2}{c}+c\int_0^1d\tau\left(\frac{d\xi}{d\tau}\right)^2.
\end{equation}
On the other hand, the measure for the reparameterization is
\begin{equation}
    \gamma(\xi,\xi)=-\int_0^1dt\sqrt{-g}(\delta t)^2g_{tt}=c^3\int_0^1d\tau(\delta\tau)^2.
\end{equation}
The transformation of fields under reparameterization $\delta t=\xi$ is
\begin{equation}
    \delta X=\dot{X}\xi,~\delta N=\frac{d}{dt}(N\xi),
\end{equation}
under which the action transforms as
\begin{align}
    \delta I&=\Big[\Big(-X\frac{d}{dt}\frac{\dot{X}}{N}-\frac{1}{2N}(\dot{X}^2+N^2m^2)\Big)\xi\Big]^{t=1}_{t=0}
\end{align}
which is proportional to the exact equation of motion for $X$. Here we are fixing the boundary coordinate values, but not the points themselves.

The gauge fixing $\dot{N}=0$ is equivalent to $f(\tau)=\tau$. Expanding $\xi=\sum_{n>0}\xi_n\sin(n\pi\tau)$, the gauge fixing $f(\tau)=\tau$ yields the Fadeev-Popov determinant
\begin{equation}
    \Delta_{FP}\equiv\frac{\text{det}\frac{\delta_{\xi} f}{\delta \xi_n}}{\sqrt{\gamma(\xi,\xi)}}=\frac{1}{\prod_{n>0}\sqrt{\frac{c^3}{2}}}\propto c^{3/4}.
\end{equation}
Combining the measure for $N$ and the FP determinant, we simply get the following flat measure for $c$,
\begin{equation}\label{eq:flat-c}
      dc.
\end{equation}

We decompose the scalar as $X=X_{\text{on-shell}}(t)+\delta X$ with $X_{\text{on-shell}}(t)$ being the on-shell solution satisfying the Neumann boundary condition. Then the measure for scalar 
 $\delta X=x_0+\sum_{n>0}\sqrt{2}x_n\cos(n\pi t)$ is
\begin{equation}
    ||\delta X||^2=\int_0^1dt\sqrt{-g}(\delta X)^2=c\sum_{n\geq0}x_n^2,
\end{equation}
which induces a measure $[DX]c^{1/4}$. Thus the gauge-fixed path-integral takes the form
\begin{equation}
    \int dc [DX]c^{1/4}e^{iI}. 
\end{equation}

\paragraph{Phase space.}
In the phase space, $p=N^{-1}\dot{X}$ and we have
\begin{equation}
    I_{\text{phase}}=\int_0^1 dt(p\dot{X}-N\mathcal{H})+[-Xp]^1_0,~\mathcal{H}=\frac{p^2+m^2}{2}.
\end{equation}
The $X$ equation of motion is $\dot{p}=0$.
The reparameterization transformation in the phase space is
\begin{equation}
    \delta X=Np\xi,~\delta N=\frac{d}{dt}(N\xi),~\delta p=0.
\end{equation}
For arbitrary $\xi$, we have
\begin{equation}
    \delta I=[-N\mathcal{H}\xi]^1_0.
\end{equation}
The momentum $p$ has {Dirichlet} boundary condition while $X$ is free at the boundary. Thus these fields can be expanded as
\begin{equation}
    p=p_i+(p_f-p_i)t+\sum_{n>0}\sqrt{2}p_n\sin(n\pi\tau),~X=x_0+\sum_{n>0}\sqrt{2}x_n\cos(n\pi\tau),
\end{equation}
where $p_*$ being time independent. Integrating out $p_n$ yields the volume density $\prod_{n>0}\frac{1}{\sqrt{c}}\propto c^{1/4}$. This reproduces the measure in configuration space by using \eqref{eq:flat-c}.

\section{Hamiltonian BRST/BFV formulation}\label{appendix:BRST}

Starting from the phase space path-integral in $PSL(2,\mathbb{R})$ gauge theory description of \eqref{eq:fullpathintegral}, we can construct Hamiltonian BRST/BFV path-integral which will be important to rigorously construct the gravity Hilbert space. In the BRST formulation, we rewrite gauge-fixed gauge theory in terms of a theory with fermionic global BRST symmetry $Q_B$. In order to take advantage of the $PSL(2,\mathbb{R})$ gauge theory description, we consider the following identifications between ghost fields and gauge transformation parameters:
\begin{equation}\label{eq:identificationghost}
    \begin{cases}
        \tilde{\xi}^s=s&\leftrightarrow c^s,\\
        \tilde{\xi}^0=\xi=N\delta t&\leftrightarrow c=Nc^t,\\
        \tilde{\xi}^\perp=a\xi_\perp&\leftrightarrow c^\perp=ac^{\perp\text{Rescaled}},
    \end{cases}
\end{equation}
as well as the redefinitions of constraints
\begin{equation}
    \mathcal{\tilde{H}}_s=-\mathcal{H}_s,~ \mathcal{\tilde{H}}=\mathcal{H},~\mathcal{\tilde{H}}_\perp=\frac{\mathcal{H}_\perp}{a}.
\end{equation}
The metric of ghost fields is inherited from that of the transformation parameters
\begin{equation}
    g((\xi,\cdots),(\xi,\cdots))=\int dtdx (Na-MaN_\perp)(-(\tilde{\xi}^0)^2+(\tilde{\xi}^\perp)^2+(\tilde{\xi}^s)^2).
\end{equation}
We have other fields as follows:
\begin{itemize}
    \item Conjugate ghost momenta $(\rho_s,~\rho,~\rho_\perp)$ with ghost number $g=-1$.
    \item Nakanishi-Lautrup auxiliary fields $(\Pi_s,~\Pi,~\Pi_\perp)$.
    \item Anti-ghost $(\bar{c}_s,~\bar{c},~\bar{c}_\perp),~(g=-1)$ and their conjugate momenta $(\bar{\rho}^s,~\bar{\rho},~\bar{\rho}^\perp),~(g=1)$
\end{itemize}

 In this paper we define the BRST transformation by the left-derivative $\delta G=\Lambda\delta_B G$, where $\Lambda$ is fermionic infinitesimal c-number. The BRST transformation which satisfies the Leibniz rule
 \begin{equation}
     \delta_B(FG)=(\delta_BF)G+(-1)^{|F|}F\delta_B G.
 \end{equation}
 Using the indices $a=(s,0,\perp)$ with $\mathcal{H}^0=\mathcal{H},~c^0=c$, as well as $\tilde{N}^s=-v,~\tilde{N}^0=N,~\tilde{N}^\perp=aN_\perp$, the BRST charge is given by
\begin{equation}
    Q_B=\int\frac{d\theta}{2\pi}\Big[-c^a\mathcal{\tilde{H}}_a+\frac{1}{2} f_{bc}{}^{a}\, c^b c^c \rho_a-\Pi_a \bar{\rho}^{\,a}\Big],
\end{equation}
with nonzero components
\begin{equation}
    f^{0}_{\perp s}=f^{\perp}_{0s}=1,~f^{s}_{\perp 0}=\pm1.
\end{equation}
We define the graded Poisson bracket, 
\begin{equation}
\begin{cases}
    \{a(\theta),~p(\theta)\}_P&=\{\Phi(\theta),~k(\theta)\}_P=\{M(\theta),~X^0(\theta)\}_P=2\pi\delta(\theta-\theta'),\\
    \{c^a(\theta),~\rho_b(\theta)\}_P&=\{\bar{c}_b(\theta),~\bar{\rho}^a(\theta)\}_P=\{\tilde{N}^a(\theta),~\Pi_b(\theta)\}_P=2\pi\delta(\theta-\theta')\delta^{a}_b,
\end{cases}
\end{equation}
with grading $\{F,~G\}_P=-(-1)^{|F||G|}\{G,~F\}_P$, and the Leibniz rule
\begin{equation}
    \{F,~GH\}_P=\{F,~G\}_PH+(-1)^{|G||F|}G\{F,~H\}_P,
\end{equation}   
and the Jacobi identity
\begin{equation}
    (-1)^{|F||H|}\{\{F,~G\}_P,~H\}_P+(-1)^{|G||F|}\{\{G,~H\}_P,~F\}_P+(-1)^{|H||G|}\{\{H,~F\}_P,~G\}_P=0.
\end{equation}    
Here we denote the fermion number by $|F|$. The BRST transformation is defined by the left-derivative
\begin{equation}
    \delta_BF=\{Q_B,~F\}_P.
\end{equation}
Note that constraints
\begin{equation}
    \begin{cases}
        \mathcal{\tilde{H}}&=-pk\mp a\Phi-X^0{}',\\
        \mathcal{\tilde{H}}_\perp&=-X^0k-p'\pm \Phi M,\\
        \mathcal{\tilde{H}}_s&=-\Phi'-aX^0-pM,
    \end{cases}
\end{equation}
satisfy
\begin{equation}
     \{\mathcal{\tilde{H}}_a(\theta),~\mathcal{\tilde{H}}_b(\theta')\}_P=f^c_{ab}\mathcal{\tilde{H}}_c(\theta')2\pi\delta(\theta-\theta').
\end{equation}
The BRST transformation satisfies $\{Q_B,~Q_B\}_P=0$ and therefore the nilpotency $\delta_B^2F=0$ due to the Jacobi identity.

The BRST transformation on phase space variables is indeed the gauge transformation with the replacement \eqref{eq:identificationghost}
\begin{equation}
\begin{cases}
    \delta_B a&=-kc+c_\perp'-Mc_s
    ,\\   
    \delta_B p&=X^0c_s\pm\Phi c,\\    
    \delta_B \Phi&=-pc-X^0c_\perp,\\
    \delta_B k&=-c_s'\pm ac\mp Mc_\perp,
    \\
    \delta_B M&=c'-ac_s-kc_\perp,
    \\
    \delta_B X^0&=pc_s\mp \Phi c_\perp.
    \end{cases}
\end{equation}
For other fields, the BRST acts as 
\begin{equation}
\begin{cases}
    \delta_B\tilde{N}^a&=\bar{\rho}^a
    ,\\    
    \delta_B\Pi_a&=0,\\
    \delta_Bc^a&=\frac{1}{2}f^a_{bc}c^bc^c,\\
    \delta_B\bar{c}_a&=-\Pi_a,
    \\
    \delta_B \rho_a&=-\mathcal{\tilde{H}}_a+f^{c}_{ab}c^b\rho_c,\\
    \delta_B\bar{\rho}^a&=0.
    \end{cases}
\end{equation}
On the constraints, 
\begin{equation}
    \delta_B\mathcal{\tilde{H}}_a=f^{c}_{ab}c^b\mathcal{\tilde{H}}_c.
\end{equation}
The original action is
\begin{equation}
\begin{split}
    I&=\frac{1}{2\pi}\int dtd\theta  \Big(X^0\dot{M}+p\dot{a}-\Phi\dot{k}-v\mathcal{H}_s-N\mathcal{H}-N_\perp\mathcal{H}_\perp\Big)\\
    &=\int dt\frac{d\theta}{2\pi} \Big(X^0\dot{M}+p\dot{a}-\Phi\dot{k}-\tilde{N}^a\mathcal{\tilde{H}}_a\Big).
    \end{split}
\end{equation}
The BRST transformation on the original action is
\begin{equation}\label{eq:variation}
\begin{split}
    \delta_B I&=
    \int dt\frac{d\theta}{2\pi} \delta_B\Big[X^0\dot{M}+p\dot{a}-\Phi\dot{k}-\tilde{N}^a\mathcal{\tilde{H}}_a\Big]
    =
    \int dt\frac{d\theta}{2\pi} \Big[
    \mathcal{\tilde{H}}_a(\dot{c}^a-\bar{\rho}^{a}-\tilde{N}^bf^{a}_{bc}c^c)
    \Big]    
    .\end{split}
\end{equation}
We will consider the gauge-fixing condition
\begin{equation}
    M=0,~G(a,\Phi,k,p,N,N_\perp)=G_{\perp}(a,\Phi,k,p,N,N_\perp)=0.\label{eq:generalgauge}
\end{equation}
In our gauge-fixing procedure, note that we leave finite-dimensional gauge degrees of freedom $\mathcal{H}$ unfixed. 

To match $\bar{\rho}^a$ with the gauge transformation of multiplier as well as to fix the gauge and to maintain BRST symmetry, we add the following term
\begin{equation}
\begin{split}
    I_{add}
    &=\int dt\frac{d\theta}{2\pi}
    \Big[
    \rho_a(\dot{c}^a-\bar{\rho}^a-\tilde{N}^bf^{a}_{bc}c^c)
    -\delta_B\Big(
    \bar{c}_sM
    +\bar{c} G+\bar{c}_{\perp}G_{\perp}\Big)\Big]
    \\
    &=\int dt\frac{d\theta}{2\pi}
    \Big[\rho_a(\dot{c}^a-\bar{\rho}^a-\tilde{N}^bf^{a}_{bc}c^c)
    +\Pi_sM
    +\Pi G
    +\Pi_{\perp} G_{\perp}
    +
    \bar{c}_s\delta_BM
    +\bar{c}\delta_B G+\bar{c}_{\perp}\delta_B G_{\perp}\Big].
    \end{split}
\end{equation}
Here $\sum_{\text{NR}}$ is the sum over non-residual ghost modes, which ensures the residual ghost modes are contained only in $\rho_a(\dot{c}^a-\bar{\rho}^a-\tilde{N}^bf^{a}_{bc}c^c)$ in the action.

The BRST transformation on this added term is,
\begin{equation}
\begin{split}
    \delta_B I_{add}
    &=\int dt\frac{d\theta}{2\pi}\Big[-
    \mathcal{\tilde{H}}_a(\dot{c}^a-\bar{\rho}^{a}-\tilde{N}^bf^{a}_{bc}c^c))
    \Big],
    \end{split}
\end{equation}
which precisely cancels the variation \eqref{eq:variation}, showing the BRST symmetry of the total action.

We assume the path-integral measure is the graded Liouville measure associated with the extended symplectic form,
\begin{equation}\label{eq:symplectic}
    \mathcal D\mu_{\rm BFV}=[\mathcal D a\,\mathcal D p\,\mathcal D k\,\mathcal D\Phi\,\mathcal D M\,\mathcal D X^0\,\mathcal D\tilde{N}^a\,\mathcal D\Pi_a\,\mathcal D c^a\,\mathcal D\rho_a\,\mathcal D\bar c_a\,\mathcal D\bar\rho^a]
    .
\end{equation}
This path-integral measure is BRST invariant, since the BRST transformation is now canonical on this graded symplectic space, whose Jacobian is unity. 

For the boundary condition, we can again insert the delta function $\delta(O(\varphi|_{t=0,1})-O(\varphi_{i,f}))$ at the initial and the final timeslices. Since $O(\varphi)$ is gauge-invariant, it is also BRST invariant.

Finally, we will confirm that after integrating out all ghost fields, we return to the original path-integral \eqref{eq:fullpathintegral} with the correct measure. Let us first integrate out $\rho_a$. It imposes a delta-function constraint which yields $\bar{\rho}^a=\dot{c}^a-\tilde{N}^bf^{a}_{bc}c^c$, namely
\begin{equation}
\begin{cases}
    \bar{\rho}^s=\dot{c}^s\pm(\tilde{N}^0c^\perp-\tilde{N}^\perp c)
    &\leftrightarrow \dot{s}\pm Na\xi_\perp\mp aN_\perp\xi=\delta \tilde{N}^s ,\\   
    \bar{\rho}^0
    =\dot{c}^0-\tilde{N}^\perp c^s-vc^\perp
    &\leftrightarrow 
    \dot{\xi}-aN^\perp s-av\xi_\perp=\delta \tilde{N},
    \\
    \bar{\rho}^\perp=\dot{c}^\perp+\tilde{N}^sc^0-\tilde{N}^0c^s
    &\leftrightarrow
    \dot{a}\xi^\perp+a\dot{\xi}^\perp-v\xi-Ns=\delta \tilde{N}^\perp,
    \end{cases}
\end{equation}
which are precisely the gauge transformation of multipliers \eqref{eq:diffgeneralmultipliers} under the identification \eqref{eq:identificationghost}, consistent with $\delta_B\tilde{N}^a=\bar{\rho}^a$. After imposing the delta-function constraint by integrating out $\bar{\rho}^a_n~(n\neq0)$, integrating out $\Pi$s introduces the gauge-fixing delta functions $\delta(M)\delta(G)\delta(G_\perp)$. Finally, by noting $\tilde{\xi}_\perp=a\xi_\perp$, integration over ghost fields $c,~\bar{c}$ yields the determinant 
\begin{equation}
    \Big{|}\det{}{\left.
\frac{\delta(M,~G)^A[\varphi^\epsilon]}
{\delta\epsilon^\alpha}
\right|_{\epsilon=0}}\frac{1}{\prod_{(t,\theta )}a}\Big{|}.
\end{equation}
We note that we implicitly assumed the determinant from the ghost path-integral is \emph{positive}, see \cite{Held:2024rmg} for how non-positivity of determinants affects inner products. Under this positivity assumption, \emph{due to the important cancellation of volume forms \eqref{eq:cancellation}} which was also stated in \cite{Witten:1991we}, and the transformation of measures $[D\tilde{N}_\perp]=|\prod_{(t,\theta )}a|[DN_\perp]$, we arrive at the original gauge-fixed path-integral \eqref{eq:fullpathintegral}. Thus our original gauge-fixed path-integral can be cast into canonical form with Liouville measure \eqref{eq:symplectic}.

\subsection{Canonical transformation to distinct coordinate system}

Instead of the original canonical variables $(a,p),(c^\perp,\rho_\perp),(N_\perp,\Pi^\perp)$, we consider new set of canonical variables
\begin{equation}
    (a,p_{\rm new}\equiv p+\frac{\mu_\perp\gamma^\perp+\pi_{N_\perp}N_\perp}{a}),~(\gamma^\perp\equiv\frac{c^\perp}{a},\mu_\perp\equiv a\rho_\perp),~(N_\perp,~\pi_{N_\perp}\equiv a\Pi^\perp),
\end{equation}
with rest of canonical variables unchanged. Indeed, this transformation preserves the canonical one-form:
\begin{equation}
    p\,\delta a+\rho_\perp\,\delta c^\perp
       +\Pi_\perp\,\delta\widetilde N^\perp
    =
    p_{\rm new}\,\delta a
       +\mu_\perp\,\delta\gamma^\perp
       +\pi_{N_\perp}\,\delta N_\perp.
\end{equation}
The BRST charge, gauge-fixing terms, and boundary terms are now
expressed in the new variables. After the Lorentz boost gauge-fixing and integrating $v$, the reduced constraints are
\begin{equation}
    \mathcal H_{\rm new}
    =-kp_{\rm new}\mp a\Phi
       +\left(\frac{\Phi'}{a}\right)',
    \qquad
    \mathcal H_{\perp,\rm new}
    =k\Phi'-ap_{\rm new}',
\end{equation}
which takes exactly the same form as in the original reduction.
The associated bosonic gauge transformations are
\begin{equation}
\begin{cases}
    \delta a&=-k\xi+\Big(a\xi_\perp\Big)'
    ,~~\delta p_{\rm new}=\pm\Phi\xi-\frac{\Phi'}{a^2}\xi'+p_{\rm new}'\xi_\perp,\\
    \delta \Phi&=-p_{\rm new}\xi+\Phi'\xi_\perp,
    ~~~~\delta k=\Big(\pm a\xi-(\frac{\xi'}{a})'\Big)+(k\xi_\perp)',\\
    \delta N&=\dot{\xi}-N_\perp\xi'+N'\xi_\perp,
    \\
    \delta N_\perp&=\Big(\frac{-N\xi'+N'\xi}{a^2}\Big)+\Big(\dot{\xi}_\perp+N_\perp'\xi_\perp-N_\perp\xi_\perp'\Big),   
    \end{cases}
\end{equation}
The ghost--parameter correspondence is
\begin{equation}
    c\longleftrightarrow\xi,
    \qquad
    \gamma^\perp\longleftrightarrow\xi_\perp.
\end{equation}

For the lapse and shift, the displayed transformations are obtained
after eliminating the auxiliary ghost momenta using the transformed
gauge-fixed action. No residual gravitational constraint or
Hamilton equation is imposed in obtaining these bosonic
transformations.

\section{Alternative gauge fixings}
\label{app:zero-mode-gauge-equivalence}

We collect alternative gauge choices for the zero-mode path integral of Section~\ref{sec:gauge-fixed-path-integral}.
The nonzero-mode reduction and the general BRST construction are those of Subsection~\ref{subsec:nonzero-mode-reduction} and Section~\ref{sec:zero-mode-brst} and will not be repeated.
We work on regular connected components and use the boundary-state normalization of the main text.

\subsection{The $q(t)=q_{\rm gf}(t)$ gauge}
\label{subsec:complete-zero-mode}

We fix $G_{1i},G_{1f}$ and select endpoint representatives $q_i,q_f$.
The lapse remains free at the endpoints, and the remaining normal gauge transformations obey $\xi(0)=\xi(1)=0$.
We impose
\begin{align}
    \chi_0(t)&=q(t)-q_{\rm gf}(t)=0,
    &
    q_{\rm gf}(0)&=q_i,
    &
    q_{\rm gf}(1)&=q_f,
\end{align}
where $q_{\rm gf}(t)$ is a fixed external function.
The canonical one-form \eqref{eq:complete-gauge-canonical-form} implies
\begin{align}
    \{\chi_0(t),\mathcal H_0(t')\}_{\rm P}
    &=-\delta(t-t'),
    &
    \Delta_\chi&=1.
\end{align}
The unrestricted lapse integral with the action \eqref{eq:zero-mode-fp-common-action} produces $\delta[\mathcal H_0]$.
Thus the gauge-fixed path integral is
\begin{align}
    Z(G_{1f},q_f;G_{1i},q_i)
    &={}
    \int
    \mathcal DG_1\,\mathcal D\Pi_{G_1}\,
    \mathcal Dq\,\mathcal D\mathcal H_0\,
    \delta[\mathcal H_0]\,\delta[\chi_0]\,\Delta_\chi
    \notag\\
    &\qquad\times
    \exp\left[
        i\int_0^1dt\,
        \left(\Pi_{G_1}\dot G_1-\mathcal H_0\dot q\right)
    \right].
    \label{eq:complete-zero-mode-path-integral}
\end{align}
The $\mathcal H_0$ and $q$ integrations leave the reduced canonical integral, and integrating $\Pi_{G_1}$ imposes $\dot G_1=0$.
With the normalization used in the main text and the component labels restored, the result is
\begin{align}
    Z(G_{1f},q_f;G_{1i},q_i)
    =\delta(G_{1f}-G_{1i})\delta_{\alpha_i\alpha_f}.
    \label{eq:complete-zero-mode-kernel}
\end{align}
This is the kernel in \eqref{eq:intro-reduced-kernel}.
Its expressions in the length coordinates and the possible discrete identifications are discussed in Section~\ref{sec:perturbative-hilbert}.
The AdS coordinate is periodic, so this calculation is understood on a regular gauge patch, whereas the global group average is taken over one period.
As in the main text, the exceptional $n$-dS BH loci are excluded.
The interpretation in terms of invariant or co-invariant boundary states is the one specified in Section~\ref{sec:zero-mode-brst}.

\paragraph{Comparison with the constant-lapse gauge.}
\label{subsec:q-gauge-constant-lapse-comparison}

The equivalence can also be seen directly before evaluating the constraint integral.
With the same fixed endpoint representatives, \eqref{eq:physicalpathintegral} reduces the constant-lapse integral to
\begin{align}
    Z_{\dot N}
    =\mathcal N_0
    \int dT\,\mathcal Dq\,\mathcal D\mathcal H_0\,
    e^{-i\int_0^1dt\,\mathcal H_0(\dot q+T)}.
    \label{eq:constant-lapse-gauge-path-integral}
\end{align}
Here we suppress the spectator pair $(G_1,\Pi_{G_1})$ and $\mathcal N_0$ is field independent.
Write
\begin{align}
    q(t)&=q_{\rm gf}(t)+\zeta(t),
    &
    \zeta(0)&=\zeta(1)=0,
\end{align}
and define
\begin{align}
    \tilde N_0(t)=T+\dot\zeta(t).
    \label{eq:combined-lapse}
\end{align}
This change of variables is one-to-one, with inverse
\begin{align}
    T&=\int_0^1dt\,\tilde N_0(t),
    &
    \zeta(t)&=\int_0^t ds\,[\tilde N_0(s)-T].
\end{align}
Its Jacobian is field independent.
In particular, $T$ is the constant mode of $\tilde N_0$, so $dT\,\mathcal Dq=dT\,\mathcal D\zeta\propto\mathcal D\tilde N_0$.
Thus the $T$ integration is included in the functional integral over $\tilde N_0$ in \eqref{eq:constant-lapse-to-constraint}.
The unrestricted $\tilde N_0$ integral then gives
\begin{align}
    Z_{\dot N}
    &=\mathcal N
    \int\mathcal D\tilde N_0\,\mathcal D\mathcal H_0\,
    e^{-i\int_0^1dt\,\mathcal H_0(\dot q_{\rm gf}+\tilde N_0)}
    \notag\\
    &=\mathcal N
    \int\mathcal D\mathcal H_0\,
    \delta[\mathcal H_0]\,
    e^{-i\int_0^1dt\,\mathcal H_0\dot q_{\rm gf}}.
    \label{eq:constant-lapse-to-constraint}
\end{align}
The last integral is the gauge sector of \eqref{eq:complete-zero-mode-path-integral} after integrating $q$.
Denoting that integral by $Z_\chi$, we obtain
\begin{align}
    Z_{\dot N}=\mathcal N Z_\chi.
    \label{eq:zero-mode-gauge-equivalence}
\end{align}
This is an equality of the full gauge-fixed integrals, not a pointwise identity between their delta-function insertions.
In particular, $\dot N_0=0$ does not form a second-class pair with $\mathcal H_0$.
The constraint is imposed after the lapse modulus and the fluctuations of $q$ have been combined into $\tilde N_0$ and integrated out.
The derivation above uses the universal-cover variables $q\in\mathbb R$ and $T\in\mathbb R$.
For the compact AdS orbit in the $(a_0,k_0)$ description, the global quotient and the corresponding one-period group average are treated as in Subsection~\ref{subsec:canonical-zero-mode-variables}.

\subsection{An alternative gauge for the invariant inner product}
\label{app:alternative-invariant-fp}

We now keep the physical boundary data fixed and choose a different lapse representative,
\begin{align}
    G_1(0)&=G_{1i},
    &
    G_1(1)&=G_{1f},
    \notag\\
    N_0(0)&=0,
    &
    N_0(1)&=0.
    \label{eq:app-invariant-boundary-data}
\end{align}
We work on a regular connected component with $q\in\mathbb R$.
The endpoint values of $q$ and $\mathcal H_0$ are integrated, and we use the action $I_0$ in \eqref{eq:zero-mode-fp-common-action} without an additional boundary term.
In particular, $\mathcal H_0=0$ is not imposed independently at either endpoint.
The transformations in \eqref{eq:zero-mode-fp-common-transformations} preserve the lapse boundary values when
\begin{align}
    \dot\xi(0)=\dot\xi(1)=0.
    \label{eq:app-invariant-gauge-domain}
\end{align}
There is no restriction on $\xi(0)$ or $\xi(1)$ because the endpoint values of $q$ are integrated.
We choose
\begin{align}
    F_{\rm inv}[q,N_0](t)
    :=\dot N_0(t)+q(t)+\int_0^t ds\,N_0(s)-q_*=0,
    \label{eq:app-invariant-gauge}
\end{align}
where $q_*$ is a fixed real constant.
Its variation is
\begin{align}
    \delta_\xi F_{\rm inv}(t)
    &=\ddot\xi(t)-\xi(t)+\int_0^t ds\,\dot\xi(s)
    \notag\\
    &=\ddot\xi(t)-\xi(0)
    =:(M_{\rm inv}\xi)(t).
    \label{eq:app-invariant-fp-operator}
\end{align}
The operator $M_{\rm inv}$ is invertible on the domain in \eqref{eq:app-invariant-gauge-domain}.\footnote{
If $M_{\rm inv}\xi=0$, setting $a=\xi(0)$ gives $\ddot\xi=a$.
The two boundary conditions on $\dot\xi$ imply $a=0$ and hence $\xi=0$.
For any smooth function $f$, set $a_f=-\int_0^1ds\,f(s)$ and $\xi_f(t)=a_f(1+t^2/2)+\int_0^t ds\,(t-s)f(s)$.
Then $\dot\xi_f(0)=\dot\xi_f(1)=0$ and $M_{\rm inv}\xi_f=f$.
}
Thus this single condition also fixes the constant transformation.
Its Faddeev--Popov determinant $\Delta_{\rm inv}=|\det M_{\rm inv}|$ is a nonzero field-independent factor.
The gauge-fixed integral is
\begin{align}
    K_{\rm inv}^{\rm FP}(G_{1f},G_{1i})
    &={}
    \mathcal N_{\rm inv}
    \int_{\mathfrak B_{\rm inv}}
    \mathcal DG_1\,\mathcal D\Pi_{G_1}\,
    \mathcal Dq\,\mathcal D\mathcal H_0\,\mathcal DN_0
    \notag\\
    &\qquad\times
    \Delta_{\rm inv}\,\delta[F_{\rm inv}]\,e^{iI_0},
    \label{eq:app-invariant-fp-integral}
\end{align}
where $\mathfrak B_{\rm inv}$ imposes only \eqref{eq:app-invariant-boundary-data}.

Define
\begin{align}
    \tilde q(t):=q(t)+\int_0^t ds\,N_0(s).
    \label{eq:app-invariant-coordinate}
\end{align}
This triangular change of variables has unit time-sliced Jacobian, and both endpoints of $\tilde q$ remain integrated.
The action, gauge condition, and transformed gauge variation are
\begin{align}
    I_0
    &=\int_0^1dt\,
    \left[\Pi_{G_1}\dot G_1-\mathcal H_0\dot{\tilde q}\right],
    \notag\\
    F_{\rm inv}[\tilde q,N_0](t)
    &=\dot N_0(t)+\tilde q(t)-q_*,
    \notag\\
    \delta_\xi\tilde q(t)&=-\xi(0),
    \qquad
    \delta_\xi N_0(t)=\dot\xi(t).
    \label{eq:app-invariant-transformed-action}
\end{align}
Representing the delta functional by a Lagrange multiplier gives
\begin{align}
    \delta[F_{\rm inv}]
    =\mathcal C_\Pi\int\mathcal D\Pi_0\,
    e^{i\int_0^1dt\,\Pi_0F_{\rm inv}}.
\end{align}
The resulting first-order action is
\begin{align}
    I_{\rm inv}^{\rm FP}
    =\int_0^1dt\,
    \left[
        \Pi_{G_1}\dot G_1
        -\mathcal H_0\dot{\tilde q}
        +\Pi_0\dot N_0
        -\Pi_0(q_*-\tilde q)
    \right].
    \label{eq:app-invariant-fp-action}
\end{align}
The multiplier has no fixed endpoint values.
The field-independent factors $\Delta_{\rm inv}$ and $\mathcal C_\Pi$ can be obsorbed into $\mathcal N_{\rm inv}$.

Integrating $N_0$ with its fixed zero endpoint values gives
\begin{align}
    \int_{N_0(0)=N_0(1)=0}\mathcal DN_0\,
    e^{i\int_0^1dt\,\Pi_0\dot N_0}
    \propto\delta[\dot\Pi_0],
\end{align}
so $\Pi_0(t)=\lambda$ is constant.
Decompose
\begin{align}
    \tilde q(t)
    &=(1-t)\tilde q_i+t\tilde q_f+\tilde q_D(t),
    &
    \tilde q_D(0)&=\tilde q_D(1)=0.
\end{align}
Integration by parts gives
\begin{align}
    &-\int_0^1dt\,\mathcal H_0\dot{\tilde q}
    +\lambda\int_0^1dt\,(\tilde q-q_*)
    \notag\\
    &\qquad=
    \int_0^1dt\,(\dot{\mathcal H}_0+\lambda)\tilde q
    +\mathcal H_{0i}\tilde q_i
    -\mathcal H_{0f}\tilde q_f
    -\lambda q_*.
    \label{eq:app-invariant-endpoint-action}
\end{align}
The integral over $\tilde q_D$ imposes
\begin{align}
    \dot{\mathcal H}_0+\lambda&=0\Rightarrow
    \mathcal H_0(t)=\mathcal H_{0i}-\lambda t.
\end{align}
The bulk term in \eqref{eq:app-invariant-endpoint-action} then vanishes, and the endpoint integrals give
\begin{align}
    \int_{\mathbb R}d\tilde q_i\int_{\mathbb R}d\tilde q_f\,
    e^{i\mathcal H_{0i}\tilde q_i-i\mathcal H_{0f}\tilde q_f}
    =(2\pi)^2\delta(\mathcal H_{0i})\delta(\mathcal H_{0f}).
    \label{eq:app-invariant-endpoint-constraints}
\end{align}
These Fourier integrals imply $\lambda=0$ and $\mathcal H_0(t)=0$, including at the endpoints.\footnote{
The bosonic Fourier integrals are understood distributionally.
Normalized smooth representatives can be used before taking the delta-function limit in physical pairings.
The Fourier normalization is the same in the Faddeev--Popov and BFV representations.
}
Only the reduced integral remains, giving
\begin{align}
    K_{\rm inv}^{\rm FP}(G_{1f},G_{1i})
    =\delta(G_{1f}-G_{1i})
    \label{eq:app-invariant-fp-kernel}
\end{align}
on the chosen component, with the same reduced-state normalization as in the main text.
The endpoint prescription therefore realizes the invariant inner product, with $\hat{\mathcal H}_0\Psi_{\rm inv}=0$.

\subsection{BFV--BRST representation of the alternative invariant gauge}
\label{app:alternative-invariant-bfv}

We use the BRST system and boundary pairing of Subsection~\ref{subsubsec:zero-mode-full-complex}.
To retain the ghost convention of this alternative calculation, write
\begin{align}
    c'_0=-c_0,
    \qquad
    \rho'_0=-\rho_0,
    \qquad
    Q_{B,0}=c'_0\mathcal H_0-\Pi_0\bar\rho_{\,0}.
    \label{eq:app-invariant-ghost-convention}
\end{align}
The unprimed minimal ghosts on the right are those of the main text, and the non-minimal pair is unchanged.
The simultaneous sign reversal preserves the canonical brackets.
In this notation the basis in \eqref{eq:zero-mode-ghost-basis} has $|\uparrow\downarrow\rangle=c'_0|\downarrow\downarrow\rangle$ and $|\downarrow\uparrow\rangle=-\bar\rho_{\,0}|\downarrow\downarrow\rangle$, with the same pairing.

The bosonic boundary conditions are \eqref{eq:app-invariant-boundary-data}.
The integrated endpoints of $\tilde q$ insert the constant wavefunction $\langle\tilde q|\mathcal H_0=0\rangle=1$ of the invariant boundary state in \eqref{eq:zero-mode-inv-state}, with lapse representative $\nu=0$.
We neither impose an additional endpoint delta function for $\mathcal H_0$ nor add $[\mathcal H_0\tilde q]_0^1$.
The ghost momenta obey
\begin{align}
    \rho'_0=\bar\rho_{\,0}=0
    \qquad(t=0,1),
\end{align}
with boundary term
\begin{align}
    I_{\partial,\rm gh}
    =-[\rho'_0c'_0+\bar\rho_{\,0}\bar c_0]_0^1.
\end{align}
The variables $\Pi_0,c'_0,\bar c_0$ are free at the endpoints.
As in the main text, BRST closure applies to the complete boundary state.

For
\begin{align}
    \chi_*&=q_*-\tilde q,
    &
    \hat{\mathcal H}_0&=i\partial_{\tilde q},
    &
    \hat\Psi_*&=\hat{\bar c}_0\hat\chi_*,
    \label{eq:app-invariant-fermion}
\end{align}
the BRST-exact operator is
\begin{align}
    [\hat Q_{B,0},\hat\Psi_*]_+
    =-i\hat\Pi_0\hat\chi_*
      +i\hat{\bar c}_0\hat c'_0.
\end{align}
Its phase-space representation has action
\begin{align}
    I_{\rm inv}^{\rm BFV}
    =I_{\rm inv}^{\rm FP}
    +\int_0^1dt\,
    \left[
        \rho'_0\dot c'_0
        +\bar\rho_{\,0}\dot{\bar c}_0
        +\bar c_0c'_0
    \right]
    +I_{\partial,\rm gh},
    \label{eq:app-invariant-bfv-action}
\end{align}
where $I_{\rm inv}^{\rm FP}$ is \eqref{eq:app-invariant-fp-action}.
Thus the bosonic action agrees with the preceding Faddeev--Popov calculation, including its boundary prescription.
Reversing \eqref{eq:app-invariant-coordinate} gives
\begin{align}
    I_{\rm inv,bos}^{\rm BFV}
    =I_0+\int_0^1dt\,\Pi_0F_{\rm inv}[q,N_0].
    \label{eq:app-invariant-bfv-fp-comparison}
\end{align}

The ghost integral reproduces the same Faddeev--Popov operator.
Integrating $\rho'_0$ first gives $c'_0(t)=c'_{\rm const}$.
After integration by parts, the remaining ghost action is
\begin{align}
    I_{\rm gh}
    =\int_0^1dt\,
    \bar c_0\left(\dot{\bar\rho}_{\,0}+c'_{\rm const}\right).
\end{align}
Introduce
\begin{align}
    \gamma(t)=-c'_{\rm const}+\int_0^t ds\,\bar\rho_{\,0}(s).
\end{align}
The ghost boundary conditions imply
\begin{align}
    \dot\gamma(0)&=\dot\gamma(1)=0,
    \notag\\
    \dot{\bar\rho}_{\,0}+c'_{\rm const}
    &=\ddot\gamma-\gamma(0)
    =M_{\rm inv}\gamma.
    \label{eq:app-invariant-ghost-fp-operator}
\end{align}
This is the same as \eqref{eq:app-invariant-fp-operator} including the domain.
Including the field-independent Jacobians, the ghost integral supplies $\Delta_{\rm inv}$, with its phase fixed by the boundary pairing.
The bosonic integrations have already been evaluated in Subsection~\ref{app:alternative-invariant-fp} and give the kernel in \eqref{eq:app-invariant-fp-kernel}.

\paragraph{The gauge-fixing map.}

The map $\kappa$ follows by instead integrating $\mathcal H_0$ first, which imposes $\dot{\tilde q}=0$.
The lapse integral gives $\Pi_0(t)=\lambda$, while integrating both ghost momenta leaves constant $c'_0$ and $\bar c_0$.
The remaining ghost factor is
\begin{align}
    {}_{\rm gh}\langle\downarrow\uparrow|
    e^{i\hat{\bar c}_0\hat c'_0}
    |\downarrow\uparrow\rangle_{\rm gh}
    =i\int d\bar c_0\,dc'_0\,e^{i\bar c_0c'_0}=1,
\end{align}
where $\int d\bar c_0\,dc'_0\,c'_0\bar c_0=1$ and the prefactor follows from the ghost pairing.
The orbit integral is therefore
\begin{align}
    \int_{\mathbb R}d\tilde q
    \int_{\mathbb R}\frac{d\lambda}{2\pi}\,
    e^{-i\lambda(q_*-\tilde q)}
    =\int_{\mathbb R}d\tilde q\,
    \delta(\tilde q-q_*)=1.
    \label{eq:app-invariant-orbit-factor}
\end{align}
The corresponding gauge-fixing map is
\begin{align}
    (\kappa_*\psi)(G_1,\tilde q)
    =\delta(\tilde q-q_*)\psi(G_1).
    \label{eq:app-invariant-kappa}
\end{align}
This is the sharp-gauge form of the construction in \cite{Held:2025mai}.
Indeed,
\begin{align}
    [\hat{\mathcal H}_0,e^{-i\lambda\hat\chi_*}]
    &=-\lambda e^{-i\lambda\hat\chi_*},
    \notag\\
    \kappa_*
    &=-\frac{1}{2\pi}\int_{\mathbb R}\frac{d\lambda}{\lambda}\,
    [\hat{\mathcal H}_0,e^{-i\lambda\hat\chi_*}]
    \notag\\
    &=\frac{1}{2\pi}\int_{\mathbb R}d\lambda\,
    e^{-i\lambda(q_*-\tilde q)}
    =\delta(\tilde q-q_*).
    \label{eq:app-invariant-kappa-HM}
\end{align}
With normalized smooth representatives before the sharp limit, this is a special case of the map in \cite[Eq.~(6.14)]{Held:2025mai}.
Its co-invariant class is independent of the normalized profile and its location, as shown in Subsection~\ref{subsubsec:zero-mode-full-complex}.
Hence the same canonical pairing gives the physical inner product in \eqref{eq:zero-mode-invariant-kernel}, without repeating the proof of the invariant/co-invariant correspondence.


\section{Zero-mode reduction with positive lapse}
\label{app:positive-lapse-reduction}

We evaluate the gauge-fixed phase-space path integral with the pointwise restriction $N(t,\theta )>0$ in the Lorentz gauge $M=0$.
We use the measure in \eqref{eq:secondorderZ}, the same gauge-volume normalization as in the main text, and fixed boundary representatives.
When integrating out the nonzero spatial Fourier modes, we work on a patch where $a_0(t)\neq0$ and
\begin{align}
    D_n(t):=n^2+k_0(t)^2\pm a_0(t)^2\neq0,
    \qquad n\neq0.
\end{align}
Here $a_0$ and $k_0$ are the spatial averages of $a$ and $k$.
These conditions ensure that the nonzero-mode FP operator is well defined and invertible.
The upper and lower signs correspond to AdS and dS, respectively.

\subsection{Nonzero spatial modes.}

We start from the phase-space action appearing in \eqref{eq:secondorderZ},
\begin{align}
    I_{\pm,\text{phase}}
    &=
    \int_0^1dt\int_0^{2\pi}\frac{d\theta }{2\pi}
    \left[
        p\dot a-\Phi\dot k
        -N\mathcal H-N_\perp\mathcal H_\perp
    \right],
    \\
    \mathcal H
    &=
    -kp\mp a\Phi+\left(\frac{\Phi'}{a}\right)',
    \qquad
    \mathcal H_\perp=k\Phi'-ap'.
\end{align}
Dots and primes denote derivatives with respect to $t$ and $\theta $, respectively.
For $\theta \sim \theta +2\pi$, write each real field as $X(t,\theta )=\sum_nX_n(t)e^{in\theta }$, with $X_{-n}=X_n^*$.
In the unrestricted calculation, the nonzero lapse and shift modes are integrated first.
The pointwise condition $N(t,\theta )>0$ couples the Fourier coefficients of the lapse, so these integrals can no longer be performed independently over the full real axes.
We therefore integrate the nonzero modes of $p$ and $\Phi$ first.
For each $n>0$, their real and imaginary parts are integrated independently over $\mathbb R$.

In the gauge $a'=k'=0$, their kinetic terms vanish, and their contribution to the action is
\begin{align}
    I_{n\neq0}
    &=
    \int_0^1dt\sum_{n>0}
    2\operatorname{Re}\bigl(p_n^*A_n+\Phi_n^*B_n\bigr),
    \\
    \begin{pmatrix}A_n\\ B_n\end{pmatrix}
    &=
    C_n\begin{pmatrix}N_n\\ N_{n\perp}\end{pmatrix},
    \qquad
    C_n=
    \begin{pmatrix}
        k_0 & -in a_0\\
        \displaystyle\pm a_0+\frac{n^2}{a_0} & in k_0
    \end{pmatrix}.
\end{align}
Thus the momentum integrals produce $\delta^{(2)}(A_n)\delta^{(2)}(B_n)$ for each $n>0$, where $\delta^{(2)}(z):=\delta(\operatorname{Re}z)\delta(\operatorname{Im}z)$.
Since $\det C_n=inD_n$, the Jacobian for the four independent real components is $|\det C_n|^2=n^2D_n^2$.
Consequently,
\begin{align}
    \delta^{(2)}(A_n)\delta^{(2)}(B_n)
    &=
    \frac{\delta^{(2)}(N_n)\delta^{(2)}(N_{n\perp})}
    {n^2D_n^2}.
    \label{eq:app-positive-lapse-jacobian}
\end{align}
The nonzero-mode FP factor \eqref{fpnz} is proportional to $\prod_t\prod_{n>0}D_n^2$.
It therefore cancels the field-dependent denominator in \eqref{eq:app-positive-lapse-jacobian}.
The remaining powers of $n$ are field independent and are absorbed into the same normalization as in the unrestricted calculation.

To define these Fourier integrals, we introduce a finite spatial mode cutoff and a time slicing, and regulate the real momentum integrals by Gaussians.
For each real integration variable, we use the distributional limit
\begin{align}
    \int_{\mathbb R}\frac{dy}{2\pi}\,
    e^{iyz-\varepsilon y^2/2}
    &=
    \frac{e^{-z^2/(2\varepsilon)}}{\sqrt{2\pi\varepsilon}}
    \longrightarrow\delta(z),
    \qquad \varepsilon\to0^+.
\end{align}
We remove the momentum regulators at fixed cutoffs before taking the continuum limits.
The resulting delta functions set $N_n=N_{n\perp}=0$ for every $n\neq0$.
For fixed $N_0>0$, this support lies in the interior of the positive-lapse domain, whereas $N_0<0$ is incompatible with positivity because $N_0$ is the spatial average.
The pointwise restriction therefore reduces to $N_0(t)>0$, without an additional $N_0$-dependent measure factor.

\subsection{The positive lapse modulus.}

We treat the shift zero mode and the residual spatial translations with the same normalization as in the main text.
The remaining canonical action is \eqref{eq:zero-mode-fp-common-action},
\begin{align}
    I_0
    &=
    \int_0^1dt\,
    \bigl[\Pi_{G_1}\dot G_1-\mathcal H_0(\dot q+N_0)\bigr].
\end{align}
With the endpoint representatives fixed, the remaining gauge transformations obey
\begin{align}
    \delta_\xi q=-\xi,
    \qquad
    \delta_\xi N_0=\dot\xi,
    \qquad
    \xi(0)=\xi(1)=0.
\end{align}
For a smooth positive lapse, define $T:=\int_0^1dt\,N_0(t)>0$.
It is invariant under these endpoint-fixed transformations.
The parameter $\xi(t)=Tt-\int_0^t ds\,N_0(s)$ takes $N_0(t)$ to the constant $T$ and vanishes at both endpoints.
This transformation can be performed along the positive interpolation $(1-s)N_0(t)+sT$, $0\leq s\leq1$.

In the canonical FP prescription, the gauge $\dot N_0=0$ gives $\delta_\xi\dot N_0=\ddot\xi$.
The corresponding operator $\partial_t^2$ has Dirichlet boundary conditions, no zero mode, and a field-independent determinant.
Decomposing the flat lapse measure into its constant and nonconstant parts also has a field-independent Jacobian.
In particular, up to a common normalization,
\begin{align}
    \int_{N_0(t)>0}\mathcal D N_0\,
    \delta[\dot N_0]
    \bigl|\det\nolimits_{\mathrm{DD}}\partial_t^2\bigr|\,
    F\!\left(\int_0^1dt\,N_0\right)
    &\propto
    \int_0^\infty dT\,F(T),
\end{align}
where $F$ is a test function and $\mathrm{DD}$ denotes the endpoint conditions on $\xi$.
The remaining modulus measure is therefore $dT$ on $T>0$, with no additional power of $T$.

\subsection{The boundary kernel.}

For the boundary kernel, we assume that the reduction applies on a component with a complete noncompact gauge orbit $q\in\mathbb R$, and take both endpoints in that component.
At fixed $T$, the integrals over $\Pi_{G_1}$ and $\mathcal H_0$ impose $\dot G_1=0$ and $\dot q=-T$.
Integrating the intermediate values of $G_1$ and $q$ then gives
\begin{align}
    K_T(G_{1f},q_f;G_{1i},q_i)
    &=
    \delta(G_{1f}-G_{1i})\delta(q_f-q_i+T).
\end{align}
The normalization is fixed by the identity kernel at $T=0$.
Integrating the positive modulus yields
\begin{align}
    K_+(G_{1f},q_f;G_{1i},q_i)
    &=
    \delta(G_{1f}-G_{1i})
    \int_0^\infty dT\,\delta(q_f-q_i+T)
    \nonumber\\
    &=
    \delta(G_{1f}-G_{1i})\Theta(q_i-q_f).
    \label{eq:app-positive-lapse-kernel}
\end{align}
Since the momentum conjugate to $q$ is $-\mathcal H_0$, we have $\hat{\mathcal H}_0=i\partial_q$ and hence
\begin{align}
    \hat{\mathcal H}_{0,f}K_+
    &=
    -i\,\delta(G_{1f}-G_{1i})\delta(q_f-q_i).
\end{align}
Thus $K_+$ is a Wheeler--DeWitt Green function, with a contact source from the endpoint $T=0$ \cite{Casali:2021ewu}.
It satisfies the homogeneous constraint away from this source.
The step function depends on the ordering of the chosen boundary representatives, so $K_+$ does not define a pairing invariant under arbitrary changes of those representatives.
Integrating over the full real line instead removes the step function and reproduces \eqref{eq:co-invariant-result}.
All kernel identities are understood distributionally.


\bibliographystyle{jhep}
\bibliography{references}

@article{Marolf:2026rzv,
    author = "Marolf, Donald and Yu, Xuyang",
    title = "{On the Positivity of de Sitter Group Averaging}",
    eprint = "2610.24073",
    archivePrefix = "arXiv",
    primaryClass = "hep-th",
    month = "10",
    year = "2026"
}

@article{Louko:2003cn,
    author = "Louko, Jorma and Molgado, Alberto",
    title = "{Group averaging in the (p,q) oscillator representation of SL(2,R)}",
    eprint = "gr-qc/0312014",
    archivePrefix = "arXiv",
    doi = "10.1063/1.1689001",
    journal = "J. Math. Phys.",
    volume = "45",
    pages = "1919--1943",
    year = "2004"
}

@article{Maxfield:2026ynw,
    author = "Maxfield, Henry",
    title = "{A menagerie of Schwarzians: coadjoint orbits of Virasoro and near-dS$_2$ quantum gravity}",
    eprint = "2603.18154",
    archivePrefix = "arXiv",
    primaryClass = "hep-th",
    month = "3",
    year = "2026"
}

@article{Alonso-Monsalve:2025lvt,
    author = "Alonso-Monsalve, Elba",
    title = "{The Hilbert space of gauge theories: group averaging and the quantization of Jackiw-Teitelboim gravity}",
    eprint = "2512.03030",
    archivePrefix = "arXiv",
    primaryClass = "hep-th",
    reportNumber = "MIT-CTP/5969",
    month = "12",
    year = "2025"
}

@article{Boruch:2020wax,
    author = "Boruch, Jan and Caputa, Pawel and Takayanagi, Tadashi",
    title = "{Path-Integral Optimization from Hartle-Hawking Wave Function}",
    eprint = "2011.08188",
    archivePrefix = "arXiv",
    primaryClass = "hep-th",
    reportNumber = "YITP-20-147, IPMU20-0119",
    doi = "10.1103/PhysRevD.103.046017",
    journal = "Phys. Rev. D",
    volume = "103",
    number = "4",
    pages = "046017",
    year = "2021"
}

@article{Boruch:2021hqs,
    author = "Boruch, Jan and Caputa, Pawel and Ge, Dongsheng and Takayanagi, Tadashi",
    title = "{Holographic path-integral optimization}",
    reportNumber = "YITP-21-25, IPMU21-0022",
    doi = "10.1007/JHEP07(2021)016",
    journal = "JHEP",
    volume = "07",
    pages = "016",
    year = "2021",
    note = "[Erratum: JHEP 09, 111 (2022)]"
}

@article{Caputa:2017urj,
    author = "Caputa, Pawel and Kundu, Nilay and Miyaji, Masamichi and Takayanagi, Tadashi and Watanabe, Kento",
    title = "{Anti-de Sitter Space from Optimization of Path Integrals in Conformal Field Theories}",
    eprint = "1703.00456",
    archivePrefix = "arXiv",
    primaryClass = "hep-th",
    reportNumber = "YITP-17-20, IPMU17-0039, YITP-17-20, IPMU17-0039",
    doi = "10.1103/PhysRevLett.119.071602",
    journal = "Phys. Rev. Lett.",
    volume = "119",
    number = "7",
    pages = "071602",
    year = "2017"
}

@article{Caputa:2017yrh,
    author = "Caputa, Pawel and Kundu, Nilay and Miyaji, Masamichi and Takayanagi, Tadashi and Watanabe, Kento",
    title = "{Liouville Action as Path-Integral Complexity: From Continuous Tensor Networks to AdS/CFT}",
    eprint = "1706.07056",
    archivePrefix = "arXiv",
    primaryClass = "hep-th",
    reportNumber = "YITP-17-65, IPMU17-0091, YITP-17-65, IPMU17-0091",
    doi = "10.1007/JHEP11(2017)097",
    journal = "JHEP",
    volume = "11",
    pages = "097",
    year = "2017"
}

@article{Hikida:2021ese,
    author = "Hikida, Yasuaki and Nishioka, Tatsuma and Takayanagi, Tadashi and Taki, Yusuke",
    title = "{Holography in de Sitter Space via Chern-Simons Gauge Theory}",
    eprint = "2110.03197",
    archivePrefix = "arXiv",
    primaryClass = "hep-th",
    reportNumber = "YITP-21-105; IPMU21-0059",
    doi = "10.1103/PhysRevLett.129.041601",
    journal = "Phys. Rev. Lett.",
    volume = "129",
    number = "4",
    pages = "041601",
    year = "2022"
}

@article{Maldacena:2002vr,
    author = "Maldacena, Juan Martin",
    title = "{Non-Gaussian features of primordial fluctuations in single field inflationary models}",
    eprint = "astro-ph/0210603",
    archivePrefix = "arXiv",
    doi = "10.1088/1126-6708/2003/05/013",
    journal = "JHEP",
    volume = "05",
    pages = "013",
    year = "2003"
}

@article{Hikida:2022ltr,
    author = "Hikida, Yasuaki and Nishioka, Tatsuma and Takayanagi, Tadashi and Taki, Yusuke",
    title = "{CFT duals of three-dimensional de Sitter gravity}",
    eprint = "2203.02852",
    archivePrefix = "arXiv",
    primaryClass = "hep-th",
    reportNumber = "YITP-22-20, IPMU22-0006",
    doi = "10.1007/JHEP05(2022)129",
    journal = "JHEP",
    volume = "05",
    pages = "129",
    year = "2022"
}

@article{Doi:2024nty,
    author = "Doi, Kazuki and Ogawa, Naoki and Shinmyo, Kotaro and Suzuki, Yu-ki and Takayanagi, Tadashi",
    title = "{Probing de Sitter space using CFT states}",
    eprint = "2405.14237",
    archivePrefix = "arXiv",
    primaryClass = "hep-th",
    reportNumber = "YITP-24-66",
    doi = "10.1007/JHEP02(2025)093",
    journal = "JHEP",
    volume = "02",
    pages = "093",
    year = "2025"
}

@article{Ailiga:2023wzl,
    author = {Ailiga, Manishankar and Mallik, Shubhashis and Narain, Gaurav},
    title = {{Lorentzian Robin Universe}},
    journal = {JHEP},
    volume = {01},
    pages = {124},
    year = {2024},
    doi = {10.1007/JHEP01(2024)124},
    eprint = {2308.01310},
    archivePrefix = {arXiv},
    primaryClass = {gr-qc},
}

@article{Almheiri:2014cka,
    author = {Almheiri, Ahmed and Polchinski, Joseph},
    title = {{Models of AdS$_{2}$ backreaction and holography}},
    journal = {JHEP},
    volume = {11},
    pages = {014},
    year = {2015},
    doi = {10.1007/JHEP11(2015)014},
    eprint = {1402.6334},
    archivePrefix = {arXiv},
    primaryClass = {hep-th},
}

@article{Alonso-Monsalve:2024oii,
    author = {Alonso-Monsalve, Elba and Harlow, Daniel and Jefferson, Patrick},
    title = {{Phase space of Jackiw-Teitelboim gravity with positive cosmological constant}},
    journal = {JHEP},
    volume = {03},
    pages = {008},
    year = {2026},
    doi = {10.1007/JHEP03(2026)008},
    eprint = {2409.12943},
    archivePrefix = {arXiv},
    primaryClass = {hep-th},
    reportNumber = {MIT-CTP/5769},
}

@article{Arnowitt:1962hi,
    author = {Arnowitt, Richard L. and Deser, Stanley and Misner, Charles W.},
    title = {{The Dynamics of general relativity}},
    journal = {Gen. Rel. Grav.},
    volume = {40},
    pages = {1997--2027},
    year = {2008},
    doi = {10.1007/s10714-008-0661-1},
    eprint = {gr-qc/0405109},
    archivePrefix = {arXiv},
}

@article{Banihashemi:2024aal,
    author = {Banihashemi, Batoul and Jacobson, Ted},
    title = {{On the lapse contour in the gravitational path integral}},
    journal = {Phys. Rev. D},
    volume = {111},
    number = {6},
    pages = {066014},
    year = {2025},
    doi = {10.1103/PhysRevD.111.066014},
    eprint = {2405.10307},
    archivePrefix = {arXiv},
    primaryClass = {hep-th},
}

@article{Banihashemi:2024weu,
    author = {Banihashemi, Batoul and Jacobson, Ted},
    title = {{The enigmatic gravitational partition function}},
    journal = {Gen. Rel. Grav.},
    volume = {57},
    number = {2},
    pages = {43},
    year = {2025},
    doi = {10.1007/s10714-024-03347-0},
    eprint = {2411.00267},
    archivePrefix = {arXiv},
    primaryClass = {hep-th},
}

@article{Batalin:1977pb,
    author = {Batalin, I. A. and Vilkovisky, G. A.},
    title = {{Relativistic S Matrix of Dynamical Systems with Boson and Fermion Constraints}},
    journal = {Phys. Lett. B},
    volume = {69},
    pages = {309--312},
    year = {1977},
    doi = {10.1016/0370-2693(77)90553-6},
}

@article{Batalin:1994rd,
    author = {Batalin, Igor and Marnelius, Robert},
    title = {{Solving general gauge theories on inner product spaces}},
    journal = {Nucl. Phys. B},
    volume = {442},
    pages = {669--696},
    year = {1995},
    doi = {10.1016/0550-3213(95)00141-E},
    eprint = {hep-th/9501004},
    archivePrefix = {arXiv},
    reportNumber = {GOTEBORG-94-32, ITP-G{\textbackslash}''OTEBORG-94-32, ITP-G{\textbackslash}''oteborg 94-32},
}

@article{Blommaert:2025rgw,
    author = {Blommaert, Andreas and Levine, Adam},
    title = {{Sphere amplitudes and observing the universe{\textquoteright}s size}},
    journal = {JHEP},
    volume = {07},
    pages = {136},
    year = {2026},
    doi = {10.1007/JHEP07(2026)136},
    eprint = {2505.24633},
    archivePrefix = {arXiv},
    primaryClass = {hep-th},
}

@article{Casali:2021ewu,
    author = {Casali, Eduardo and Marolf, Donald and Maxfield, Henry and Rangamani, Mukund},
    title = {{Baby universes and worldline field theories}},
    journal = {Class. Quant. Grav.},
    volume = {39},
    number = {13},
    pages = {134004},
    year = {2022},
    doi = {10.1088/1361-6382/ac37cd},
    eprint = {2101.12221},
    archivePrefix = {arXiv},
    primaryClass = {hep-th},
}

@article{Cattaneo:2015vsa,
    author = {Cattaneo, Alberto S. and Mnev, Pavel and Reshetikhin, Nicolai},
    title = {{Perturbative quantum gauge theories on manifolds with boundary}},
    journal = {Commun. Math. Phys.},
    volume = {357},
    number = {2},
    pages = {631--730},
    year = {2018},
    doi = {10.1007/s00220-017-3031-6},
    eprint = {1507.01221},
    archivePrefix = {arXiv},
    primaryClass = {math-ph},
}

@article{Chamseddine:1989yz,
    author = {Chamseddine, Ali H. and Wyler, D.},
    title = {{Gauge Theory of Topological Gravity in (1+1)-Dimensions}},
    journal = {Phys. Lett. B},
    volume = {228},
    pages = {75--78},
    year = {1989},
    doi = {10.1016/0370-2693(89)90528-5},
    reportNumber = {Print-89-0614 (ZURICH)},
}

@article{Chaudhuri:2024yau,
    author = {Chaudhuri, Soumyadeep and Ferrari, Frank},
    title = {{Finite cut-off JT and Liouville quantum gravities on the disk at one loop}},
    journal = {JHEP},
    volume = {01},
    pages = {160},
    year = {2025},
    doi = {10.1007/JHEP01(2025)160},
    eprint = {2404.03748},
    archivePrefix = {arXiv},
    primaryClass = {hep-th},
}

@article{Chen:2020tes,
    author = {Chen, Yiming and Gorbenko, Victor and Maldacena, Juan},
    title = {{Bra-ket wormholes in gravitationally prepared states}},
    journal = {JHEP},
    volume = {02},
    pages = {009},
    year = {2021},
    doi = {10.1007/JHEP02(2021)009},
    eprint = {2007.16091},
    archivePrefix = {arXiv},
    primaryClass = {hep-th},
}

@article{Chen:2025jqm,
    author = {Chen, Yiming and Stanford, Douglas and Tang, Haifeng and Yang, Zhenbin},
    title = {{On the phase of the de Sitter density of states}},
    journal = {JHEP},
    volume = {05},
    pages = {068},
    year = {2026},
    doi = {10.1007/JHEP05(2026)068},
    eprint = {2511.01400},
    archivePrefix = {arXiv},
    primaryClass = {hep-th},
}

@article{ChouNishimura:2024MonteCarlo,
    author = {Chou, Chien-Yu and Nishimura, Jun},
    title = {{Monte Carlo studies of quantum cosmology by the generalized Lefschetz thimble method}},
    journal = {JHEP},
    volume = {05},
    pages = {142},
    year = {2025},
    doi = {10.1007/JHEP05(2025)142},
    eprint = {2407.17724},
    archivePrefix = {arXiv},
    primaryClass = {gr-qc},
}

@article{Cohen:1985sm,
    author = {Cohen, Andrew G. and Moore, Gregory W. and Nelson, Philip C. and Polchinski, Joseph},
    title = {{An Off-Shell Propagator for String Theory}},
    journal = {Nucl. Phys. B},
    volume = {267},
    pages = {143--157},
    year = {1986},
    doi = {10.1016/0550-3213(86)90148-3},
    reportNumber = {HUTP-85/A058, UTTG-16-85},
}

@article{Cotler:2019nbi,
    author = {Cotler, Jordan and Jensen, Kristan and Maloney, Alexander},
    title = {{Low-dimensional de Sitter quantum gravity}},
    journal = {JHEP},
    volume = {06},
    pages = {048},
    year = {2020},
    doi = {10.1007/JHEP06(2020)048},
    eprint = {1905.03780},
    archivePrefix = {arXiv},
    primaryClass = {hep-th},
}

@article{Cotler:2024xzz,
    author = {Cotler, Jordan and Jensen, Kristan},
    title = {{Non-perturbative de Sitter Jackiw-Teitelboim gravity}},
    journal = {JHEP},
    volume = {12},
    pages = {016},
    year = {2024},
    doi = {10.1007/JHEP12(2024)016},
    eprint = {2401.01925},
    archivePrefix = {arXiv},
    primaryClass = {hep-th},
}

@article{DeWitt:1967yk,
    author = {DeWitt, Bryce S.},
    editor = {Fang, Li-Zhi and Ruffini, R.},
    title = {{Quantum Theory of Gravity. 1. The Canonical Theory}},
    journal = {Phys. Rev.},
    volume = {160},
    pages = {1113--1148},
    year = {1967},
    doi = {10.1103/PhysRev.160.1113},
}

@article{Dey:2025osp,
    author = {Dey, Indranil and Nanda, Kanhu Kishore and Roy, Akashdeep and Sake, Sunil Kumar and Trivedi, Sandip P.},
    title = {{JT gravity in de Sitter space and its extensions}},
    journal = {JHEP},
    volume = {03},
    pages = {196},
    year = {2026},
    doi = {10.1007/JHEP03(2026)196},
    eprint = {2501.03148},
    archivePrefix = {arXiv},
    primaryClass = {hep-th},
}

@article{DiazDorronsoro:2017hti,
    author = {Diaz Dorronsoro, Juan and Halliwell, Jonathan J. and Hartle, James B. and Hertog, Thomas and Janssen, Oliver},
    title = {{Real no-boundary wave function in Lorentzian quantum cosmology}},
    journal = {Phys. Rev. D},
    volume = {96},
    number = {4},
    pages = {043505},
    year = {2017},
    doi = {10.1103/PhysRevD.96.043505},
    eprint = {1705.05340},
    archivePrefix = {arXiv},
    primaryClass = {gr-qc},
}

@article{DiazDorronsoro:2018wro,
    author = {Diaz Dorronsoro, Juan and Halliwell, Jonathan J. and Hartle, James B. and Hertog, Thomas and Janssen, Oliver and Vreys, Yannick},
    title = {{Damped perturbations in the no-boundary state}},
    journal = {Phys. Rev. Lett.},
    volume = {121},
    number = {8},
    pages = {081302},
    year = {2018},
    doi = {10.1103/PhysRevLett.121.081302},
    eprint = {1804.01102},
    archivePrefix = {arXiv},
    primaryClass = {gr-qc},
}

@article{DiTucci:2019bui,
    author = {Di Tucci, Alice and Lehners, Jean-Luc and Sberna, Laura},
    title = {{No-boundary prescriptions in Lorentzian quantum cosmology}},
    journal = {Phys. Rev. D},
    volume = {100},
    number = {12},
    pages = {123543},
    year = {2019},
    doi = {10.1103/PhysRevD.100.123543},
    eprint = {1911.06701},
    archivePrefix = {arXiv},
    primaryClass = {hep-th},
}

@article{DiTucci:2019dji,
    author = {Di Tucci, Alice and Lehners, Jean-Luc},
    title = {{No-Boundary Proposal as a Path Integral with Robin Boundary Conditions}},
    journal = {Phys. Rev. Lett.},
    volume = {122},
    number = {20},
    pages = {201302},
    year = {2019},
    doi = {10.1103/PhysRevLett.122.201302},
    eprint = {1903.06757},
    archivePrefix = {arXiv},
    primaryClass = {hep-th},
}

@article{Engelsoy:2016xyb,
    author = {Engels\"oy, Julius and Mertens, Thomas G. and Verlinde, Herman},
    title = {{An investigation of AdS$_{2}$ backreaction and holography}},
    journal = {JHEP},
    volume = {07},
    pages = {139},
    year = {2016},
    doi = {10.1007/JHEP07(2016)139},
    eprint = {1606.03438},
    archivePrefix = {arXiv},
    primaryClass = {hep-th},
}

@article{FanarasVilenkin:2021JTKS,
    author = {Fanaras, Georgios and Vilenkin, Alexander},
    title = {{Jackiw-Teitelboim and Kantowski-Sachs quantum cosmology}},
    journal = {JCAP},
    volume = {03},
    pages = {056},
    year = {2022},
    doi = {10.1088/1475-7516/2022/03/056},
    eprint = {2112.00919},
    archivePrefix = {arXiv},
    primaryClass = {gr-qc},
}

@article{FanarasVilenkin:2022TunnelingKS,
    author = {Fanaras, Georgios and Vilenkin, Alexander},
    title = {{The tunneling wavefunction in Kantowski-Sachs quantum cosmology}},
    journal = {JCAP},
    volume = {08},
    pages = {069},
    year = {2022},
    doi = {10.1088/1475-7516/2022/08/069},
    eprint = {2206.05839},
    archivePrefix = {arXiv},
    primaryClass = {gr-qc},
}

@article{Feldbrugge:2017fcc,
    author = {Feldbrugge, Job and Lehners, Jean-Luc and Turok, Neil},
    title = {{No smooth beginning for spacetime}},
    journal = {Phys. Rev. Lett.},
    volume = {119},
    number = {17},
    pages = {171301},
    year = {2017},
    doi = {10.1103/PhysRevLett.119.171301},
    eprint = {1705.00192},
    archivePrefix = {arXiv},
    primaryClass = {hep-th},
}

@article{Feldbrugge:2017kzv,
    author = {Feldbrugge, Job and Lehners, Jean-Luc and Turok, Neil},
    title = {{Lorentzian Quantum Cosmology}},
    journal = {Phys. Rev. D},
    volume = {95},
    number = {10},
    pages = {103508},
    year = {2017},
    doi = {10.1103/PhysRevD.95.103508},
    eprint = {1703.02076},
    archivePrefix = {arXiv},
    primaryClass = {hep-th},
}

@article{Feldbrugge:2017mbc,
    author = {Feldbrugge, Job and Lehners, Jean-Luc and Turok, Neil},
    title = {{No rescue for the no boundary proposal: Pointers to the future of quantum cosmology}},
    journal = {Phys. Rev. D},
    volume = {97},
    number = {2},
    pages = {023509},
    year = {2018},
    doi = {10.1103/PhysRevD.97.023509},
    eprint = {1708.05104},
    archivePrefix = {arXiv},
    primaryClass = {hep-th},
}

@article{Ferrari:2024kpz,
    author = {Ferrari, Frank},
    title = {{Jackiw-Teitelboim gravity, random disks of constant curvature, self-overlapping curves, and Liouville CFT1}},
    journal = {Phys. Rev. D},
    volume = {111},
    number = {6},
    pages = {L061901},
    year = {2025},
    doi = {10.1103/PhysRevD.111.L061901},
    eprint = {2402.08052},
    archivePrefix = {arXiv},
    primaryClass = {hep-th},
}

@article{Ferrari:2024ndr,
    author = {Ferrari, Frank},
    title = {{Random disks of constant curvature: The lattice story}},
    journal = {Phys. Rev. D},
    volume = {113},
    number = {2},
    pages = {026012},
    year = {2026},
    doi = {10.1103/3gb4-y8r8},
    eprint = {2406.06875},
    archivePrefix = {arXiv},
    primaryClass = {hep-th},
}

@article{Fradkin:1975cq,
    author = {Fradkin, E. S. and Vilkovisky, G. A.},
    title = {{QUANTIZATION OF RELATIVISTIC SYSTEMS WITH CONSTRAINTS}},
    journal = {Phys. Lett. B},
    volume = {55},
    pages = {224--226},
    year = {1975},
    doi = {10.1016/0370-2693(75)90448-7},
}

@article{Fumagalli:2024msi,
    author = {Fumagalli, Alessandro and Gorbenko, Victor and Kames-King, Joshua},
    title = {{De Sitter Bra-Ket wormholes}},
    journal = {JHEP},
    volume = {05},
    pages = {074},
    year = {2025},
    doi = {10.1007/JHEP05(2025)074},
    eprint = {2408.08351},
    archivePrefix = {arXiv},
    primaryClass = {hep-th},
}

@article{Giulini:1998rk,
    author = {Giulini, Domenico and Marolf, Donald},
    title = {{On the generality of refined algebraic quantization}},
    journal = {Class. Quant. Grav.},
    volume = {16},
    pages = {2479--2488},
    year = {1999},
    doi = {10.1088/0264-9381/16/7/321},
    eprint = {gr-qc/9812024},
    archivePrefix = {arXiv},
    reportNumber = {ZU-TH-98-20, SU-GP-98-10-1},
}

@article{Goel:2020yxl,
    author = {Goel, Akash and Iliesiu, Luca V. and Kruthoff, Jorrit and Yang, Zhenbin},
    title = {{Classifying boundary conditions in JT gravity: from energy-branes to $\alpha$-branes}},
    journal = {JHEP},
    volume = {04},
    pages = {069},
    year = {2021},
    doi = {10.1007/JHEP04(2021)069},
    eprint = {2010.12592},
    archivePrefix = {arXiv},
    primaryClass = {hep-th},
}

@article{Grumiller:2002nm,
    author = {Grumiller, D. and Kummer, W. and Vassilevich, D. V.},
    title = {{Dilaton gravity in two-dimensions}},
    journal = {Phys. Rept.},
    volume = {369},
    pages = {327--430},
    year = {2002},
    doi = {10.1016/S0370-1573(02)00267-3},
    eprint = {hep-th/0204253},
    archivePrefix = {arXiv},
    reportNumber = {TUW-02-01},
}

@article{Halliwell:1988wc,
    author = {Halliwell, Jonathan J.},
    title = {{Derivation of the Wheeler-De Witt Equation from a Path Integral for Minisuperspace Models}},
    journal = {Phys. Rev. D},
    volume = {38},
    pages = {2468},
    year = {1988},
    doi = {10.1103/PhysRevD.38.2468},
    reportNumber = {NSF-ITP-88-25},
}

@article{Halliwell:2018ejl,
    author = {Halliwell, Jonathan J. and Hartle, James B. and Hertog, Thomas},
    title = {{What is the No-Boundary Wave Function of the Universe?}},
    journal = {Phys. Rev. D},
    volume = {99},
    number = {4},
    pages = {043526},
    year = {2019},
    doi = {10.1103/PhysRevD.99.043526},
    eprint = {1812.01760},
    archivePrefix = {arXiv},
    primaryClass = {hep-th},
}

@article{Harlow:2018tqv,
    author = {Harlow, Daniel and Jafferis, Daniel},
    title = {{The Factorization Problem in Jackiw-Teitelboim Gravity}},
    journal = {JHEP},
    volume = {02},
    pages = {177},
    year = {2020},
    doi = {10.1007/JHEP02(2020)177},
    eprint = {1804.01081},
    archivePrefix = {arXiv},
    primaryClass = {hep-th},
}

@article{Harlow:2023hjb,
    author = {Harlow, Daniel and Numasawa, Tokiro},
    title = {{Gauging spacetime inversions in quantum gravity}},
    journal = {JHEP},
    volume = {01},
    pages = {098},
    year = {2026},
    doi = {10.1007/JHEP01(2026)098},
    eprint = {2311.09978},
    archivePrefix = {arXiv},
    primaryClass = {hep-th},
    reportNumber = {MIT-CTP/5647},
}

@article{Harlow:2025pvj,
    author = {Harlow, Daniel and Usatyuk, Mykhaylo and Zhao, Ying},
    title = {{Quantum mechanics and observers for gravity in a closed universe}},
    year = {2025},
    month = {1},
    eprint = {2501.02359},
    archivePrefix = {arXiv},
    primaryClass = {hep-th},
    reportNumber = {MIT-CTP/5824},
}

@article{Hartle:1983ai,
    author = {Hartle, J. B. and Hawking, S. W.},
    editor = {Fang, Li-Zhi and Ruffini, R.},
    title = {{Wave Function of the Universe}},
    journal = {Phys. Rev. D},
    volume = {28},
    pages = {2960--2975},
    year = {1983},
    doi = {10.1103/PhysRevD.28.2960},
    reportNumber = {PRINT-83-0937 (CAMBRIDGE)},
}

@article{Hartle:1997dc,
    author = {Hartle, James B. and Marolf, Donald},
    title = {{Comparing formulations of generalized quantum mechanics for reparametrization - invariant systems}},
    journal = {Phys. Rev. D},
    volume = {56},
    pages = {6247--6257},
    year = {1997},
    doi = {10.1103/PhysRevD.56.6247},
    eprint = {gr-qc/9703021},
    archivePrefix = {arXiv},
    reportNumber = {UCSBTH-96-15, NSF-ITP-96-59, UCSBTH-96-15, NSF-ITP-96-59},
}

@article{Held:2024rmg,
    author = {Held, Jesse and Maxfield, Henry},
    title = {{The Hilbert space of de Sitter JT: a case study for canonical methods in quantum gravity}},
    year = {2024},
    month = {10},
    eprint = {2410.14824},
    archivePrefix = {arXiv},
    primaryClass = {hep-th},
}

@article{Held:2025mai,
    author = {Held, Jesse and Maxfield, Henry},
    title = {{Gravitational Hilbert spaces: invariant and co-invariant states, inner products, gauge-fixing, and BRST}},
    journal = {JHEP},
    volume = {07},
    pages = {021},
    year = {2026},
    doi = {10.1007/JHEP07(2026)021},
    eprint = {2509.05412},
    archivePrefix = {arXiv},
    primaryClass = {hep-th},
}

@article{Henneaux:1985nw,
    author = {Henneaux, M.},
    title = {{QUANTUM GRAVITY IN TWO-DIMENSIONS: EXACT SOLUTION OF THE JACKIW MODEL}},
    journal = {Phys. Rev. Lett.},
    volume = {54},
    pages = {959--962},
    year = {1985},
    doi = {10.1103/PhysRevLett.54.959},
}

@book{Henneaux:1992ig,
    author = {Henneaux, M. and Teitelboim, C.},
    title = {{Quantization of gauge systems}},
    year = {1992},
    isbn = {978-0-691-03769-1},
}

@article{Hoehn:2019fsy,
    author = {Hoehn, Philipp A. and Smith, Alexander R. H. and Lock, Maximilian P. E.},
    title = {{Trinity of relational quantum dynamics}},
    journal = {Phys. Rev. D},
    volume = {104},
    number = {6},
    pages = {066001},
    year = {2021},
    doi = {10.1103/PhysRevD.104.066001},
    eprint = {1912.00033},
    archivePrefix = {arXiv},
    primaryClass = {quant-ph},
}

@article{Honda:2024hdr,
    author = {Honda, Masazumi and Matsui, Hiroki and Numajiri, Kota and Okabayashi, Kazumasa},
    title = {{Jackiw-Teitelboim gravity and Lorentzian quantum cosmology}},
    journal = {Phys. Rev. D},
    volume = {111},
    number = {12},
    pages = {126022},
    year = {2025},
    doi = {10.1103/vrcy-y61l},
    eprint = {2412.20398},
    archivePrefix = {arXiv},
    primaryClass = {gr-qc},
    reportNumber = {RIKEN-iTHEMS-Report-24, STUPP-24-272, YITP-24-134},
}

@article{Honda:2024Resurgence,
    author = {Honda, Masazumi and Matsui, Hiroki and Okabayashi, Kazumasa and Terada, Takahiro},
    title = {{Resurgence in Lorentzian quantum cosmology: No-boundary saddles and resummation of quantum gravity corrections around tunneling saddle points}},
    journal = {Phys. Rev. D},
    volume = {110},
    number = {8},
    pages = {083508},
    year = {2024},
    doi = {10.1103/PhysRevD.110.083508},
    eprint = {2402.09981},
    archivePrefix = {arXiv},
    primaryClass = {gr-qc},
}

@article{Honda:2026Dimensional,
    author = {Honda, Masazumi and Matsui, Hiroki and Numajiri, Kota and Okabayashi, Kazumasa and Terada, Takahiro},
    title = {{Dimensional (In)dependence of Lorentzian Quantum Cosmology}},
    year = {2026},
    month = {9},
    eprint = {2609.33452},
    archivePrefix = {arXiv},
    primaryClass = {gr-qc},
    reportNumber = {RIKEN-iTHEMS-Report-26, STUPP-26-304, YITP-26-126, OCU-PHYS-631, AP-GR-215},
}

@article{HondaShinmyo:2026Stokes,
    author = {Honda, Masazumi and Shinmyo, Kotaro},
    title = {{Stokes Phenomena between AdS/CFT and dS/CFT}},
    year = {2026},
    month = {9},
    eprint = {2609.10677},
    archivePrefix = {arXiv},
    primaryClass = {hep-th},
    reportNumber = {RIKEN-iTHEMS-Report-26, STUPP-26-301, YITP-26-123},
}

@article{Isler:1989hq,
    author = {Isler, K. and Trugenberger, C. A.},
    title = {{A Gauge Theory of Two-dimensional Quantum Gravity}},
    journal = {Phys. Rev. Lett.},
    volume = {63},
    pages = {834},
    year = {1989},
    doi = {10.1103/PhysRevLett.63.834},
    reportNumber = {MIT-CTP-1739},
}

@article{Ivo:2025yek,
    author = {Ivo, Victor and Maldacena, Juan and Sun, Zimo},
    title = {{Physical instabilities and the phase of the Euclidean path integral}},
    journal = {JHEP},
    volume = {04},
    pages = {118},
    year = {2026},
    doi = {10.1007/JHEP04(2026)118},
    eprint = {2504.00920},
    archivePrefix = {arXiv},
    primaryClass = {hep-th},
}

@article{Jackiw:1984je,
    author = {Jackiw, Roman},
    editor = {Baier, R. and Satz, H.},
    title = {{Lower Dimensional Gravity}},
    journal = {Nucl. Phys. B},
    volume = {252},
    pages = {343--356},
    year = {1985},
    doi = {10.1016/0550-3213(85)90448-1},
    reportNumber = {MIT-CTP-1203},
}

@article{Jafferis:2022uhu,
    author = {Jafferis, Daniel Louis and Kolchmeyer, David K. and Mukhametzhanov, Baur and Sonner, Julian},
    title = {{Matrix Models for Eigenstate Thermalization}},
    journal = {Phys. Rev. X},
    volume = {13},
    number = {3},
    pages = {031033},
    year = {2023},
    doi = {10.1103/PhysRevX.13.031033},
    eprint = {2209.02130},
    archivePrefix = {arXiv},
    primaryClass = {hep-th},
}

@article{Jafferis:2022wez,
    author = {Jafferis, Daniel Louis and Kolchmeyer, David K. and Mukhametzhanov, Baur and Sonner, Julian},
    title = {{Jackiw-Teitelboim gravity with matter, generalized eigenstate thermalization hypothesis, and random matrices}},
    journal = {Phys. Rev. D},
    volume = {108},
    number = {6},
    pages = {066015},
    year = {2023},
    doi = {10.1103/PhysRevD.108.066015},
    eprint = {2209.02131},
    archivePrefix = {arXiv},
    primaryClass = {hep-th},
}

@article{Janssen:2019sex,
    author = {Janssen, Oliver and Halliwell, Jonathan J. and Hertog, Thomas},
    title = {{No-boundary proposal in biaxial Bianchi IX minisuperspace}},
    journal = {Phys. Rev. D},
    volume = {99},
    number = {12},
    pages = {123531},
    year = {2019},
    doi = {10.1103/PhysRevD.99.123531},
    eprint = {1904.11602},
    archivePrefix = {arXiv},
    primaryClass = {gr-qc},
}

@article{Kitaev:2017awl,
    author = {Kitaev, Alexei and Suh, S. Josephine},
    title = {{The soft mode in the Sachdev-Ye-Kitaev model and its gravity dual}},
    journal = {JHEP},
    volume = {05},
    pages = {183},
    year = {2018},
    doi = {10.1007/JHEP05(2018)183},
    eprint = {1711.08467},
    archivePrefix = {arXiv},
    primaryClass = {hep-th},
}

@article{Kontsevich:2021dmb,
    author = {Kontsevich, Maxim and Segal, Graeme},
    title = {{Wick Rotation and the Positivity of Energy in Quantum Field Theory}},
    journal = {Quart. J. Math. Oxford Ser.},
    volume = {72},
    number = {1-2},
    pages = {673--699},
    year = {2021},
    doi = {10.1093/qmath/haab027},
    eprint = {2105.10161},
    archivePrefix = {arXiv},
    primaryClass = {hep-th},
}

@article{Kugo:1977zq,
    author = {Kugo, Taichiro and Ojima, Izumi},
    title = {{Manifestly Covariant Canonical Formulation of Yang-Mills Field Theories: Physical State Subsidiary Conditions and Physical S Matrix Unitarity}},
    journal = {Phys. Lett. B},
    volume = {73},
    pages = {459--462},
    year = {1978},
    doi = {10.1016/0370-2693(78)90765-7},
    reportNumber = {Print-77-0717 (KYOTO)},
}

@article{Kugo:1979gm,
    author = {Kugo, Taichiro and Ojima, Izumi},
    title = {{Local Covariant Operator Formalism of Nonabelian Gauge Theories and Quark Confinement Problem}},
    journal = {Prog. Theor. Phys. Suppl.},
    volume = {66},
    pages = {1--130},
    year = {1979},
    doi = {10.1143/PTPS.66.1},
    reportNumber = {KUNS-493},
}

@article{Louis-Martinez:1993bge,
    author = {Louis-Martinez, D. and Gegenberg, J. and Kunstatter, G.},
    title = {{Exact Dirac quantization of all 2-D dilaton gravity theories}},
    journal = {Phys. Lett. B},
    volume = {321},
    pages = {193--198},
    year = {1994},
    doi = {10.1016/0370-2693(94)90463-4},
    eprint = {gr-qc/9309018},
    archivePrefix = {arXiv},
    reportNumber = {WIN-93-06, UNB-93-03},
}

@article{Maldacena:2001kr,
    author = {Maldacena, Juan Martin},
    title = {{Eternal black holes in anti-de Sitter}},
    journal = {JHEP},
    volume = {04},
    pages = {021},
    year = {2003},
    doi = {10.1088/1126-6708/2003/04/021},
    eprint = {hep-th/0106112},
    archivePrefix = {arXiv},
    reportNumber = {NSF-ITP-01-59},
}

@article{Maldacena:2016upp,
    author = {Maldacena, Juan and Stanford, Douglas and Yang, Zhenbin},
    title = {{Conformal symmetry and its breaking in two dimensional Nearly Anti-de-Sitter space}},
    journal = {PTEP},
    volume = {2016},
    number = {12},
    pages = {12C104},
    year = {2016},
    doi = {10.1093/ptep/ptw124},
    eprint = {1606.01857},
    archivePrefix = {arXiv},
    primaryClass = {hep-th},
}

@article{Maldacena:2019cbz,
    author = {Maldacena, Juan and Turiaci, Gustavo J. and Yang, Zhenbin},
    title = {{Two dimensional Nearly de Sitter gravity}},
    journal = {JHEP},
    volume = {01},
    pages = {139},
    year = {2021},
    doi = {10.1007/JHEP01(2021)139},
    eprint = {1904.01911},
    archivePrefix = {arXiv},
    primaryClass = {hep-th},
}

@article{Marolf:1995cn,
    author = {Marolf, Donald},
    title = {{Refined algebraic quantization: Systems with a single constraint}},
    year = {1995},
    month = {8},
    eprint = {gr-qc/9508015},
    archivePrefix = {arXiv},
    reportNumber = {UCSBTH-95-16},
}

@inproceedings{Marolf:2000iq,
    author = {Marolf, Donald},
    title = {{Group averaging and refined algebraic quantization: Where are we now?}},
    booktitle = {{9th Marcel Grossmann Meeting on Recent Developments in Theoretical and Experimental General Relativity, Gravitation and Relativistic Field Theories (MG 9)}},
    year = {2000},
    month = {7},
    eprint = {gr-qc/0011112},
    archivePrefix = {arXiv},
    reportNumber = {SU-GP-11-2},
}

@article{Matsui:2023tkw,
    author = {Matsui, Hiroki},
    title = {{Summing over non-singular paths in quantum cosmology}},
    journal = {Class. Quant. Grav.},
    volume = {41},
    number = {5},
    pages = {055005},
    year = {2024},
    doi = {10.1088/1361-6382/ad1fc9},
    eprint = {2304.12024},
    archivePrefix = {arXiv},
    primaryClass = {gr-qc},
    reportNumber = {YITP-23-55},
}

@article{Matsui:2025Determinants,
    author = {Matsui, Hiroki},
    title = {{Functional determinants for constrained path integrals in minisuperspace Jackiw-Teitelboim gravity}},
    journal = {Phys. Rev. D},
    volume = {113},
    number = {6},
    pages = {066003},
    year = {2026},
    doi = {10.1103/jspm-lswp},
    eprint = {2512.21549},
    archivePrefix = {arXiv},
    primaryClass = {gr-qc},
}

@article{Matsui:2026pax,
    author = {Matsui, Hiroki},
    title = {{Exact DeWitt kernels in Lorentzian quantum cosmology}},
    year = {2026},
    month = {9},
    eprint = {2609.21585},
    archivePrefix = {arXiv},
    primaryClass = {gr-qc},
}

@article{Miyaji:2021lcq,
    author = {Miyaji, Masamichi},
    title = {{Island for gravitationally prepared state and pseudo entanglement wedge}},
    journal = {JHEP},
    volume = {12},
    pages = {013},
    year = {2021},
    doi = {10.1007/JHEP12(2021)013},
    eprint = {2109.03830},
    archivePrefix = {arXiv},
    primaryClass = {hep-th},
}

@article{Moitra:2019bub,
    author = {Moitra, Upamanyu and Sake, Sunil Kumar and Trivedi, Sandip P. and Vishal, V.},
    title = {{Jackiw-Teitelboim Gravity and Rotating Black Holes}},
    journal = {JHEP},
    volume = {11},
    pages = {047},
    year = {2019},
    doi = {10.1007/JHEP11(2019)047},
    eprint = {1905.10378},
    archivePrefix = {arXiv},
    primaryClass = {hep-th},
    reportNumber = {TIFR/TH/19-17},
}

@article{Moitra:2021uiv,
    author = {Moitra, Upamanyu and Sake, Sunil Kumar and Trivedi, Sandip P.},
    title = {{Jackiw-Teitelboim gravity in the second order formalism}},
    journal = {JHEP},
    volume = {10},
    pages = {204},
    year = {2021},
    doi = {10.1007/JHEP10(2021)204},
    eprint = {2101.00596},
    archivePrefix = {arXiv},
    primaryClass = {hep-th},
    reportNumber = {TIFR/TH/20-53},
}

@article{Moitra:2022glw,
    author = {Moitra, Upamanyu and Sake, Sunil Kumar and Trivedi, Sandip P.},
    title = {{Aspects of Jackiw-Teitelboim gravity in Anti-de Sitter and de Sitter spacetime}},
    journal = {JHEP},
    volume = {06},
    pages = {138},
    year = {2022},
    doi = {10.1007/JHEP06(2022)138},
    eprint = {2202.03130},
    archivePrefix = {arXiv},
    primaryClass = {hep-th},
    reportNumber = {TIFR/TH/22-1},
}

@article{Nanda:2023wne,
    author = {Nanda, Kanhu Kishore and Sake, Sunil Kumar and Trivedi, Sandip P.},
    title = {{JT gravity in de Sitter space and the problem of time}},
    journal = {JHEP},
    volume = {02},
    pages = {145},
    year = {2024},
    doi = {10.1007/JHEP02(2024)145},
    eprint = {2307.15900},
    archivePrefix = {arXiv},
    primaryClass = {hep-th},
}

@article{Penington:2023dql,
    author = {Penington, Geoff and Witten, Edward},
    title = {{Algebras and States in JT Gravity}},
    year = {2023},
    month = {1},
    eprint = {2301.07257},
    archivePrefix = {arXiv},
    primaryClass = {hep-th},
}

@article{Polchinski:1988ua,
    author = {Polchinski, Joseph},
    title = {{The phase of the sum over spheres}},
    journal = {Phys. Lett. B},
    volume = {219},
    pages = {251--257},
    year = {1989},
    doi = {10.1016/0370-2693(89)90387-0},
    reportNumber = {UTTG-31-88},
}

@article{Saad:2019lba,
    author = {Saad, Phil and Shenker, Stephen H. and Stanford, Douglas},
    title = {{JT gravity as a matrix integral}},
    year = {2019},
    month = {3},
    eprint = {1903.11115},
    archivePrefix = {arXiv},
    primaryClass = {hep-th},
}

@article{Schaller:1993np,
    author = {Schaller, Peter and Strobl, Thomas},
    title = {{Diffeomorphisms versus nonAbelian gauge transformations: An Example of (1+1)-dimensional gravity}},
    journal = {Phys. Lett. B},
    volume = {337},
    pages = {266--270},
    year = {1994},
    doi = {10.1016/0370-2693(94)90974-1},
    eprint = {hep-th/9401110},
    archivePrefix = {arXiv},
    reportNumber = {TUW-93-25},
}

@article{Strominger:2001pn,
    author = {Strominger, Andrew},
    title = {{The dS / CFT correspondence}},
    journal = {JHEP},
    volume = {10},
    pages = {034},
    year = {2001},
    doi = {10.1088/1126-6708/2001/10/034},
    eprint = {hep-th/0106113},
    archivePrefix = {arXiv},
}

@article{Teitelboim:1983fh,
    author = {Teitelboim, Claudio},
    title = {{Causality Versus Gauge Invariance in Quantum Gravity and Supergravity}},
    journal = {Phys. Rev. Lett.},
    volume = {50},
    pages = {705},
    year = {1983},
    doi = {10.1103/PhysRevLett.50.705},
    reportNumber = {Print-83-0131 (TEXAS)},
}

@article{Teitelboim:1983ux,
    author = {Teitelboim, C.},
    title = {{Gravitation and Hamiltonian Structure in Two Space-Time Dimensions}},
    journal = {Phys. Lett. B},
    volume = {126},
    pages = {41--45},
    year = {1983},
    doi = {10.1016/0370-2693(83)90012-6},
}

@article{Vilenkin:2018dch,
    author = {Vilenkin, Alexander and Yamada, Masaki},
    title = {{Tunneling wave function of the universe}},
    journal = {Phys. Rev. D},
    volume = {98},
    number = {6},
    pages = {066003},
    year = {2018},
    doi = {10.1103/PhysRevD.98.066003},
    eprint = {1808.02032},
    archivePrefix = {arXiv},
    primaryClass = {gr-qc},
}

@article{Vilenkin:2018oja,
    author = {Vilenkin, Alexander and Yamada, Masaki},
    title = {{Tunneling wave function of the universe II: the backreaction problem}},
    journal = {Phys. Rev. D},
    volume = {99},
    number = {6},
    pages = {066010},
    year = {2019},
    doi = {10.1103/PhysRevD.99.066010},
    eprint = {1812.08084},
    archivePrefix = {arXiv},
    primaryClass = {gr-qc},
}

@article{WIP,
    author = {Miyaji, M. and Sake, S. and Shinmyo, K.},
    title = {{Work in progress}},
}

@article{Witten:1991we,
    author = {Witten, Edward},
    title = {{On quantum gauge theories in two-dimensions}},
    journal = {Commun. Math. Phys.},
    volume = {141},
    pages = {153--209},
    year = {1991},
    doi = {10.1007/BF02100009},
}

@article{Witten:2021nzp,
    author = {Witten, Edward},
    title = {{A Note On Complex Spacetime Metrics}},
    year = {2021},
    month = {11},
    eprint = {2111.06514},
    archivePrefix = {arXiv},
    primaryClass = {hep-th},
}

@article{Witten:2022xxp,
    author = {Witten, Edward},
    title = {{A note on the canonical formalism for gravity}},
    journal = {Adv. Theor. Math. Phys.},
    volume = {27},
    number = {1},
    pages = {311--380},
    year = {2023},
    doi = {10.4310/ATMP.2023.v27.n1.a6},
    eprint = {2212.08270},
    archivePrefix = {arXiv},
    primaryClass = {hep-th},
}

\end{document}